\documentclass[JHEP,12pt]{article}
\pdfoutput=1
\usepackage{floatrow} 
\usepackage{subcaption}
\usepackage{caption}
\usepackage[multiple]{footmisc}
\usepackage{float}

\usepackage{graphics,color,soul}
\usepackage{graphicx}
\usepackage{amssymb}
\usepackage{bm}
\usepackage{jheppub}
\usepackage{lmodern,mathtools}
\usepackage{caption}
\usepackage {subcaption}
\usepackage{booktabs}
\usepackage[english]{babel}
\usepackage{amsmath,amssymb,amsbsy,amstext, amsthm, simplewick}
\usepackage{hyperref}
\usepackage{tikz}
\usepackage{pgfplots}
\usetikzlibrary{decorations.markings}
\usetikzlibrary{decorations.pathmorphing}
\tikzset{snake it/.style={decorate, decoration=snake}}
\usepackage{xcolor}
\hypersetup{
    colorlinks=true,
    linkcolor=orange,
    filecolor=magenta,      
    urlcolor=cyan,
    }

\usepackage{caption}
\usepackage{placeins}

\usepackage{amsfonts}
\usepackage{amssymb}
\usepackage{upgreek}
\usepackage{simplewick}
 \usepackage{exscale,relsize}
\usepackage{mathtools}
\usepackage{comment}

\usepackage[margin=1cm,labelfont={sf,bf,scriptsize},textfont={sf,scriptsize}]{caption}

\newcommand{\namedref}[2]{\hyperref[#2]{#1~\ref*{#2}}}
\newcommand{\secref}[1]{\namedref{Section}{#1}}
\newcommand{\appref}[1]{\namedref{Appendix}{#1}}
\newcommand{\tabref}[1]{\namedref{Table}{#1}}
\newcommand{\figref}[1]{\namedref{Figure}{#1}}

\def\cH{\mathcal{H}}

\def\cN{\mathcal{N}}
\def\cO{\mathcal{O}}

\def\cP{\mathcal{P}}

\def\tr{{\rm tr}}

\def\z{\bm{z}}
\def\zb{\bar{\bm{z}}}
\def\OO{O}

\def \la {\langle}
\def \ra {\rangle}

\def\nn{{\nonumber}}

\def\be#1\ee{\begin{align}#1\end{align}}

\DeclareMathOperator{\SU}{SU}
\DeclareMathOperator{\SO}{SO}

\let\Re\relax
\DeclareMathOperator{\Re}{Re}

\def\gym{\text{g}_{\text{YM}}}

\def\st{\text{s.t. }}

\begin{document}

\title{Conformal collider bootstrap in ${\mathcal N}=4$ SYM
}

\author{Ross Dempsey,$^{a}$}
\author{Robin Karlsson,$^{b}$}
\author{Silviu S. Pufu,$^{c,d}$}
\author{Zahra Zahraee,$^{e}$}
\author{Alexander Zhiboedov$^{f}$}

\affiliation{$^{a}$Center for Theoretical Physics -- a Leinweber Institute, Massachusetts Institute of Technology, Cambridge, MA 02139, USA}
\affiliation{$^{b}$Mathematical Institute, University of Oxford, Andrew Wiles Building, Radcliffe Observatory Quarter, Woodstock Road, Oxford, OX2 6GG, UK}
\affiliation{$^{c}$Joseph Henry Laboratories, Princeton University, Princeton, NJ 08544, USA}
\affiliation{$^{d}$Princeton Center for Theoretical Science, Princeton University, Princeton, NJ 08544, USA}
\affiliation{$^{e}$Perimeter Institute for Theoretical Physics, Waterloo, Ontario N2L 2Y5, Canada}
\affiliation{$^{f}$CERN, Theoretical Physics Department, CH-1211 Geneva 23, Switzerland}

\abstract{
We use a combination of perturbation theory, holography, supersymmetric localization, integrability, and numerical conformal bootstrap methods to constrain the energy-energy correlator in $\SU(N_c)$ ${\mathcal N}=4$ SYM at finite coupling. For finite $N_c$, we derive lower bounds on the second and fourth multipoles of the energy-energy correlator at different couplings, along with a smeared energy-energy correlator as a function of the angle between the two detectors. We present evidence that our lower bounds on the multipoles are nearly saturated by the ${\cal N} = 4$ SYM theory. In the planar limit, we further use dispersive functionals to obtain tight two-sided bounds on both the first three non-trivial multipoles and on the angular dependence of the energy-energy correlator.  As the coupling is varied from weak to strong, the energy-energy correlator exhibits a transition from single-trace to double-trace operator dominance in the collinear limit, which we characterize quantitatively. A similar phenomenon occurs in QCD, 
where a parton-hadron transition is observed as detectors are brought closer together.
}

\begin{flushleft}
\hfill \parbox[c]{40mm}{CERN-TH-2025-212}\\
\hfill \parbox[c]{40mm}{MIT-CTP/5979} \\
\hfill \parbox[c]{40mm}{PUPT-2659}
\end{flushleft}
\maketitle

\section{Introduction}

The energy-energy correlator (EEC) \cite{Basham:1978zq,Basham:1978bw} in a high-energy state in Quantum Chromodynamics (QCD) exhibits a remarkable transition as a function of the angle $\theta$ between the two detectors. For small angles $\theta \ll 1$ that are nevertheless much larger than $\Lambda_\text{QCD} / Q$, with $\Lambda_\text{QCD}$ being the QCD mass scale and $Q$ the jet momentum, the EEC has a power-like dependence on $\theta$ that is controlled by the asymptotically-free theory of quarks and gluons. As the angle becomes of order $\Lambda_\text{QCD}/Q$, the angle-dependent effective Yang-Mills coupling  grows, and the theory undergoes confinement. Eventually, for angles  $\theta \ll \Lambda_\text{QCD}/Q$, the EEC becomes flat, with the relevant degrees of freedom being essentially free hadrons \cite{Komiske:2022enw,Lee:2022uwt,Electron-PositronAlliance:2025fhk}. 
The details of the transition between the power-law and flat behaviors are not available at present, because they require understanding QCD physics at intermediate and strong effective couplings.  To understand this transition better, in this work we study a related problem:  we explore the EEC in a finite-energy state of ${\cal N} = 4$ super-Yang-Mills (SYM) theory as a function of the Yang-Mills coupling. 

A significant difference between ${\cal N} = 4$ SYM theory and QCD is that ${\cal N} = 4$ SYM theory is a conformal field theory (CFT) \cite{Grisaru:1980nk, Grisaru:1980jc, Caswell:1980yi, Caswell:1980ru, Sohnius:1981sn,Brink:1982wv,Mandelstam:1982cb}, while QCD is not.  Consequently, in ${\cal N} = 4$ SYM, the Yang-Mills coupling $g_\text{YM}$ is an exactly marginal parameter of the theory that does not undergo renormalization group running.  Thus, one cannot observe a transition analogous to that in QCD by simply changing the angle between the detectors, because such a change in the angle does not change the effective coupling.  Nevertheless, as observed in \cite{Hofman:2008ar,Kologlu:2019mfz} in the planar limit, a similar transition between power-law and flat dependence on the angle $\theta$ does occur in ${\cal N} = 4$ SYM theory as we dial the 't Hooft coupling $\lambda = g_\text{YM}^2 N_c$ by hand from weak to strong values.  This similarity between the physics of the EEC between ${\cal N}=4$ SYM and QCD is part of the motivation for our present work.  Additional motivation stems from the recent rich interplay between the study of energy correlations in QCD and general conformal field theory (CFT) developments---see, for example,~\cite{Kravchuk:2018htv,Kologlu:2019bco,Kologlu:2019mfz,Chen:2020vvp,Lee:2022ige,Chen:2022jhb,Chen:2023zzh,Chen:2024nyc,Chen:2025rjc,Chang:2025zib,Chang:2025kgq} (for a recent review, see~\cite{Moult:2025nhu}).

\begin{figure}
    \centering    \includegraphics[width=0.8\textwidth]{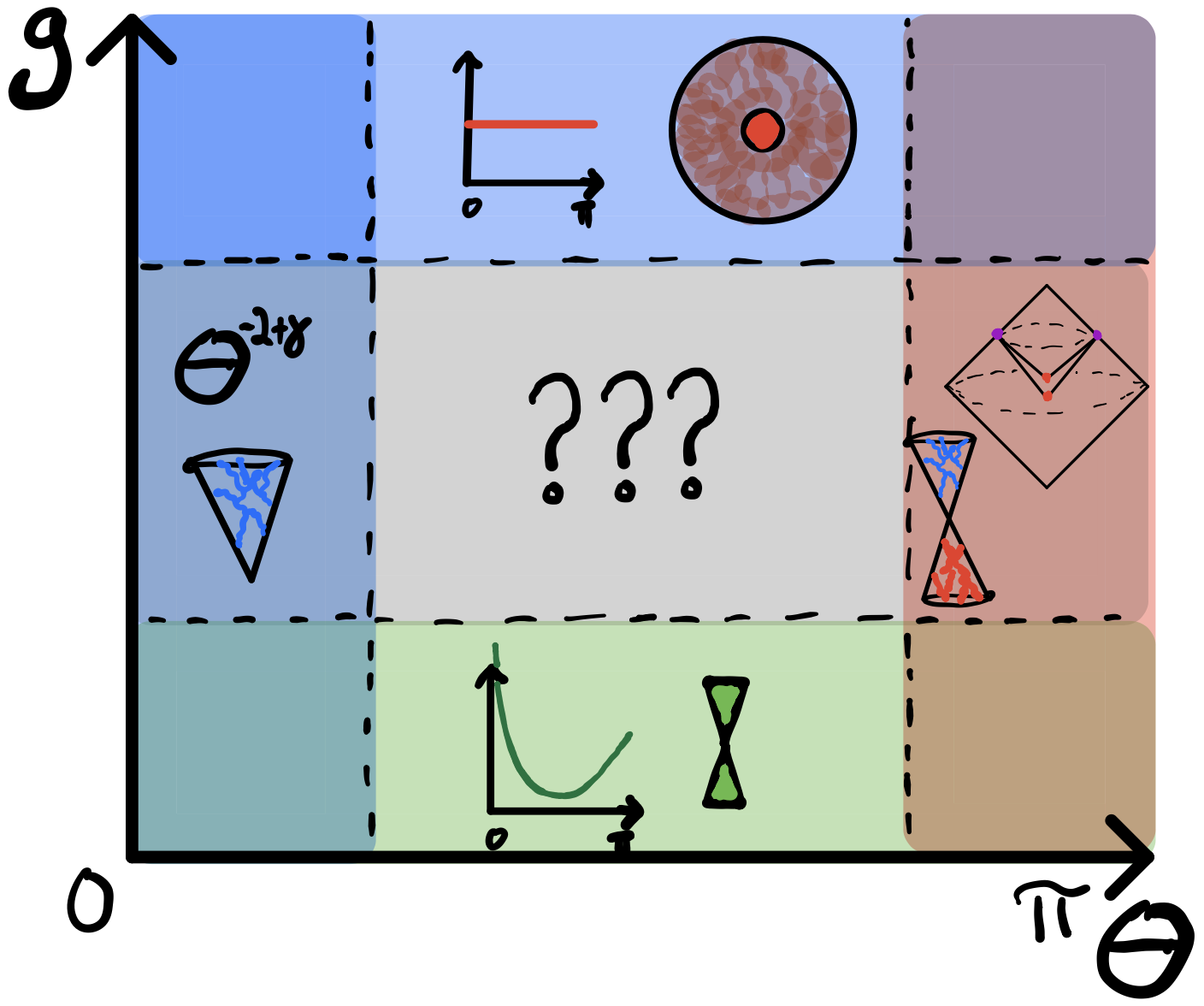}
    \caption{Schematic overview of the different regimes of the EEC in a conformal collider experiment in the planar limit of $\mathcal{N}=4$ SYM. At weak coupling (the bottom region of the diagram), the EEC is known to three loops  \cite{Engelund:2012re,Belitsky:2013xxa,Belitsky:2013ofa,Henn:2019gkr}, 
    while at strong coupling (the top region of the diagram) there is a homogeneous distribution with small inhomogeneities on top of it \cite{Hofman:2008ar}. At small angles (the left region of the diagram), the light-ray OPE predicts a power-law behavior $\theta^{-2+\gamma}$ with a power determined by the anomalous dimension of a spin-$3$ twist-$2$ operator \cite{Hofman:2008ar,Kologlu:2019mfz}. The value of $\gamma$ is known from integrability, and in particular $\gamma>2$ for $g>g_{\text{cr}}\sim 0.81$. Above $g_{\text{cr}}$, there is a transition where the twist-$4$ double-trace operator dominates and contributes analytic terms. In the back-to-back limit (the right region of the diagram), large logarithms invalidate the perturbative expansion and require resummation that renders the back-to-back limit finite. This regime was studied in \cite{Korchemsky:2019nzm,Chen:2023wah,Chen:2023amz} using the known behavior of the local correlator in the double-lightcone limit.  Notice that as the coupling constant increases, the contribution of twist-4 and higher operators becomes more important rendering the analysis of the papers above inaccurate. In this work, we explore the previously inaccessible central region using the conformal bootstrap, supplemented by input from integrability and supersymmetric localization.}
    \label{fig:review}
\end{figure}

In more detail, in the planar limit of ${\cal N} = 4$ SYM theory, the relevant coupling constant is the 't Hooft coupling $\lambda$.  For easy comparison with the integrability literature, when discussing the planar limit, we will use a coupling constant $g$ related to $\lambda$ via $(4\pi g)^2=\lambda$.  At small $g$, the EEC is similar to the one in perturbative QCD;  it was computed up to three loops at weak coupling \cite{Engelund:2012re,Belitsky:2013xxa,Belitsky:2013ofa,Henn:2019gkr}, and it exhibits a power-like singularity at small angles \cite{Hofman:2008ar,Kologlu:2019mfz} corresponding to collimated fluxes of energy (jets).  By contrast, at large coupling, the EEC can be computed using the gauge/gravity duality \cite{Maldacena:1997re, Witten:1998qj, Gubser:1998bc} in the supergravity approximation, and it becomes flat \cite{Hofman:2008ar}, corresponding to an approximately homogeneous distribution of energy with small inhomogeneities suppressed by $\sim 1/g^2$.  The transition between the two behaviors can be understood from the light-ray operator product expansion (OPE)\@.  In the light-ray OPE, the relevant operators that control the weak-coupling behavior are single-trace operators, while the strong-coupling behavior is controlled by double-trace operators.\footnote{In the source-detector OPE channel, the strong coupling result is dominated by low-spin EEC multipoles, whereas the weak coupling result is dominated by higher-spin EEC multipoles.}
At intermediate couplings, single-trace and double-trace operators both contribute to the light-ray OPE, and no perturbative methods are applicable, just as is the case in the transition region in QCD\@. The relationship between the single-trace/double-trace transition and the parton/hadron transition was elucidated by Polchinski and Strassler in the context of deep inelastic scattering in \cite{Polchinski:2002jw}. For a schematic compact summary of the different regimes for the EEC in the planar limit of ${\cal N} = 4$ SYM, see~\figref{fig:review}.\footnote{For a discussion of various physical regimes of the energy correlators in QCD, see \cite{Electron-PositronAlliance:2025fhk,ianupcoming}.} 

Our setup is as follows.  The energy detector is defined by placing the stress-energy tensor $T_{\mu \nu}(t, \vec x)$ at null infinity, and integrating it over the retarded time $u=t-r$ (see \cite{Sveshnikov:1995vi,Korchemsky:1999kt}): 
\be\label{eq:DefDetector}
{\mathcal E}(\vec{n}) = \frac{1}{4}  \int_{-\infty}^\infty du\, \underset{r\to\infty}{\lim} r^{2}\, T_{\mu\nu}(u+r,r\vec{n})\bar{n}^{\mu}\bar{n}^\nu \,,
\ee 
where $\vec{n}$ (with $\vec{n}^2=1$) is a unit vector in $\mathbb{R}^3$ pointing in the direction in which the energy flux is being measured, and $\bar{n}^{\mu}=(1,-\vec{n})$.\footnote{We work in mostly plus signature.}  The energy detector \eqref{eq:DefDetector} measures the total energy $dE / d\Omega$ per unit solid angle that arrives at future infinity in the direction $\vec{n}$. We will consider two such energy detectors whose corresponding unit vectors $\vec{n}_1$ and $\vec{n}_2$ make an angle $\theta$ with each other.  

The state in which we measure the correlations between energy fluxes will be taken to be a state of definite four-momentum $p^\mu$ created by acting on the vacuum with an appropriate Fourier mode of a local operator.  The local operator ${\cal O}(x)$ we consider is a dimension-two scalar operator in the same superconformal multiplet as the stress-energy tensor, normalized so that its two-point Wightman function in the vacuum is 
 \be \label{eq:opNormalization}
   \langle 0 | {\cal O}^\dagger(x) {\cal O}(0) | 0 \rangle 
    = \frac{1}{\left[\vec{x}^2 - (t - i \epsilon)^2 \right]^2} \,,
 \ee
where $x^\mu = (t, \vec{x})$.  We will specify the precise form of the source operator ${\cal O}$ in the next section.  We now consider the state $|{\mathcal O}(p)\rangle$ of four-momentum $p^\mu$ defined as
\be\label{eq:DefState}
|{\mathcal O}(p)\rangle &\equiv \frac{1}{\sqrt{2\pi^3}}\int d^4 x\, e^{ip \cdot x}{\mathcal O}(x)|0\ra \,.
\ee
It can be verified that such a state has non-zero norm provided that $p^0 > |\vec{p}|$, which implies that the energy $p^0$ is positive and that the four-momentum $p^\mu$ is timelike.\footnote{Using the identity, ${1 \over 2 \pi^3} \int d^4x {e^{- i p \cdot x} \over (-(t-i\epsilon)^2 + \vec x^2)^2} = \theta(p^0)\theta(-p^2)$, it can be checked that  $\langle {\mathcal O}(p') | {\mathcal O}(p)\rangle = (2 \pi)^4 \delta^4(p-p') \theta(p^0) \theta(-p^2)$. }  Since $p^\mu$ is timelike, we can pass to a frame where $p^\mu = (p^0, \vec{0})$;  we will refer to this frame as the rest frame of the source. 
In this frame, the object of study, the energy-energy correlator, is defined by 
\be\label{eq:DefEEC}
\langle {\mathcal E}(\vec{n}_1){\mathcal E}(\vec{n}_2) \rangle \equiv {\la {\mathcal O}(p)|{\mathcal E}(\vec{n}_1){\mathcal E}(\vec{n}_2)|\mathcal{O}(p)\ra \over \la {\mathcal O}(p) |\mathcal{O}(p)\ra} \equiv \left( \frac{p^0}{4\pi} \right)^2   \text{EEC}(\theta) \,,
\ee 
where $\theta = \arccos (\vec{n}_1 \cdot \vec{n}_2)$ is the angle between the two detectors.\footnote{Since we take the initial state to be created by a scalar operator, our choice $p^\mu = (p^0, \vec{0})$ guarantees that the state $|{\cal O}(p)\rangle$ is rotationally invariant. That is why the energy-energy correlator depends only on the relative angle between detectors.} 
In \eqref{eq:DefEEC}, we defined the dimensionless quantity $\text{EEC}(\theta)$ by extracting a factor of $p^0 / (4 \pi)$ from $\langle {\mathcal E}(\vec{n}_1){\mathcal E}(\vec{n}_2) \rangle$ for each detector operator. As any function of the relative angle $\theta$, the EEC can be expanded in Legendre polynomials (or ``zonal spherical harmonics'').  The Ward identities fix the coefficients of the terms proportional to the degree $0$ and $1$ Legendre polynomials to be $1$ and $0$, respectively, and the Legendre decomposition of the EEC takes the form
\be\label{eq:leg}
	{\rm EEC}(\theta) = 1+\sum_{s=2}^\infty c_s P_s(\cos\theta) \geq 0\,, \qquad c_s\geq 0 \,,
\ee 
with coefficients $c_s$ that can be shown to be non-negative (for instance, as a consequence of \eqref{eq:csIntro} below).  We refer to this decomposition as the \emph{multipole representation} of the EEC\@. Let us emphasize that the multipole expansion \eqref{eq:leg} (including the positivity condition $c_s \geq 0$) is valid in any relativistic unitary QFT (such as QCD) because it is based only upon rotation symmetry and on conservation of energy and momentum (see  \cite{Fox:1978vw,Fox:1978vu} and \appref{app:MultipoleBounds}).\footnote{Some consequences of the positivity of multi-point energy correlators are explored in \cite{Mecaj2025,BelinTOA}.}

In ${\cal N} = 4$ SYM theory, using the superconformal Ward identities \cite{Belitsky:2013xxa,Belitsky:2013bja,Korchemsky:2015ssa}, the EEC can be conveniently related to  the four-point function of scalar half-BPS operators. This four-point function has been actively studied recently using a combination of bootstrap and supersymmetric localization methods \cite{Binder:2019jwn, Chester:2019jas, Chester:2020vyz, Chester:2020dja, Chester:2021aun, Chester:2023ehi, Alday:2023pet, Caron-Huot:2024tzr}, building on earlier work \cite{Beem:2013qxa, Beem:2016wfs, Caron-Huot:2022sdy}. Part of our goal in the present work is to extend these studies to include the EEC and its multipole moments.

A starting point of the conformal bootstrap analysis is the (super)conformal block decomposition of correlation functions.  Thus, an important step in our analysis is to also expand the EEC and the moments $c_s$ in appropriate conformal blocks. As we will show in \secref{sec:OPE}, the conformal-block decomposition of the energy multipoles can be written as
\be\label{eq:csIntro}
 c_s = \sum_{\tau,J} \lambda^2_{\tau,J}\sin^2\left(\frac{\pi\tau}{2}\right)\alpha(\tau,J)f_{J+2,s}(\tau)\geq 0 \,,
 \ee  
where the sum is over long multiplets that appear in the OPE of two stress tensor multiplet operators.\footnote{The factor $\sin^2 \left(\frac{\pi\tau}{2}\right)$ reflects the Lorentzian nature of the observable and follows from the identity
\begin{equation}
    \langle {\cal O}^\dagger {\cal E}(\vec n_1) {\cal E}(\vec n_2) {\cal O} \rangle 
    = \langle [{\cal O}^\dagger , {\cal E}(\vec n_1)] \, [{\cal E}(\vec n_2), {\cal O}] \rangle \,,
\end{equation}
which holds because ${\cal E}(\vec n)$ annihilates the vacuum \cite{Kravchuk:2018htv}.}  These long multiplets are characterized by the twist $\tau = \Delta - J$ and spin $J$ of their superconformal primaries.  In \eqref{eq:csIntro}, $\lambda^2_{\tau,J} \geq 0$ are squared OPE coefficients, and $\alpha(\tau,J),f_{J+2,s}(\tau)\geq 0$ are kinematical non-negative functions given in \eqref{eq:superblock} below.  The conformal block decomposition for the EEC can be obtained from combining \eqref{eq:csIntro} with \eqref{eq:leg}.
With \eqref{eq:csIntro} in hand, in~\secref{sec:bootstrap} we use the methods of \cite{Caron-Huot:2022sdy, Caron-Huot:2024tzr} to derive bounds on $c_s(g)$ and on the EEC in the planar limit, and we also build on the work of \cite{Chester:2021aun,Chester:2023ehi} to derive bounds on $c_s(g)$ in the non-planar regime. 

In addition to the kinematical decomposition of the EEC, we input several kinds of constraints into our bootstrap analysis. At finite $N_c$, we use the crossing symmetry of the four-point correlator of the superconformal primary mentioned above, as well as integral constraints derived using supersymmetric localization. We also use a constraint derived from the averaged null energy condition (ANEC), which we introduce in this work. In the planar limit, we use the same integral constraints, a dispersive form of crossing symmetry, and additional dispersive sum rules; we also use spectral data for single-trace operators obtained from integrability studies.

\begin{figure}
    \centering
    \includegraphics[width=\textwidth]{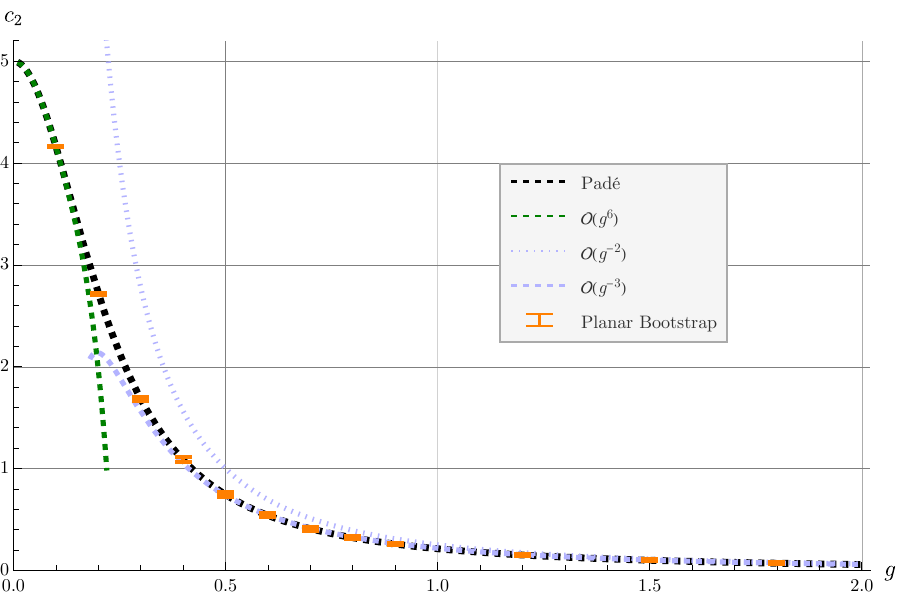}
    \caption{In orange we plot the bootstrap bounds on $c_2(g)$ in the planar limit, as obtained in \secref{sec:planarbootstrap}.  We also plot the perturbative $3$-loop result (dashed green), the leading strong coupling result (dotted blue), as well as the strong coupling result \eqref{eq:c2SubStrongF} up to subleading order in the strong coupling expansion (dashed blue). In black we plot a Pad\'e approximation based on the $3$-loop result at weak coupling and the leading strong coupling result. In particular, the subleading strong prediction \eqref{eq:c2SubStrongF} (in dashed blue) improves the agreement with the bootstrap bounds significantly relative to the leading result (in dotted blue). 
    }
    \label{fig:c2Intro}
\end{figure}

\begin{figure}
    \centering
    \includegraphics[width=\textwidth]{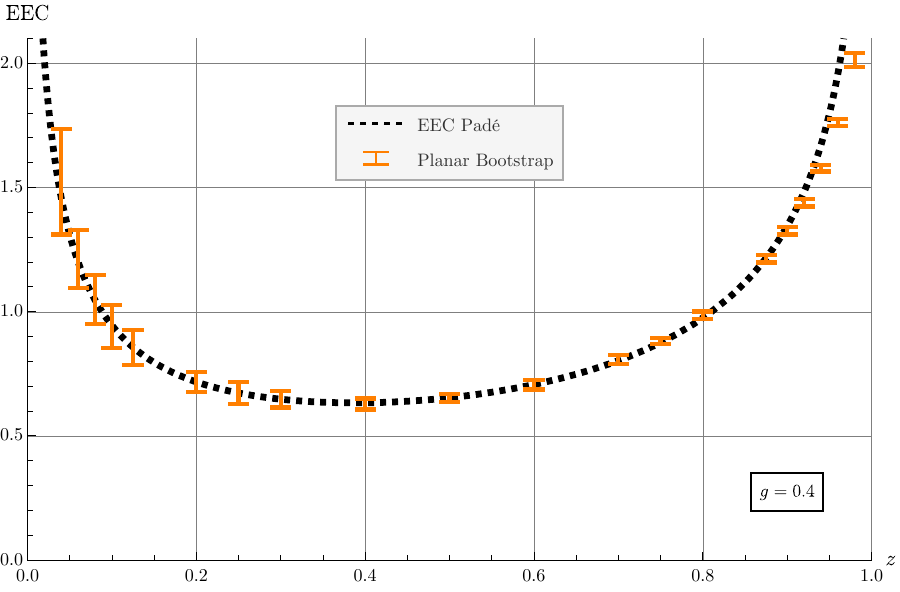}
    \caption{In orange, we show upper and lower bootstrap bounds for $\text{EEC}(z)$ in the planar limit at $g=0.4$, as obtained in \secref{sec:planarbootstrap}. This value of the $g$ is not accessible in perturbation theory, nor using the strong coupling expansion. In black, we also plotted a Pad\'e approximation based on the three-loop result at weak coupling and the leading  $\OO(1/g^2)$ behavior at strong coupling. Away from the end-points, we see that the Pad\'e approximation does a good job at reconstructing the $\text{EEC}(z)$ for any $z$, while towards the endpoints we do not expect the Pad\'e approximation to be accurate because the perturbative expansion that it is based on breaks down.}
    \label{fig:eecg04Intro}
\end{figure}

\begin{figure}
    \centering
    \includegraphics[width=\textwidth]{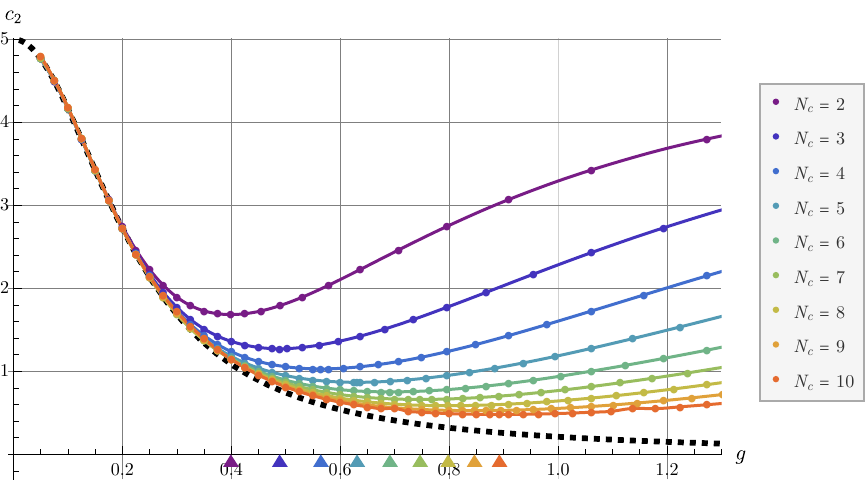}
    \caption{Bootstrap lower bounds for the second EEC multipole, $c_2$, as a function of coupling $g$ for $N_c = 2,3,\ldots,10$. For each value of $N_c$, we calculate bounds up to the self-dual point $g = \sqrt{N_c/4\pi}$ (indicated with the colored triangles) and then use $S$-duality to continue to larger $g$. We see that for each value of $N_c$, our lower bounds match the value of $c_2$ in the planar limit (shown as the Pad\'e approximant in the black dashed line; see \figref{fig:c2Intro}) for sufficiently small $g$, and the range of such values of $g$ increases with $N_c$. }
    \label{fig:c2Intro_nonplanar}
\end{figure}

The paper contains both novel analytical and numerical results.  Our main analytical results are:
 \begin{enumerate}
     \item In~\secref{sec:OPE}, we derive the conformal block decomposition \eqref{eq:csIntro}, which serves as a building block for both the analytical analysis and the numerical bootstrap.
     \item In the planar strong-coupling limit, which we discuss in~\secref{sec:Strong}, we reproduce the leading stringy correction to the EEC found in \cite{Hofman:2008ar,Goncalves:2014ffa} using the block decomposition together with the OPE data for stringy modes recently computed in \cite{Alday:2022uxp,Alday:2022xwz,Alday:2023mvu}.   Using the same method, we further derive the subleading $\OO(1/g^3)$ correction analytically, and we also derive the leading-order terms in all the EEC multipoles $c_s(g)\sim 1/g^s$.  We also reproduce these leading order terms independently through a bulk shockwave calculation generalizing the analysis of \cite{Hofman:2008ar}. 
     \item Assuming that the EEC is analytic, except for cuts in the collinear and back-to-back regimes, and that it is polynomially bounded, in~\secref{sec:inversion} we derive a simple inversion formula for the EEC multipoles, and we test its validity in perturbation theory and numerically at finite coupling in \secref{sec:InvertedBoots}. 
     \item In \secref{sec:InvertedBoots}, we use the inversion formula constructively to obtain the EEC at small but finite coupling for all angles; this is difficult numerically due to large-spin contributions. From the numerical bootstrap bounds we further extract the leading light-ray OPE contribution as $z\to0$ and the finite-coupling behavior in the back-to-back limit $z\to1$.
 \end{enumerate}
Our main numerical results are:
 \begin{enumerate}
   \item  In the planar limit, we obtain sharp two-sided bounds on the EEC multipoles $c_2$, $c_3$, and $c_4$. (This analysis can be straightforwardly generalized to any $c_s$.) The bounds on $c_2$ are presented in \figref{fig:c2Intro}. We have also bounded the EEC itself at various values of $g$. As an example, see \figref{fig:eecg04Intro} for $g=0.4$, which is not accessible from either perturbation theory or the strong coupling expansion. More bootstrap results in the planar theory can be found in \secref{sec:planarbootstrap}.  
   \item At finite $N_c$, we have fewer bootstrap constraints and so the bounds are correspondingly less strong. Nevertheless, we do obtain lower bounds on the EEC multipoles $c_2$ and $c_4$; our bounds on $c_2$ are presented in \figref{fig:c2Intro_nonplanar}.\footnote{We have also studied the dependence of the energy-energy correlator on the topological $\theta$-angle, but in the fundamental domain of the complexified coupling, we find that our bounds barely depend on this angle.} We see evidence that our bounds are nearly saturated. Indeed, in \figref{fig:c2Intro_nonplanar}, we see that at weak coupling the bounds are nearly independent of $N_c$ and closely follow the Pad\'e approximant to $c_2$ obtained in the $N_c\to\infty$ limit. We also obtain lower bounds on a smeared version of the EEC itself in the range $0.8<z<1$; although these bounds exhibit some of the expected qualitative features of the EEC, they appear not to be saturated as we increase $N_c$.
 \end{enumerate}

The rest of the paper is organized as follows.  In \secref{sec:Review}, we begin by giving an overview of the energy-energy correlator $\mathcal{N}=4$ SYM that includes the perturbative weak coupling expansion, the strong coupling expansion (including new subleading corrections obtained in this work), the collinear light-ray OPE and the back-to-back limit of the correlator. In \secref{sec:OPE}, we present the OPE of the EEC in the detector-source channel which is the basis for the bootstrap implementations. In \secref{sec:Strong}, we obtain new results on the planar EEC at strong coupling from stringy modes, and use these results to obtain Padé approximations as analytical models at finite coupling. In \secref{sec:inversion}, we derive the large-spin behaviour of the Legendre decomposition of the EEC using dispersive arguments and the knowledge of the EEC in the small-angle and back-to-back limit.  In~\secref{sec:bootstrap}, we describe our numerical bootstrap results, with the results at finite $N_c$ in~\secref{sec:nonplanarbootstrap} and those in the planar limit in~\secref{sec:planarbootstrap}.  In \secref{sec:InvertedBoots}, we extract physics in the back-to-back limit from the numerical bootstrap bounds and both test and use the inversion formula obtained in~\secref{sec:inversion}. We end with a discussion and future directions in \secref{sec:discussion}. Various technical details are collected in the appendices.

\section{Energy-energy correlator in ${\mathcal N}=4$: an overview}\label{sec:Review}

In this section, we begin with an overview of the EEC in ${\mathcal N}=4$ SYM theory in various analytically-tractable limits.  Many of the results presented here can already be found elsewhere in the literature, but we also include some new results that are derived in \secref{sec:OPE}--\secref{sec:inversion} below.  To set the stage, we begin by defining our conventions in \secref{sec:defEEC}. In \secref{subsec:weakReview} and \secref{subsec:strongReview}, we give the   weak and strong coupling expansions of the EEC\@. We then discuss the small angle and the back-to-back limits in \secref{subsec:LROPE} and \secref{subsec:BB}, respectively. A schematic summary of the various regimes is shown in~\figref{fig:review}.  

\subsection{Definition of the EEC}\label{sec:defEEC}

We are interested in studying the correlations of two energy detectors \eqref{eq:DefDetector} in a momentum eigenstate created by the local scalar operator $\mathcal{O}(p)$ \eqref{eq:DefState}, as defined in \eqref{eq:DefEEC}. The dependence on the angle $\theta$ between the detectors is commonly encoded in the cross-ratio $z \equiv {p^2 (n_1 \cdot n_2) \over 2 (p \cdot n_1) (p \cdot n_2)}$, where $n_i^\mu \equiv (1, \vec n_i)$, which in the rest frame of the source takes the form 
\be 
z = {1 - \cos \theta \over  2} = \sin^2(\theta/2) \,.
\ee 
From its definition, we see that $0 \leq z \leq 1$, with $z=0$ being the collinear limit $\theta=0$, and $z=1$ being the back-to-back limit $\theta=\pi$. We will use $z$ and $\theta$ interchangeably throughout this work. 

As mentioned in the introduction, an energy detector ${\cal E}(\vec{n})$ measures the energy flux that arrives in the direction $\vec n$ at future null infinity.  When integrated over the celestial sphere with an appropriate measure, it reduces to the total four-momentum operator $P^\mu$ \cite{Hofman:2008ar}
 \begin{equation}
 \label{eq:generators}
     P^\mu = \int_{S^2} d^2\vec{n}\,  n^\mu {\cal E}(\vec{n}) = (P^0, \vec{P}) \,,
 \end{equation}
where $d^2\vec{n}$ represents the standard measure on $S^2$. In particular, this expression implies the following Ward identity constraints on $\text{EEC}(z)$, which follow from the simple fact that in the rest frame of the source we have $P^0 |{\cal O}(p) \rangle = p^0 |{\cal O}(p) \rangle$ and $\vec{P} |{\cal O}(p) \rangle = 0$. Using \eqref{eq:generators} together with $\langle {\cal E}(\vec{n}) \rangle = {p^0 \over 4 \pi}$, the energy $\langle P^0 {\cal E}(\vec{n}_2) \rangle= p^0 {p^0 \over 4 \pi}$ and the spatial momentum $\langle \vec{P} {\cal E}(\vec{n}_2) \rangle=0$ Ward identities translate into
\be 
&\int_0^1 dz \, \text{EEC}(z)= 1 \,, \qquad \int_0^1 dz \, (1-2 z)\,  \text{EEC}(z) = 0 \label{eq:sumrules} \,.
\ee 
Being a function on the unit sphere that depends only on the relative angle $\theta$ through the variable $z$, the function $\text{EEC}(z)$ can be expanded in Legendre polynomials, with coefficients $c_s$, $s = 0, 1, \ldots$: 
 \begin{equation}
  \text{EEC}(z) = \sum_{s=0}^\infty c_s P_s (1 - 2z) \,, \qquad c_s = (2s+1) \int_0^1 dz\, P_s(1-2z) \text{EEC}(z)  \,. \label{eq:Legendre}
 \end{equation}
Here, the inverse formula was obtained using the orthogonality of the Legendre polynomials. The sum rules \eqref{eq:sumrules} imply that $c_0 = 1$ and $c_1 = 0$, yielding the Legendre decomposition \eqref{eq:leg} quoted in the Introduction.   We will refer to the expansion \eqref{eq:Legendre} as the EEC multipole expansion, and of the index $s$ as the multipole spin.

Let us now restrict to ${\mathcal N}=4$ SYM theory with $\SU(N_c)$ gauge group and establish some theory-specific conventions. We will study two cases: the planar limit, in which $N_c \to \infty$ while the 't Hooft coupling $\lambda = \gym^2N_c$ is held fixed, and the non-planar regime in which both $N_c$ and $\lambda$ are finite: 
\be 
\text{Planar}:&\qquad N_c\to \infty\qquad\,\;\text{with}\qquad  \lambda = (4\pi g)^2\quad\text{fixed} \,, \cr
\text{Non-planar}:&\qquad N_c, \lambda \quad \text{fixed} \,.
\ee 
Instead of the 't Hooft coupling $\lambda$, we will mostly use the coupling constant $g$ that is commonly encountered in the integrability literature and that is related to $\lambda$ via $\lambda = (4 \pi g)^2$.

The stress tensor of ${\cal N} = 4$ SYM theory belongs to the same superconformal multiplet as dimension-two scalar operators that are usually easier to work with.  We will both use these scalar operators as a source and also as detectors that are related to the energy detectors defined in the Introduction through supersymmetric Ward identities.  These scalar operators transform in the $\bm{20}'$ representation of the $\SU(4) \cong \SO(6)$ R-symmetry.  Since the $\bm{20}'$ can be constructed as the symmetric traceless product  of two $\bm{6}$ representations, we write  the ${\bf 20}'$ operators as symmetric traceless tensors $\cO_{IJ}(x)$, where $I = 1,\ldots,6$.  To simplify the subsequent formulas, it is customary to introduce null polarization vectors $y^I$ obeying $y \cdot y = 0$ and to denote
\begin{equation}
    \cO(x,y) = \cO_{IJ}(x) y^I y^J \,.
\end{equation}
We normalize ${\cal O}_{IJ}$ by requiring that at space-like separations (or in Euclidean signature), the vacuum two-point function takes the form 
\begin{equation} \label{eq:normalization}
    \langle \cO(x_1,y_1)\cO(x_2,y_2)\rangle= \left(\frac{y_{12}}{x_{12}^2}\right)^2\,,
\end{equation}
where $x_{ij} \equiv x_i - x_j$ and $y_{ij} \equiv y_i \cdot y_j$.  

For the source operator, in order to obey the normalization condition \eqref{eq:opNormalization}, we should take 
 \begin{equation} \label{eq:OOdDefs}
     {\cal O}(x) = {\cal O}(x, y_S) \,, \qquad {\cal O}^\dagger(x) = {\cal O}(x, y_S^*) \,,
 \end{equation}
with $y_S \cdot y_S^* = 1$.  
Due to supersymmetry, the energy detectors ${\cal E}(\vec{n})$ are related to scalar detectors, defined by taking the light transform of the ${\bf 20}'$ operators. It is simplest to focus on a specific such operator
 \begin{equation} \label{OtildeDef}
     \widetilde{{\cal O}}(x) = {\cal O}(x, y_D) \,,
 \end{equation}
corresponding to the detector polarization vector $y_D$, and its light transform
 \begin{equation} \label{eq:light}
  \widetilde{{\cal O}}(\vec{n}) 
    = \int_{-\infty}^\infty du\underset{r\to\infty}{\lim}r^{2} \widetilde{{\cal O}} (u+r,r\vec{n}) \,.
 \end{equation}
It can be shown \cite{Belitsky:2013xxa,Belitsky:2013bja,Korchemsky:2015ssa} (see also \cite{Chen:2024iuv} which we follow closely) that in a state created by the operator ${\cal O}(x)$ as above, the energy-energy correlator can be equivalently written as
\be
\label{eq:SWI}
\la {\cal E}(\vec{n}_1) {\cal E}(\vec{n}_2)\ra= {(p^0)^4 \over z^2} \la \widetilde {\cal O}(\vec{n}_1) \widetilde {\cal O}(\vec{n}_2)\ra \,,  
\ee
where the source and detector polarizations obey
 \begin{equation} \label{eq:yConditions}
     y_D \cdot y_S = y_D \cdot y_S^* = \left( {(N_c^2 - 1)\over 8 (2 \pi)^4 }\right)^{1/4} \,,
 \end{equation}
in addition to $y_S \cdot y_S^* = 1$ as mentioned above. The relative factor $(p^0)^4$ in the formula above follows from dimensional analysis: the stress-energy tensor $T_{\mu \nu}$ appearing in the energy detector has dimension $4$, whereas $\cO(x,y_D)$ appearing in the scalar detector has dimension $2$. On the other hand, the fact that the angular dependence is identical up to a factor of $1/z^2$ is a nontrivial consequence of the superconformal Ward identities.

It may be useful to provide a description of the source and detector operators in terms of a standard description of the ${\cal N} = 4$ SYM theory as a gauge theory with six adjoint-valued canonically normalized scalar fields $\phi_I^a$, with $a = 1, \ldots, N_c^2 - 1$, a gauge field $A_\mu^a$, and fermions.  In terms of these fields, the operators ${\cal O}_{IJ}$ normalized as in \eqref{eq:normalization} can be written as
 \begin{equation}
   {\cal O}_{IJ} = \frac{\sqrt{2} (4 \pi^2)}{\sqrt{N_c^2 - 1}}  \tr \left( \phi_{I} \phi_{J} - \frac{1}{6} \delta_{IJ} \phi_K \phi_K \right) \,,
 \end{equation}  
where $\phi_I = \phi_I^a T^a$, with $T^a$ being the $\SU(N_c)$ generators obeying $\tr (T^a T^b) = \delta^{ab}/2$ in the fundamental representation.  Simple choices for the source and detector operators are given by
 \begin{equation}
  \begin{aligned} \label{eq:opchoice}
     {\cal O}(x) &= \sqrt{2} {\cal O}_{12}(x) = \frac{8 \pi^2}{\sqrt{N_c^2 - 1}} \tr (\phi_1 \phi_2)  \,, \\
     \widetilde{\cal O}(x) &= \frac{i \sqrt{N_c^2-1}}{ (4 \pi)^2} 
  \left( {\cal O}_{11} - {\cal O}_{22} + 2 i {\cal O}_{12} \right)
  = \frac{i \sqrt{2}}{4} \tr (\phi_1 + i \phi_2)^2 \,. 
   \end{aligned}
 \end{equation}
The choices \eqref{eq:opchoice} can be obtained by taking linear combinations of the polarizations $y_S$ and $y_D$ obeying the requirements listed above in and around \eqref{eq:yConditions}.

The key takeaway is that the EEC we are studying is governed by the four-point function of $\bm{20}'$ half-BPS scalar operators, which, to our advantage, is a particularly well-studied observable. We now move on to a review of what is known about the EEC in various limits, in both coupling and kinematic space.

\subsection{Weak coupling}\label{subsec:weakReview}

The weak-coupling expansion of the EEC takes the form
\be  \label{eq:EECExpansion}
\text{EEC}(z) = \sum_{k=0}^\infty g^{2k}\text{EEC}^{(k)}(z) + \OO\left(N_c^{-2}\right)\,.
\ee 
As we will review, the coefficients $\text{EEC}^{(k)}(z)$ are known up to three loops, and they contain distributional terms localized at $z = 0$ and $z = 1$ in addition to smooth contributions on the interval $z \in (0, 1)$.  Moreover, in ${\mathcal N}=4$ SYM the EEC is obtained from the local four-point correlator of ${\bf 20}'$ operators, which is known not to receive non-planar corrections up to three loops \cite{Eden:2011we}. Consequently, the three-loop result for the EEC is identical in the planar and non-planar theories.

\begin{figure}
    \centering
    \includegraphics[width=0.7\linewidth]{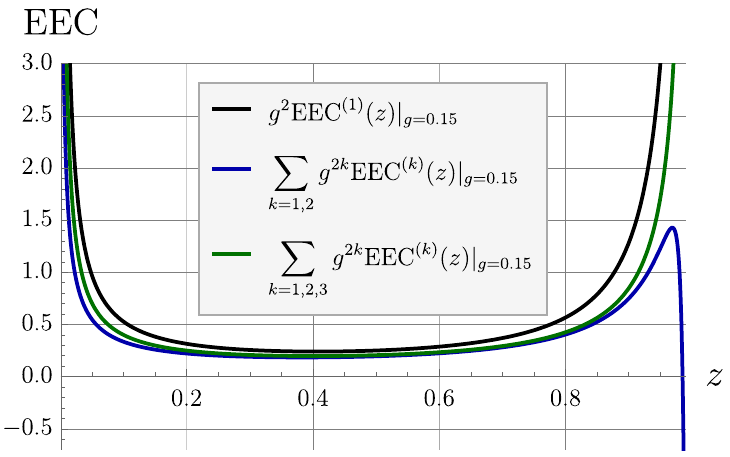}
    \caption{Here we show the weak-coupling expansion of the energy-energy correlator evaluated at LO, NLO and NNLO order at fixed value of the coupling $g=0.15$. In the bulk the perturbative corrections are small while they become large at the endpoints due to large logarithms. To explore the $z\to0,1$ regions we thus need to resum the perturbative expansion. 
    }
    \label{fig:pert}
\end{figure}

At zero coupling, the state created by the source operator ${\cal O}(x)$ consists of two back-to-back particles, resulting in a delta-function-localized energy flux\footnote{In the rest frame of the source, the state is a superposition of two-particle states $a^\dagger_{\vec p}\, a^{\dagger}_{- \vec p}|0 \rangle$. One can either detect the same particle with both detectors, producing $\delta(z)$, or detect the two opposite particles, producing $\delta(1-z)$.}  \cite{Basham:1978bw,Basham:1978zq}
\be  \label{eq:EEC0}
    \text{EEC}^{(0)}(z)= \frac{1}{2}\Big(\delta(z)+\delta(1-z)\Big)\,.
\ee 
Turning on a small coupling, the one-loop correction spreads the energy into narrow cones around $z = 0$ and $z = 1$. The $1$-loop result is given by \cite{Engelund:2012re,Belitsky:2013xxa}
\be\label{eq:1loopNonCont}
    \text{EEC}^{(1)}(z)= -\frac{2\ln(1-z)}{z^2(1-z)} \,,\qquad 0<z<1 \,,
\ee 
which is shown in \figref{fig:pert}. The expression \eqref{eq:1loopNonCont} is not a well defined distribution on the interval $[0, 1]$, but it can be uniquely extended to a well-defined distribution on $[0, 1]$ that satisfies the Ward identities \eqref{eq:sumrules}. These distributional terms were derived in  \cite{Korchemsky:2019nzm,Kologlu:2019mfz}, and they are \emph{essential} for obtaining the EEC multipole coefficients $c_s$ from the expansion \eqref{eq:EECExpansion}.   For further details, see \appref{app:contacterms}.\footnote{These distributional terms are absent at finite coupling and effectively arise from expanding integrable functions of the form ${\alpha}/{z^{1-\alpha}}$ in a series around $\alpha = 0$.} 

The two-loop result, originally obtained in \cite{Belitsky:2013ofa}, is:
\be\label{eq:twoloopNonCont}
\text{EEC}^{(2)}(z) = \frac{8{\cal B}^{(2)}(z)}{z^2(1-z)}\,, \qquad 0<z<1,
\ee 
where
\be
{\cal B}^{(2)}(z) &= (1-z){\cal B}_1(z)+{\cal B}_2(z) \,, \cr 
{\cal B}^{(2)}_1(z) &= 4\sqrt{z}\Big[\text{Li}_2(-\sqrt{z})-\text{Li}_2(\sqrt{z})+\frac{\ln z}{2}\ln\Big(\frac{1+\sqrt{z}}{1-\sqrt{z}}\Big)\Big]+(1+z)(2\text{Li}_2(z)
\cr&+\ln^2(1-z))+2\ln(1-z)\ln(\frac{z}{1-z})+z\frac{\pi^2}{3} \,, \cr
{\cal B}^{(2)}_2(z) &=\frac{1}{4}\Big\{(1-z)(1+2z)\Big[\ln^2(\frac{1+\sqrt{z}}{1-\sqrt{z}})\ln(\frac{1-z}{z})-8\text{Li}_3(\frac{\sqrt{z}}{\sqrt{z}-1})-8\text{Li}_3(\frac{\sqrt{z}}{\sqrt{z}+1})\Big]\cr
&-4(z-4)\text{Li}_3(z)+6(3+3z-4z^2)\text{Li}_3(\frac{z}{z-1})-2z(1+4z)\zeta_3\cr
&+2[2(2z^2-z-2)\ln(1-z)+(3-4z)z\ln z]\text{Li}_2(z)\cr 
&+\frac{1}{3}\ln^2(1-z)[4(3z^2-2z-1)\ln(1-z)+3(3-4z)z\ln z]\cr
&+\frac{\pi^2}{3}[2z^2\ln z
-(2z^2+z-2)\ln(1-z)]\Big\} \,.
\ee 
As in the one-loop case, the expression \eqref{eq:twoloopNonCont} must be supplemented by contact terms to be integrable and to obey the Ward identities \eqref{eq:sumrules}; these are provided in Eq.~\eqref{eq:twoloop}.

Finally, the three-loop result was derived in \cite{Henn:2019gkr} and takes the form
\be\label{eq:threeloopNonCont}
\text{EEC}^{(3)}(z) = \frac{32{\cal B}^{(3)}(z)}{z^2(1-z)} \,, \qquad 0<z<1 \,, 
\ee 
where ${\cal B}^{(3)}(z)$ consists of a sum of polylogarithms and a two-fold integral
\be 
{\cal B}^{(3)}(z) =& f_{\text{HPL}}(z)+\int_0^1 d\bar{z}\int_0^{\bar{z}}  dt\frac{z-1}{t(z-\bar{z})+(1-z)\bar{z}}\Big[R_1(\zeta,\bar{z})P_1(\zeta,\bar{z})+R_2(\zeta,\bar{z})P_2(\zeta,\bar{z})\Big] \,, \cr 
   R_1(\zeta,\bar{z}) =& \frac{\zeta\bar{z}}{1-\zeta-\bar{z}} \,,  \qquad  
   R_2(\zeta,\bar{z}) = \frac{\zeta^2\bar{z}}{(1-\zeta)^2(1-\zeta\bar{z})},\qquad \zeta = \frac{zt(t-\bar{z})}{t(z-\bar{z})+(1-z)\bar{z}} \,.
\ee 
Here, $P_i$ denote harmonic polylogarithms of weight three in the variables $\zeta$ and $\bar{z}$. Further details can be found in \cite{Henn:2019gkr}. As before, the expression \eqref{eq:threeloopNonCont} must be supplemented by contact terms to render it integrable over $z$. This completion was carried out in \cite{Kologlu:2019mfz,Korchemsky:2019nzm}, and the fully integrable result can be extracted, for example, from Eq.~(7.88) in \cite{Kologlu:2019mfz}.

The behavior of the perturbative expansion at fixed coupling is illustrated in~\figref{fig:pert}, where, for $g = 0.15$, we plot the expansion of $\text{EEC}(z)$ truncated to successive orders in the small $g$ expansion. As can be seen from this figure, the perturbative series breaks down near the endpoints.  There, it is important to resum the perturbative expansion, as will be discussed in \secref{subsec:LROPE} and \secref{subsec:BB}.

From the results presented above and the associated contact terms given in \appref{app:contacterms}, it is straightforward to obtain the perturbative expansion of the EEC multipoles $c_s$ using \eqref{eq:Legendre}. For the lowest multipoles we find\footnote{The two-loop results were obtained numerically to high precision and are consistent with the exact expressions, as verified using \texttt{FindIntegerNullVector} in \textit{Mathematica}. At three loops, the coefficients can again be extracted by explicit numerical integration, although achieving high precision is more challenging. Moments of the three-loop EEC were computed exactly in \cite{Lance:TA} in terms of zeta values up to weight six, from which the Legendre coefficients can be determined analytically. Here we report the numerical values.} 
\be\label{eq:csPert}
c_2 &= 5-10\pi^2g^2+(80\pi^2+720\zeta_3+\pi^4)g^4-29528.972g^6+\cdots \,,\\
c_3 &= 0+\frac{70}{3} \left(12-\pi ^2\right) g^2-\frac{7}{27} \left(420+260 \pi ^2+360 \zeta_3-9 \pi ^4\right)g^4+39566.852g^6+\cdots \,,\cr
c_4 &= 9-15 \left(4 \pi ^2-21\right) g^2+(2765+\frac{950 \pi ^2}{3}+120 \zeta_3+6 \pi ^4)g^4-140071.82g^6+\cdots \,.\nonumber
\ee 
In the free theory limit, it is straightforward to check that the sum of the two delta functions in \eqref{eq:EEC0} yields $c_{2k}=4k+1$ and $c_{2k+1}=0$ for $k=0,1,\ldots$. The one-loop contribution arises from twist-2 operators, as shown explicitly in \secref{sec:FreeTheory}.

\subsection{Strong coupling}\label{subsec:strongReview}

At strong ’t Hooft coupling, the EEC can be computed using the AdS/CFT correspondence \cite{Maldacena:1997re, Gubser:1998bc, Witten:1998qj}.   In the planar limit, the first two orders in $1/g$ arise from the  supergravity limit and the first stringy correction, obtained in \cite{Hofman:2008ar} (see also \cite{Goncalves:2014ffa}). As explained in~\cite{Hofman:2008ar}, the bulk picture is that of a string falling through AdS, probed by shockwaves at the AdS horizon. The leading-order gravity result is an angle-independent homogeneous distribution arising from the repeated branching of particles at strong coupling \cite{Polchinski:2002jw,Hatta:2008tx}. On top of this distribution, there are small inhomogeneities suppressed by inverse powers of the coupling $1/g$.

In \secref{sec:Strong}, we will also derive this result using the detector-source OPE, which also enables us to determine the next term in the strong-coupling expansion. The result reads 
\be \label{eq:strongFinal}
\text{EEC}(z) =1 &+\Big[\frac{1}{4g^2}+ \underbrace{\frac{\pi^2-15\zeta_3}{8\pi^3g^3}}_{\text{new}}  +\OO(g^{-4})\Big]P_2(1-2z) + \Big[  \underbrace{\frac{45\zeta_3}{8\pi^3 g^3}}_{\text{new}} +\OO(g^{-4}) \Big]  P_3(1-2z)\nonumber \\
&+\sum_{s=4}^\infty c_s(g)  P_s(1-2z) ] \,,
\ee 
where the leading stringy correction $O(1/g^2)$ was found in \cite{Hofman:2008ar}, while the subleading $O(1/g^3)$ terms are new. The higher-spin multipoles, suppressed as $c_s \sim 1/g^s$, are also new, and to leading order they are given by
\be\label{eq:csAllStrong}
c_s(g)= \underbrace{{d_s \over g^s}\frac{\Gamma (s+1)^2 \Gamma (s+3)^2}{2^{s+2}\pi^s\Gamma (2 s+1)}}_{\text{new}} +\OO(g^{-s-1}) \,,
\ee 
with $d_s$ defined implicitly in \eqref{eq:ds} below. 

We derive the expression \eqref{eq:csAllStrong} independently by two methods: by generalizing the shockwave argument of \cite{Hofman:2008ar} and by employing the detector–source OPE, with perfect agreement between the two. In contrast, the $\OO(1/g^3)$ contribution to $c_2$ probes effects beyond the shockwave limit and is obtained using the detector-source OPE\@. The relevant contributions originate from stringy modes with twist $\tau\sim g^{1/2}$, whose OPE data was computed in the context of the AdS Virasoro-Shapiro amplitude in \cite{Alday:2022uxp,Alday:2022xwz,Alday:2023mvu}.

As emphasized in \cite{Hofman:2008ar,Goncalves:2014ffa}, extracting the large-$g$ corrections to the energy correlators requires special care. The difficulty arises because a naive expansion of the underlying four-point function in powers of $1/g$ leads to a divergent result.
The divergence originates from the fact that energy correlators probe the Regge limit of the four-point function, it is soft enough for the light-transform integrals to converge at finite $g$, but divergent when expanded in $1/g$. A simple illustration of this phenomenon is provided by the toy integral 
\be\label{eq:enhancement}
\int^\infty {d s \over s^3} s^{2-{c_0 \over g}} \sim g \,,\qquad c_0>0 \,,
\ee
which diverges term by term in a $1/g$ expansion but remains finite at fixed $g$. This mechanism leads to an enhancement effect: while the leading stringy correction to the four-point function of half-BPS operators scales as $g^{-3}$, the corresponding correction to the EEC scales as $g^{-2}$. This enhancement was analyzed in detail in \cite{Chen:2024iuv} for both stringy and quantum-gravity corrections.

\subsection{Small angles and the light-ray OPE}\label{subsec:LROPE}

At small angles, perturbation theory produces large logarithms, as can be verified in \secref{subsec:weakReview}.
These logarithms are resummed by the light-ray OPE  \cite{Hofman:2008ar,Kologlu:2019mfz,Chang:2020qpj,Korchemsky:2019nzm}. 

The light-ray OPE is an expansion in the small angle between detectors ($z\to0$). Its leading behavior is governed by the plus-signature (i.e.~analytically continued from even spins)  Regge trajectory analytically continued to spin $J=3$ \cite{Hofman:2008ar}.\footnote{In its simplest form this statement follows from matching the boost $(y^+,y^-) \to (\lambda^{-1} y^+,\lambda y^-)$ quantum numbers in $\lim_{\vec y_2 \to \vec y_1}\int dy^- T_{--}(0,y^-,\vec y_1)\int dy^- T_{--}(0,y^-,\vec y_2) \sim \int dy^- \cO_{---}(0,y^-,\vec y_1)$.} More generally, in the OPE of light-ray operators built from $J_i$-spin local operators only $J_1+J_2-1$ light-ray operators appear \cite{Kologlu:2019mfz}. This statement requires a refinement for spinning source operators \cite{Chang:2020qpj}, which is however beyond the scope of the present paper.

In ${\mathcal N}=4$ SYM, the leading small-angle behavior takes the form
\be  \label{eq:EECSmallz}
\text{EEC}(z) \underset{z\to0}{\sim} \frac{1}{z^{1-\gamma_{2,-1}^{(+)}/2}}\,,
\ee 
where $\gamma_{2,-1}^{(+)}$ is the analytic continuation to $J=-1$ of the anomalous dimensions $\gamma_{2, J}$ of the even-spin unprotected operators that belong to the twist-$2$ trajectory. The appearance of $J=-1$ light-ray operator stems from the superconformal Ward identities, which allowed us to reduce the EEC to the four-point function of scalar primaries for which $J_1=J_2=0$. The operators that belong to this trajectory are superconformal primaries of long multiplets with scaling dimensions $\Delta(J) = 2+J + \gamma(J)$.  The $(+)$ superscript in $\gamma_{2, -1}^{(+)}$ signifies that the analytic continuation is performed from considering only even values of $J$.

\begin{figure}
    \centering
    \includegraphics[width=0.7\linewidth]{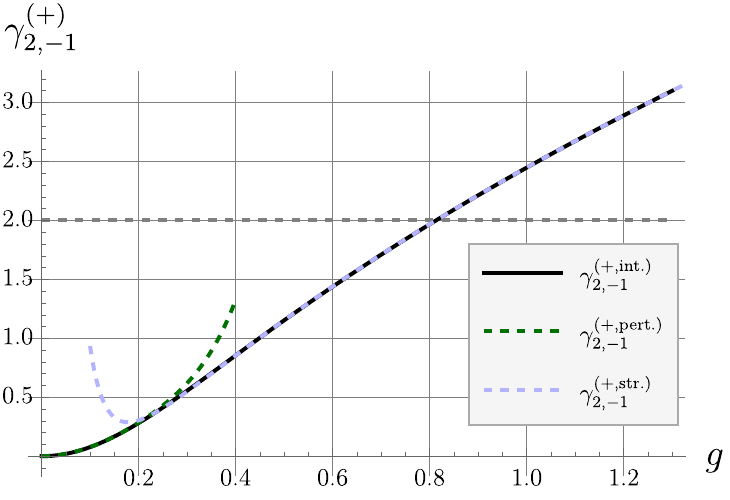}
    \caption{
    The anomalous dimension $\gamma_{2,-1}^{(+)}$ as a function of the coupling~$g$. The black solid line shows the exact integrability result (courtesy of N.~Gromov), while the green and blue dashed lines represent the weak- and strong-coupling expansions, respectively. The condition $\gamma_{2,-1}^{(+)}=2$ marks the critical coupling at which the $z \to 0$ limit of the EEC transitions from singular to regular behavior. 
    }
    \label{fig:lrOPEPred}
\end{figure}

In the planar theory, $\gamma_{2,-1}^{(+)}$ is known exactly at finite coupling from integrability~\cite{Gromov:2013pga,Gromov:2014bva}, and it is a monotonically increasing function of $g$ that behaves as $\gamma_{2,-1}^{(+)} \sim g^2$ at weak coupling and $\gamma_{2,-1}^{(+)} \sim \sqrt{g}$ at strong coupling (see, for instance, \eqref{eq:gammaExp} below). As can be seen from \eqref{eq:EECSmallz}, the EEC is regular at $z=0$ provided that $\gamma_{2,-1}^{(+)}>2$.  The condition $\gamma_{2,-1}^{(+)}(g_\text{cr}) = 2$ then defines a critical coupling above which the EEC becomes regular at $z=0$. Using the results of \cite{Gromov:2013pga,Gromov:2014bva} (see \figref{fig:lrOPEPred}), one finds 
\be
g_\text{cr} \approx 0.81 \,.
\ee 
As we go from $g < g_{\text{cr}}$ to $g > g_{\text{cr}}$, the EEC undergoes a transition from a singular small-angle behavior dominated by single-trace operators to a regular regime controlled by twist-4 double-trace operators.

In \cite{Kologlu:2019mfz}, not only the scaling behavior but also the full celestial block contribution of the leading Regge trajectory was obtained, yielding
\be\label{eq:leadtraj}
\text{EEC}(z) &=a_{2,-1}^{(+)} \frac{8\pi^4 \Gamma(3+\gamma_{2,-1}^{(+)})}{\Gamma(2+\frac{\gamma_{2,-1}^{(+)}}{2})^3\Gamma(-1-\frac{\gamma_{2,-1}^{(+)}}{2})}f^{4,4}_{5+\gamma_{2,-1}^{(+)}}(z)+\ldots \,,
 \cr
f^{4,4}_{5+\gamma_{2,-1}^{(+)}}(z) &= z^{-1+\frac{\gamma_{2,-1}^{(+)}}{2}}{}_2F_1(2+\frac{\gamma_{2,-1}^{(+)}}{2},2+\frac{\gamma_{2,-1}^{(+)}}{2},4+\gamma_{2,-1}^{(+)};z)
 \,,
\ee
where in the first line the ellipses denote contributions from higher-twist trajectories, and $a_{2,-1}^{(+)}$ is the analytically continued OPE coefficient of the even-spin operators from the leading twist-2 trajectory. In the next section we introduce the three-point couplings $\lambda^2_{\tau,J}$ such that
\be
a_{2,-1}^{(+)} = {N_c^2 - 1\over 64\pi^4}(\lambda^{(+)} _ {2, -1})^2 \,.
\ee
The $z \to 0$ limit is thus controlled by the Regge trajectories of twist-2 operators for $g < g_{\text{cr}}$.

While the anomalous dimension $\gamma_{2,-1}^{(+)}$ is known exactly at any coupling from integrability, the OPE coefficient $a_{2,-1}^{(+)}$ has been computed up to $\OO(g^6)$ at weak coupling~\cite{Eden:2012rr,Alday:2013cwa}, with its explicit expression given in Eq.~(7.98) of~\cite{Kologlu:2019mfz}. 

Expanding the prefactor in \eqref{eq:leadtraj} for small anomalous twist, $\gamma_{2,-1}^{(+)}\ll1$, we have
\be
{1 \over \Gamma(-1-\frac{\gamma_{2,-1}^{(+)}}{2}) } \approx\frac{\gamma_{2,-1}^{(+)}}{2} \,,
\ee
which gives, at weak coupling and small~$z$,
\be\label{eq:smallgsmallz}
\text{EEC}(z)\propto \frac{\gamma_{2,-1}^{(+)}}{z}(1+\frac{\gamma_{2,-1}^{(+)}}{2}\log z+\cdots)+\cdots \,,
\ee
where we kept only the leading terms as $z \to 0$ at each order in the small-coupling expansion and omitted an overall positive coefficient. Equation~\eqref{eq:smallgsmallz} makes it manifest that the two-loop truncation becomes negative at sufficiently small angles, signaling the breakdown of fixed-order perturbation theory.

\subsection{Back-to-back detectors and the double-lightcone limit}\label{subsec:BB}
\begin{figure}
    \centering
    \includegraphics[width=0.7\linewidth]{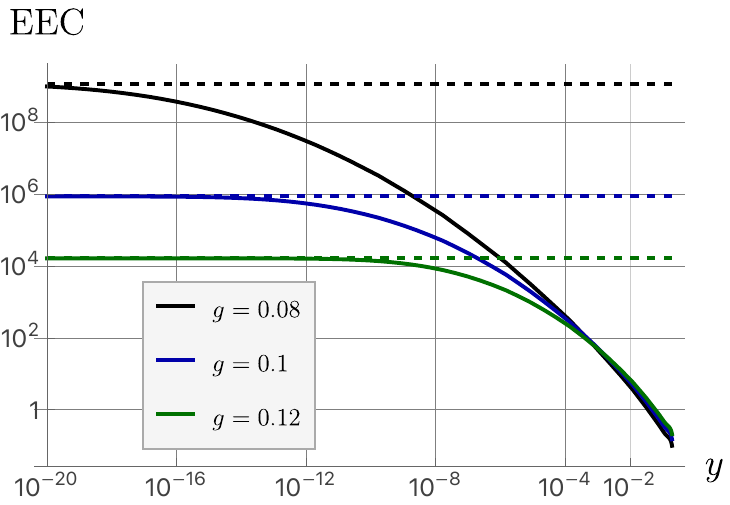}
    \caption{The resummed expression \eqref{eq:BB} evaluated for three different values of the coupling $g$. The EEC grows as the back-to-back limit is approached but saturates to a finite value, shown by the dashed lines corresponding to~\eqref{eq:bbAnalyticPred}. This contrasts with the collinear small-$z$ region, where the light-ray OPE also resums large logarithms yet remains singular as $z \to 0$.
    }
    \label{fig:bbpertExample}
\end{figure}

The back-to-back limit of detectors was studied in \cite{Belitsky:2013ofa,Korchemsky:2019nzm,Moult:2019vou,Chen:2023wah,Chen:2023amz}. In \cite{Korchemsky:2019nzm}, the leading-power behavior was derived by analyzing the double-lightcone limit of the four-point function of local operators. The resulting expression, valid at arbitrary coupling, is given by \cite{Korchemsky:2019nzm}
\be\label{eq:BB}
\text{EEC}(y \equiv 1- z)&= {H(g) \over 4y} \int_0^\infty db\,  b J_0(b)\cr
&{}\times\exp \Big[-{1 \over 2} \Gamma_{\text{cusp}}(g) \log^2 {b^2 \over y b_0^2}-\Gamma_{\text{coll}}(g) \log {b^2 \over y b_0^2} \Big]\cr
&{}+\cdots \,,
\ee
where $b_0 = 2 e^{-\gamma_E}$, where $\gamma_E \simeq 0.577$ is the Euler-Mascheroni constant, and $\Gamma_{\text{cusp}}(g)$ and $\Gamma_{\text{coll}}(g)$ denote the cusp and collinear anomalous dimensions, respectively. We show \eqref{eq:BB} for a three different values of the coupling in \figref{fig:bbpertExample}. In the formula \eqref{eq:BB}, the ellipsis denote less singular terms in the $y \to 0$ of the weak-coupling expansion. Both $\Gamma_{\text{cusp}}(g)$ and $\Gamma_{\text{coll}}(g)$ are known at finite coupling from integrability \cite{Beisert:2006ez,Freyhult:2007pz,Freyhult:2009my}, while the overall coefficient $H(g)$ is known up to three loops at weak coupling \cite{Alday:2013cwa,Eden:2012rr}. These functions can equivalently be extracted from the large-spin behavior of twist-2 operators \cite{Alday:2013cwa,Korchemsky:2019nzm} 
\be\label{eq:twist2dataLargeJ}
\gamma_J(g) &= 2  \Gamma_{\text{cusp}}(g) (\log J + \gamma_E) + \Gamma_{\text{coll}}(g) + \OO(1/J) \,, \nn \\
\lambda_{2,J}^2(g)/\lambda_{2,J}^2(0)&=H(g) 2^{-\gamma_J(g)} \Gamma^2 \Big( 1 - {1 \over 2} \gamma_J(g) \Big) e^{-\Gamma_{\text{coll}}(g)(\log J + \gamma_E)-\gamma_J(g) \gamma_E} \,.
\ee
Perturbative expressions for these quantities are collected in~\eqref{eq:GammaCWeak} below. In the back-to-back limit $z\to 1$, the expression \eqref{eq:BB} approaches 
\be\label{eq:bbAnalyticPred}
\text{EEC}(z)\underset{z\to1}{=} \frac{\sqrt{\frac{\pi }{2}} b_0^2 H(g) e^{\frac{(-1+\Gamma_{\text{coll}}(g))^2}{2 \Gamma_{\text{cusp}}(g)}}}{4 \sqrt{\Gamma_{\text{cusp}}(g)}}+\cdots \,.
\ee 
For small coupling, the prediction \eqref{eq:bbAnalyticPred} is nonperturbative in $g$ and grows like $\sim e^{\frac{1}{8g^2}}/g$. At strong coupling, the expansions of $\Gamma_{\text{cusp}}(g)$ and $\Gamma_{\text{coll}}(g)$ are also known \cite{Frolov:2002av,Freyhult:2009my}
\be 
\Gamma_{\text{cusp}}(g)&\sim 2g-\frac{3\log2}{2\pi}+\OO(1/g) \,, \cr
\Gamma_{\text{coll}}(g)&\sim -2(\gamma_E+\log g)\Gamma_{\text{cusp}}(g)-4g(1-\log2)\cr
&-(1-\frac{6\log2}{\pi}+\frac{3(\log^2g)}{\pi})+\OO(1/g)\,,
\ee 
while $\text{EEC}(z)_{g\to\infty}\sim \OO(1)$ for all $z$. This suggests that in order for \eqref{eq:bbAnalyticPred} to be applicable at strong coupling, it should be that
\be\label{eq:HStrong}
H(g)\lesssim e^{-\frac{(-1+\Gamma_{\text{coll}}(g))^2}{2 \Gamma_{\text{cusp}}(g)}} = e^{-4 g (\log (g/2)+\gamma_E +1)^2+\cdots} \,,
\ee 
up to powers in $g$. We will test this strong-coupling prediction, together with \eqref{eq:bbAnalyticPred}, numerically in the planar limit in \secref{sec:planarbootstrap}. 

In addition, the leading- and next-to-leading logarithms at leading- and next-to-leading power as $z\to1$ were resummed in perturbation theory in \cite{Chen:2023wah,Chen:2023amz}:
\be\label{eq:BBResum}
 \begin{aligned}
\text{EEC}(z=1-y) &\underset{y\to0}{=}-\frac{2 g^2 L_y e^{-2 g^2 L_y^2}}{y}-\sqrt{\frac{\pi }{2}} g\text{Erf}\left(\sqrt{2} g L_y\right)\\
&{}+\frac{1}{6} g^2 \left(-12L_y e^{-2 g^2 L_y^2}-4 e^{-2 g^2 L_y^2}+4\right)+\frac{14}{3} g^4 L_y^2 e^{-2 g^2 L_y^2}+\cdots \,, 
\end{aligned}
\ee 
where $L_y=\log y$. At leading power, $1/y$, the leading- and next-to-leading logarithms agree with those obtained from perturbatively expanding \eqref{eq:BB}. 

Remarkably, the limits $g\to0 $ and $y\to0$ in \eqref{eq:BBResum} do not commute. Expanding first in $g\to0$ gives \eqref{eq:BBResum} starting at $\OO(g^2)$, whereas expanding first in $y\to0$ and then $g\to0$ we find from \eqref{eq:BBResum}
\be
\label{eq:btbhighertwists}
\text{EEC}(z=1-y)&\underset{y\to0,g\to0}{=}\sqrt{\frac{\pi}{2}}g+\frac{2}{3}g^2+\cdots \,.
\ee 
Our interpretation of this result is as follows. The leading weak-coupling asymptotic behavior of the EEC, arising from twist-2 operators, is correctly captured by the non-perturbative contribution \eqref{eq:bbAnalyticPred}, whereas the perturbative effects of higher-twist operators are described by \eqref{eq:btbhighertwists}. We will use our finite-coupling results to get further insight into the back-to-back behavior of the EEC and importance of the higher-twist contributions at finite coupling.

\section{OPE in the source-detector channel}\label{sec:OPE}

In this section, we introduce the OPE decomposition of the EEC in the detector-source channel.  This OPE decomposition is an important ingredient in the numerical bootstrap analysis in \secref{sec:bootstrap}. 

\subsection{Review of the correlator}\label{sec:correlator_review}

We begin by briefly reviewing some properties of the vacuum four-point function of half-BPS operators $\langle \cO(x_1,y_1) \cO(x_2,y_2) \cO(x_3,y_3) \cO(x_4,y_4) \rangle$ from which the EEC is obtained. Superconformal symmetry implies that this correlator depends only on cross-ratios of the position and polarization vectors, defined as
\begin{align}
u= \frac{x_{12}^{2}x_{34}^2}{x_{13}^2x_{24}^2} = \z \bar{\z}\,, \qquad &v = \frac{x_{23}^2x_{14}^2}{x_{13}^2x_{24}^2} = (1-\z)(1-\bar{\z})\,, \label{cross-ratios u,v} \\
\sigma= \frac{y_{12} y_{34}}{y_{13} y_{24}} = \alpha \bar{\alpha}\,, \qquad& \rho = \frac{y_{23} y_{14}}{y_{13} y_{24}} = (1-\alpha)(1-\bar{\alpha}) \label{cross-ratios sigma tau}\,,
\end{align}    
where $y_{ij} \equiv y_i \cdot y_j$. The superconformal Ward identities \cite{Eden:2000bk,Dolan:2004mu, Nirschl:2004pa} further constrain the form of the four-point function: it can be written as the sum of a free-theory part and an interacting part, the latter expressed in terms of a single function $\cH(u,v)$, known as the reduced correlator. The explicit form of the correlator reads
\begin{align}
{\langle \cO(x_1,y_1)\cdots \cO(x_4,y_4)\rangle \over \frac{y_{13}^2y_{24}^2}{x_{13}^4x_{24}^4}}
&=1+\frac{\sigma^2}{u^2}+\frac{\rho^2}{v^2} +\frac{1}{c}\left(\frac{\sigma}{u}+\frac{\rho}{v}+\frac{\sigma\rho}{u v}\right)
\nonumber \\ &\quad + \frac{1}{c}(\z-\alpha) (\z - \bar{\alpha}) (\bar{\z}- \alpha) (\bar{\z} - \bar{\alpha})\mathcal{H}(u,v)\,. \label{G ansatz}
\end{align}
Here, 
\begin{equation}\label{eq:c_def}
    c = \frac{N_c^2-1}{4}
\end{equation}
denotes the $c$-anomaly coefficient. Crossing symmetry of the four-point function further requires the reduced correlator to satisfy the relations
\begin{align}\label{eq:h_crossing}
    \mathcal{H}(u,v)=\mathcal{H}(v,u)=u^{-4}\mathcal{H}(\tfrac{1}{u},\tfrac{v}{u}) \,.
\end{align}

The four-point function $\langle \cO(x_1,y_1) \cO(x_2,y_2) \cO(x_3,y_3) \cO(x_4,y_4) \rangle$ can be expanded in the $s$-channel OPE\@. This OPE contains short multiplets, whose conformal data are protected and known analytically, as well as long multiplets with spin $J=0,2,4,\ldots$ and twist $\tau \ge 2$, whose contributions are expressed in terms of superconformal blocks. Consequently, the reduced correlator admits the expansion
\begin{equation}\label{eq:h_expansion}
    \cH(u,v) = \cH^\text{protected}(N_c;u,v) + \sum_{\tau,J} \lambda^2_{\tau,J} G^{\cN = 4}_{\tau,J}(u,v) \,,
\end{equation}
where $G^{\cN = 4}_{\tau,J}(u,v)$ denotes the corresponding superconformal block.  As shown in \cite{Nirschl:2004pa}, it can be written as 
\be
G^{\cN = 4}_{\tau,J}(u,v) = u^{-4} G_{\tau+J+4,J}(u,v)  \,,
\ee
where
\begin{equation}\label{eq:block_norm}
\begin{split}
G_{\tau,J}(\z,\zb) &= \frac{1}{2^J}\frac{\z \zb}{\z-\zb}(f_{\tau+2J}(\z)f_{\tau-2}(\zb)-f_{\tau-2}(\z)f_{\tau+2J}(\zb))\,,\\
f_\beta(\z)&=\z^{\beta\over 2}{}_2F_1\left(\frac{\beta}{2},\frac{\beta}{2},\beta,\z\right) \,.
\end{split}
\end{equation}
The protected contribution $\cH^\text{protected}(N_c;u,v)$ is given explicitly in (2.31) of \cite{Beem:2016wfs}.

\subsection{Source-detector OPE decomposition}
We are interested in writing the OPE of the four-point correlator in the detector-source channel, which takes the form (here we follow closely \cite{Chen:2024iuv}), 
\be
\label{eq:OPEsourcedet}
\la {\cal O}^\dagger(x_1) \widetilde {\cal O}(x_2) \widetilde {\cal O}(x_3) {\cal O}(x_4) \ra|_{\text{long}} &={N_c^2 - 1\over 8 (2 \pi)^4} {1 \over (x_{12}^2 x_{34}^2)^2} \sum_{\tau, J} \lambda_{\tau,J}^2\, u^2 v^2 F_{\tau, J}(\z, \zb) \,,
\ee
where $J$ is even, and where we restricted the sum to the contribution from long multiplets only \cite{Beem:2013qxa,Beem:2016wfs}. With our choice of the R-symmetry polarizations described in and around \eqref{eq:yConditions} and the identifications $y_1 = y_S^*$, $y_2 = y_3 = y_D$, $y_4 = y_S$, one finds $(\sigma, \rho) = (1, 0)$, or equivalently $\alpha=\bar \alpha=1$ for the R-symmetry cross-ratios in \cite{Aprile:2017bgs}. Comparing \eqref{eq:OPEsourcedet} with \eqref{G ansatz} above, we find that the superconformal blocks $F_{\tau, J}(z, \bar z)$ that appear in \eqref{eq:OPEsourcedet} are simply related to the scalar conformal blocks \eqref{eq:block_norm} via
\be\label{eq:n4Block}
u^2 v^2 F_{\tau, J}(\z, \zb) = {v^2 \over u^2} G_{\tau+4, J}(\z, \zb) \,,
\ee
with the conformal blocks are given in \eqref{eq:block_norm}.

In order to use the result of \cite{Kologlu:2019bco} for the detector-source OPE, we decompose \eqref{eq:n4Block} in terms of conformal blocks as follows 
\be\label{eq:normBlocks}
{v^2 \over u^2} G_{\tau+4, J}(\z,\zb) = \sum_{k=0}^8\sum_{p=-4}^{4} d_{k,p}(\tau,J)G_{\tau+k-p, J+p}(\z, \zb) \,, 
\ee 
where the coefficients $d_{k,p}(\tau,J)$ can be readily found as rational functions of $(\tau,J)$ and only those with $k+p\in 2\mathbb{Z}$ are non-zero.  See \appref{app:superconformal} and also \cite{Dolan:2001tt}.  

The decomposition of the scalar detector two-point function in terms of operators in the $\widetilde {\cal O}(x_3)\times {\cal O}(x_4)$ OPE is given by Eq.~(5.163) in \cite{Kologlu:2019bco} for conformal primaries. Using that result together with \eqref{eq:normBlocks}, we find the following decomposition in terms of ${\cal N} = 4$ superconformal primaries
\be\label{eq:detectorOPE}
\la {\cal O}(p)| \widetilde {\cal O}(\vec{n}_1) \widetilde {\cal O}(\vec{n}_2)| {\cal O}(p) \ra = \frac{N_c^2-1}{2^{10}\pi^7p_0^2}\sum_{\tau,J}\lambda_{\tau,J}^2\sum_{k=0}^8\sum_{p=-4}^{4}\sum_{s=0}^{J+p} d_{k,p}(\tau,J)G^t_{\tau+k-p,J+p}(s)P_s(\cos\theta) \,, 
\ee 
where $p^\mu = (p^0,0)$ and $\vec{n}_1\cdot \vec{n}_2=\cos\theta=1-2z$ with $\theta$ the angle between detectors.\footnote{When $J+p<0$ we use $G_{\tau,J+p}(\z,\zb)=-4^{-(J+p)-1}G_{\tau+2(J+p)+2,-2-J-p}(\z,\zb)$ before applying Eq.\ (5.163) in \cite{Kologlu:2019bco}.} The block $G^t_{\tau,J}(s)$ is given in Eq.~(5.163) of \cite{Kologlu:2019bco} and $P_s(\cos\theta)$ are Legendre polynomials. In \eqref{eq:detectorOPE}, the sums over $(k,p)$ comes from decomposing the superconformal multiplets into conformal primaries while the sum over $s$ is due to the fact that a spin-$J$ primary contribute to all EEC multipoles with spin $s\leq J$. Since our setup involves only the scalar operator in the stress-tensor multiplet, it is enough to know $G^t_{\tau,J}(s)$ for $\tau_i = 2$ in $d=4$:
\be\label{eq:defGt}
G^t_{\tau,J}(s)&= \beta(\tau,J,s){}_4F_3\left(1+s,1+s,s-J,\tau+J+s-1;2+2s,\frac{\tau}{2}+s,\frac{\tau}{2}+s,1\right)^2 \,, \nonumber\\
\beta(\tau,J,s) &=\frac{\pi ^{5/2} (2 s+1) \Gamma(s+1)^4 2^{J+\tau +2} \Gamma(J+s+2)\Gamma \left(J+\frac{\tau }{2}+\frac{1}{2}\right)\Gamma (J+s+\tau -1)}{\Gamma(2s+2)^2\Gamma(J-s+1) \Gamma \left(J+\frac{\tau }{2}\right) \Gamma \left(s+\frac{\tau }{2}\right)^4}\nonumber\\
&\times\frac{\Gamma \left(\frac{\tau }{2}-1\right)^2\Gamma (\tau -2)}{\Gamma (J-s+\tau -2)}\sin ^2\left(\frac{\pi  \tau }{2}\right)\,.
\ee 
An important feature of \eqref{eq:defGt} is the presence of manifest double-zeroes at $\tau=2\mathbb{Z}$, except at the unitarity bound $\tau=2$, where only $s=J$ is allowed \cite{Kologlu:2019bco} and $G^t_{2,J}$ reduces to
\be
G^t_{2,J}(s)=\frac{2^{4+J}\pi^{9/2}\Gamma(J+3/2)}{\Gamma(J+1)}\delta_{s,J} \,.
\ee 
These double zeroes thus suppress contributions from $\tau=4,6,\ldots$ operators in the weak-coupling regime and of double-trace operators with twist $\tau_{n,J} = 4+2n+J+\OO(N_c^{-2})$ in the planar limit where $n = 0, 1, 2, \ldots$.

We are now ready to obtain the superconformal block for the energy-energy detectors in ${\mathcal N}=4$ SYM by combining \eqref{eq:SWI}, \eqref{eq:detectorOPE}, \eqref{eq:defGt}, and the values of the coefficients $d_{k, p}$ that can be obtained from \eqref{eq:normBlocks}. The final result is 
\be\label{eq:eecrep}
\text{EEC}(z)=1+\sum_{\tau,J}\lambda_{\tau,J}^2\sin^2\left(\frac{\pi\tau}{2}\right)\alpha(\tau,J){\cal P}_{J+2}(\tau,1-2z) \,,
\ee 
where the homogeneous term ($1$) is the contribution from protected operators,\footnote{Note that, at strong coupling in the planar limit, the contribution from long multiplets is suppressed. Since $\text{EEC}^{(\text{strong})}(z)=1$ \cite{Hofman:2008ar,Goncalves:2014ffa}, the claim follows.} ${\cal P}_{J+2}(\tau,1-2z)$ a sum over Legendre polynomials, and $\alpha(\tau,J)$ is given by
\be\label{eq:superblock} 
\alpha(\tau,J) &=(N_c^2-1)\frac{2^{3 J+2 \tau +6} \Gamma \left(\frac{\tau }{2}-1\right)^2 \Gamma (J+5)^2 \Gamma
	\left(J+\frac{\tau }{2}+\frac{3}{2}\right) \Gamma \left(J+\frac{\tau
	}{2}+\frac{5}{2}\right)}{\pi ^3 (2 J+\tau +4)^2 (2 J+\tau +6)^2 \Gamma
	\left(J+\frac{\tau }{2}+2\right)^4}\geq 0 \,,\cr
{\cal P}_{J+2}(\tau, 1-2z) &= \sum_{s=2}^{J+2} f_{J+2,s}(\tau)P_s(1-2z)\qquad f_{J+2,s}(\tau)\geq 0 \,.
\ee 
When writing \eqref{eq:eecrep}, we normalized ${\cal P}_{J+2}(\tau,1-2z)=z^{J+2}+\cdots$, with the ellipses denoting a lower order polynomial in $z$. Some explicit examples of low-spin ${\cal P}_{J+2}(\tau,1-2z)$ are given by 
 \be\label{eq:blockExamples}
  \begin{aligned}
    {\cal P}_{0+2}(\tau,1-2z) &= \frac 16 P_2(1-2z) \,, \\
    {\cal P}_{2+2}(\tau,1-2z) &= \frac{1}{70} P_4(1-2z)
     + \frac{(\tau + 4)^2  - 36}{2880} P_3(1-2z) \\
     &{}+  \frac{\left[ (\tau + 4)^2  - 36\right]
      \left[7 (\tau + 4)^4 - 168 (\tau + 4)^2 + 176 \right]}{3225600 \left[ (\tau + 4)^2 - 1 \right]} P_2(1-2z) \,.
  \end{aligned}
 \ee 
The representation \eqref{eq:superblock} satisfies the Ward identities \eqref{eq:sumrules}, since ${\cal P}_J(\tau,1-2z)$  does not have support on $P_0(1-2z)$ or on $P_1(1-2z)$. The blocks ${\mathcal P}_{J+2}(\tau,1-2z)$ can also be obtained from Mellin space, see \eqref{eq:alphaPMellin} and \eqref{eq:mellinopeint}.

A key feature of \eqref{eq:superblock} is that there are no contributions from $\tau=2\mathbb{Z}$ with the exception of $\tau=2$ where the double zero of $\sin^2(\pi\tau/2)$ is compensated by a double pole of $\alpha(\tau,J)$. This makes the representation \eqref{eq:eecrep} amenable to the planar bootstrap study of \cite{Caron-Huot:2022sdy,Caron-Huot:2024tzr}, as we will see in \secref{sec:bootstrap}, since double-trace operators are suppressed. We note, however, that the representation \eqref{eq:eecrep} can also be applied also in the non-planar theory. 

Since in the planar limit double-trace operators drop out from the EEC, it follows that at strong coupling only the single-trace operators dual to stringy modes contribute. In the limit of large twist and fixed spin, we find that the blocks \eqref{eq:superblock} take the following form
\be\label{eq:leadstringy}
\lim_{\tau \to \infty} \alpha(\tau,J){\cal P}_{J+2}(\tau,1-2z) = \frac{3(N_c^2-1) (J+1) 2^{J+2 \tau +17}}{\pi ^3 \tau ^{10}}P_2(1-2z)(1+\OO(\tau^{-1})) \,,
\ee
which implies that the leading correction from the stringy modes with large twist only has support on the spin-2 EEC multipole $c_2$. In particular, the blocks at large twist $\mathcal{P}_{J+2}(\tau\to\infty,1-2z)\propto P_2(1-2z)$ should be contrasted to the blocks when the twist $\tau$ approaches the unitarity bound $\mathcal{P}_{J+2}(\tau\to2,1-2z)\propto P_{J+2}(1-2z)$ (as we will see in \secref{sec:FreeTheory}). This is kinematically why one might expect a transition from a jet-like behavior to a flat behavior as the contributing operators transition from having a small to large twist when the coupling increases.

\subsection{From OPE to EEC multipoles}
Given the representation \eqref{eq:eecrep} and \eqref{eq:superblock}, we can write down an OPE for the EEC multipoles $c_s$. We find
\be\label{eq:cs1} 
c_s = \sum_{\tau}\sum_{J=s-2}^{\infty}\lambda_{\tau,J}^2\sin^2\left(\frac{\pi\tau}{2}\right)\alpha(\tau,J)f_{J+2,s}(\tau)\geq 0 \,,
\ee 
where $f_{J+2,s}(\tau)$ is defined in \eqref{eq:superblock}, and the summand is manifestly non-negative, as each term in the sum is itself non-negative. 

Instead of constructing the blocks explicitly and then decomposing them into Legendre polynomials to read off $f_{J+2,s}(\tau)$ in \eqref{eq:cs1}, we can take a more direct route. The source-(scalar) detector block in \eqref{eq:detectorOPE} is already expressed as a sum of Legendre polynomials. However, because of the factor $1/z^2$ relating scalar and energy detectors in \eqref{eq:SWI}, the decomposition is not straightforward. In particular, we have to decompose expressions of the form $\mathcal{P}_{J+2}(\tau,1-2z)=z^{-2}\sum_{r=0}^{J+4} a_{r,J}(\tau) P_r(1-2z)$ into a sum over Legendre polynomials with spin $2\leq s\leq J+2$. We define 
\be\label{eq:KDef}
K_{r,s} \equiv (2s+1)\int_0^1 &dz\, \frac{1}{z^2}(P_r(1-2z)-(1-r(r+1)z))P_s(1-2z) \,,
\ee 
where the $1/z^2$ is due to the superconformal Ward identity \eqref{eq:SWI} and we have subtracted the constant and linear terms $P_r(1-2z)=1-r(r+1)z+\ldots$ as $z\to 0$ to ensure that the integral converges. For further details see \appref{app:OPEMultipoles}.  Note that since \eqref{eq:KDef} is the decomposition of a degree-$(r-2)$ polynomial into Legendre polynomials, it is nonzero only when $s \le r-2$. We can then equivalently obtain the contribution to the EEC multipoles $c_s$ from a long superconformal primary operator with quantum numbers $(\tau,J)$ using \eqref{eq:SWI}, \eqref{eq:detectorOPE}, and \eqref{eq:KDef}:
\be\label{eq:cs2}
c_s|_{{\cal O}_{\tau,J}}=\lambda_{\tau,J}^2\frac{N_c^2-1}{64\pi^5}\sum_{k=0}^8\sum_{p=-4}^4d_{k,p}(\tau,J)\sum_{r=s+2}^{J+p} G^t_{\tau+k-p,J+p}(r)K_{r,s} \,,
\ee 
where the sums over $(k,p)$ arise from the decomposition in \eqref{eq:normBlocks}, and $G^t_{\tau,J}$ is defined in \eqref{eq:defGt}. After summing \eqref{eq:cs2} over spins and twists it is equivalent to \eqref{eq:cs1}.

\subsection{A weak coupling check}\label{sec:FreeTheory}

In this section, we perform a check of the formulas derived in the previous section (in particular, of \eqref{eq:superblock} and related expressions such as \eqref{eq:cs2}) at weak coupling, where the operator spectrum consists of operators of twist $\tau = 2k + O(g^2)$, with integer $k \geq 1$, and where we expect that the leading order EEC is a sum of two delta functions as in \eqref{eq:EEC0}.  

Naively, the factor $\sin^2(\pi\tau/2)$ in \eqref{eq:eecrep} suppresses operators with all even integer twist. However, we find that $\alpha(\tau,J)$, defined in \eqref{eq:superblock}, has a double pole at $\tau=2$, such that in the case $k=1$ we have a non-zero contribution:
\be
 \begin{aligned}
  \alpha(\tau,J)\sin^2(\pi\tau/2) &\underset{\tau\to 2}{=} (N_c^2-1)\frac{8^{J+2} \Gamma \left(J+\frac{5}{2}\right) \Gamma \left(J+\frac{7}{2}\right)}{\pi   \Gamma (J+3)^2}+\OO((\tau-2)) \,,  \label{eq:alphasin} \\
  \alpha(\tau,J)\sin^2(\pi\tau/2) &\underset{\tau\to 2k}{=}
   O(\tau - 2k)^2 \,, \qquad \text{for integer $k>1$}\,.
 \end{aligned}
\ee 
Since the spectrum consists of operators of twist $\tau = 2k + O(g^2)$, it follows that, up to order $g^2$, only operators with twist approximately equal to two contribute.  There is a unique such operator for every $J$, with twist and OPE coefficients given by (see, for example, \cite{Henriksson:2017eej}) 
\be\label{eq:freetheorydata}
\tau &= 2+8(\psi(J+3)+\gamma_\text{E})g^2+\OO(g^4) \,,\cr
\lambda_{2,J}^2 &=\frac{\sqrt{\pi }   \Gamma (J+3)}{(N_c^2-1)2^{J+1}\Gamma \left(J+\frac{5}{2}\right)}\Big\{1+\cr
&-\frac{2}{3}\left[12 H_{J+2} \left(H_{2 J+4}-H_{J+2}\right)-6 \psi ^{(1)}(J+3)+\pi ^2\right]g^2+\OO(g^4)\Big\} \,.
\ee 
Moreover, at twist $\tau = 2$ the polynomials ${\cal P}_{J+2}(\tau,1-2z)$ simplify to 
\be\label{eq:P2tau}
{\cal P}_{J+2}(2,1-2z) = \frac{\sqrt{\pi } \Gamma (J+3)}{2^{2J+4} \Gamma \left(J+\frac{5}{2}\right)} P_{J+2}(1-2z) \,,
\ee 
which shows that, at weak coupling, $\mathcal{P}_{J+2}(\tau,1-2z)$ mostly has support on $P_{J+2}(1-2z)$.

Using the equations above (\eqref{eq:freetheorydata}, \eqref{eq:alphasin} and \eqref{eq:P2tau}) at leading order in $g^2$, together with \eqref{eq:eecrep}, we find
 \be
  \text{EEC}(z) &=1+ \sum_{\text{even $J \geq 0$}} (2J+5) P_{J+2}(1 - 2z) + O(g^2) \nonumber \\
   &= \frac{\delta(z) + \delta(1-z)}{2} + O(g^2) \,,
 \ee
which reproduces the free theory result \eqref{eq:EEC0}.

Taking into account the $\OO(g^2)$ corrections to \eqref{eq:freetheorydata}, \eqref{eq:alphasin}, and \eqref{eq:P2tau}, we can obtain the $\OO(g^2)$ corrections to the EEC multipoles. In particular, at subleading order as $\tau\to2$ the blocks $\mathcal{P}_{J+2}(\tau,1-2z)$ have support on all $2\leq s\leq J+2$ multipoles, and $c_s$ get contributions from all operators with spin $J\geq s-2$. At $\OO(g^2)$, we test \eqref{eq:cs2} by truncating the sum over spin and sum over the first $J_{\text{max}}=40$ twist-two operators to find the $\OO(g^2)$ correction 
\be
c_2|_{g^2} &\approx -98.69635g^2+\OO(g^4) \,,\cr
c_3|_{g^2} &\approx 49.7071g^2+\OO(g^4) \,,
\ee 
which converges as $J_{\text{max}}$ increases and agrees with \eqref{eq:csPert}. This serves as a check of the block decomposition \eqref{eq:eecrep}. It is straightforward to verify that the one-loop result for higher-spin EEC multipoles is analogously reproduced.

\section{Stringy corrections and a finite-coupling Pad\'e model}\label{sec:Strong}
In this section, we study stringy corrections to a homogeneous energy distribution at strong coupling, extending the analysis of \cite{Hofman:2008ar}. We do so using two different methods. One extends the bulk shockwave argument of \cite{Hofman:2008ar} to compute the leading $\OO(1/\lambda^{s/2})$ contribution to all higher-spin EEC multipoles  $c_s$. The other method starts from the source-detector OPE \eqref{eq:superblock} and uses the OPE data of stringy operators that have been recently found in the study of the AdS Virasoro-Shapiro amplitude \cite{Alday:2022uxp,Alday:2022xwz,Alday:2023mvu}. We reproduce the leading stringy result for $c_s$ obtained from the shockwave calculation. Moreover, the OPE allows us to also find the subleading stringy correction to $c_2$, which scales as $\OO(1/\lambda^{3/2})$. In particular, this result implies that we obtain the full EEC at strong coupling up to $\OO(1/\lambda^{3/2})$ as given in \eqref{eq:strongFinal}.

\subsection{Strong-coupling expansion from stringy modes and OPE}
\label{sec:stringysubsection}

The calculation of the first stringy correction from the source-detector OPE requires resummation of stringy operators. They become heavy at strong coupling ($\tau \gg 1$), and for fixed $J$ their blocks are given, to leading order in $\tau$, in \eqref{eq:leadstringy}. 

The OPE data of the stringy operators can be found in \cite{Alday:2022uxp,Alday:2022xwz,Alday:2023mvu} as $\lambda\to\infty$:
\be\label{eq:stringyOps} 
\tau &= \tau_0(\delta,J)\lambda^{1/4}+\tau_1(\delta,J)+\tau_2(\delta,J)\lambda^{-1/4}+\cdots \,,\cr 
\lambda_{\tau,J}^2 &= \frac{1}{(N_c^2-1)}\frac{\pi^3}{2^{10+2\tau+J}}\frac{\tau^6}{J+1}f(\delta,J;\lambda) \,,\cr
f(\delta,J;\lambda) &= f_0(\delta,J)+f_1(\delta,J)\lambda^{-1/4}+f_2(\delta,J)\lambda^{-1/2}+\cdots \,,
\ee
where $\tau_0 =2\sqrt{\delta}$ and $\delta=n+\frac{J}{2}+1$ with $n=0,1,\ldots$ and $J=0,2,4,\ldots,2(\delta-1)$ with $n$ labeling different Regge trajectories.\footnote{For $n>0$ there are degeneracies in the spectrum (see for example \cite{Alday:2023flc,Alday:2023mvu}) and the OPE data here is understood as the average over all operators with fixed $(\delta,J)$.}

Inserting the explicit OPE coefficients \eqref{eq:stringyOps} into \eqref{eq:cs1} and expanding at large $
\tau$ using \eqref{eq:leadstringy}, we obtain the following expression for the leading correction to $c_2$ 
\be\label{eq:leadStringyc2}
    c_2 =384\sum_{\delta=1}^\infty \sum_{J=0,2,\ldots}^{2(\delta-1)} \frac{f_0(\delta,J)}{\tau(\delta,J)^4}+\cdots \,.
\ee 
In \cite{Alday:2022uxp}, $f_0(\delta,J)$ is shown to satisfy the following sum rule 
\be 
1=\sum_{J=0,2,\ldots}^{2(\delta-1)}f_0(\delta,J) \,,
\ee 
from which it immediately follows that
\be
c_2 =384 \sum_{\delta=1}^\infty {1 \over (2 \sqrt{\delta} \lambda^{1/4})^4} = \frac{4\pi^2}{\lambda}+\OO\Big(\frac{1}{\lambda^{3/2}}\Big) \,,
\ee 
as expected from \cite{Hofman:2008ar}. Taking into account corrections to \eqref{eq:leadStringyc2}, we get
\be\label{eq:StringysumRulec2}
c_2 =384 \sum_{\delta=1}^\infty \sum_{J=0,2,\ldots}^{2(\delta-1)} \frac{f(\delta,J;\lambda)}{\tau^4}\Big[1-\frac{20J+39}{2\tau}+\frac{4 J (113 J+435)+1989}{8 \tau ^2}+\OO(\tau^{-3})] \,,
\ee
which follows from the large $\tau$ expansion of the blocks. Inserting the leading correction to the OPE data from \cite{Alday:2022uxp}, $\tau_1=-2-J$ and $f_1(\delta,J)=f_0(\delta,J)\frac{3J+\frac{23}{4}}{\sqrt{\delta}}$, and expanding in $\lambda\to\infty$ one finds that the $\OO(\lambda^{-5/4})$ contribution cancels. We further show in \appref{app:stringy} that the $\OO(1/\lambda^{3/2})$ term can be found using the subleading corrections computed in \cite{Alday:2022uxp,Alday:2022xwz,Alday:2023mvu} (up-to subsubleading order in $1/\lambda^{1/4}$), and $c_2$ to this order is given by 
\be\label{eq:c2SubStrongF}
c_2 = \frac{4\pi^2}{\lambda}+\frac{8\pi^2-120\zeta_3}{\lambda^{3/2}}+\cdots \,.
\ee 
Note that knowledge of the AdS curvature corrections to the OPE data of stringy operators was crucial for obtaining the subleading term.

Likewise, we can obtain the leading order correction to all $c_s$. In particular, expanding the blocks at large $\tau$ (\eqref{eq:leadstringy} and \eqref{eq:cs1}) and using \eqref{eq:stringyOps}, the leading contribution takes the form 
\be \label{eq:cStringyOPE}
c_s = \tilde{c}_s\sum_{\delta,J}\frac{f_0(\delta,J)}{\tau^{2s}}(J+2)_{s-2} (J-s+3)_{s-2}+\cdots \,,
\ee 
where $\tilde{c}_s$ is a computable number.  See \appref{app:stringy} for the explicit computation for $c_3$ and $c_4$. It follows, in particular, that $c_s\sim 1/\lambda^{s/2}$, since $\tau\sim\lambda^{1/4}$. This result depends only on the leading-order OPE data and can thus be obtained from the flat-space limit. The sums in \eqref{eq:cStringyOPE} were computed in \cite{Alday:2022uxp,Alday:2022xwz,Alday:2023mvu}, and can be used to find, for instance,
\be\label{eq:c3c4}
c_3 = \frac{360\zeta_3}{\lambda^{3/2}}+\cdots \,, \qquad
c_4 = \frac{684\pi^4}{7\lambda^2}+\cdots \,.
\ee
As we show below, these results for $c_3$ and $c_4$ can be independently reproduced from a shockwave calculation, with the upshot being that it is easier to use the shockwave method to find a closed-form expression for $c_s$ for all $s$. Like in \eqref{eq:StringysumRulec2}, it is straightforward to calculate the kinematical corrections appearing in \eqref{eq:cStringyOPE} for any $c_{s\geq 2}$. However, the subleading corrections to \eqref{eq:c3c4} turn out to lead to divergent sums indicating that the large $\lambda$ expansion and the summation do not commute, which is the reason we have been unable to calculate further subleading corrections. This could also indicate appearance of more complicated coupling-dependent terms in the strong-coupling expansion of $c_s$, see e.g.\ \cite{Chen:2024iuv}.

\subsection{Higher-spin multipoles from flat-space shockwave}

It is instructive to re-derive the leading stringy correction to $c_s$ using a different method.  Instead of using the OPE, let us recall the dual AdS calculation of the energy correlators \cite{Hofman:2008ar}. The boundary insertion of the energy calorimeter ${\cal E}(\vec n)$ is dual to a shockwave perturbation of the AdS geometry. This shockwave is localized at the AdS horizon, reflecting the fact that on the boundary the operator is inserted at future null infinity. The source ${\cal O}(p)$ is dual to a particle propagating through AdS\@. Eventually, this particle reaches the AdS horizon and crosses the shockwave, acquiring a Shapiro time delay. We then compute the overlap between the state after crossing the shockwaves dual to the energy calorimeters and the state before the shockwaves. This overlap is dual to the energy correlators, and in the supergravity approximation it is simply given by the product of the Shapiro time delays acquired when crossing multiple shockwaves. Causality in the bulk implies that the Shapiro time delay is positive. In the dual CFT, this positivity is the familiar ANEC.

In this calculation, the only nontrivial dynamics takes place when the particle crosses the shockwaves at the AdS horizon. Moreover, one can show that the wavefunction of the particle dual to ${\cal O}(p)$, as it crosses the AdS horizon, is localized in the transverse space. Effectively, this means that the interaction between the particle and the shockwave is localized in AdS and can be approximated by a flat-space calculation. At this point, to calculate stringy corrections, we can capitalize on the fact that the propagation of strings on shockwave backgrounds in flat space is exactly solvable \cite{Amati:1988ww}. We can then obtain the leading stringy corrections to the energy–energy correlator by computing the corresponding stringy corrections to the Shapiro time delay. It is convenient to first perform the flat-space calculation and then uplift it to the AdS result.

Here we are interested in calculating the energy-energy correlator. In flat space, each energy calorimeter is mapped to a shockwave vertex operator, which in light-cone gauge takes the form $\int_0^\pi d\sigma\, e^{i \vec k \cdot \vec X(0,\sigma)}$, where $\vec X(0,\sigma)$ is evaluated at zero light-cone time $\tau = 0$ and at $\vec k^{2} = 0$. In this way, the relevant worldsheet correlator becomes
\be
\label{eq:correlatorwsstring}
\la 0 |  e^{i \vec{k}_1 \cdot \vec X(\sigma, 0)}  e^{i \vec{k}_2 \cdot \vec X(\sigma', 0)}   | 0 \ra = | 2 \sin (\sigma - \sigma')  |^{{1 \over 2} \alpha' \vec{k}_1 \cdot \vec {k}_2} \,.
\ee
To get the energy-energy correlator, we need to integrate this worldsheet correlator over the positions of the vertex operators.  The relevant worldsheet integral for the pair of detector operators takes the form
\be\label{eq:ds}
\int_0^{2 \pi}{d \sigma \over 2 \pi} \Big| 2 \sin {\sigma \over 2} \Big|^{\alpha' k_1 \cdot k_2} &=\frac{2^{\alpha' k_1 \cdot k_2} \Gamma \left(\frac{\alpha' k_1 \cdot k_2}{2}+\frac{1}{2}\right)}{\sqrt{\pi } \Gamma
	\left(\frac{\alpha' k_1 \cdot k_2}{2}+1\right)} \cr &= 1+{\pi^2 \over 24} (\alpha' k_1 \cdot k_2)^2 - {\zeta_3 \over 4} (\alpha' k_1 \cdot k_2)^3 + {19 \pi^4 \over 5760} (\alpha' k_1 \cdot k_2)^4 + ... \ \nn \\ 
&= 1 + \sum_{s=2}^\infty (-1)^s d_s (\alpha' k_1 \cdot k_2)^s \,, \qquad d_s >0 \,.
\ee
For more details, see Eq.~(4.5) and the related discussion in \cite{Hofman:2008ar}.
The small-$\alpha'$ corrections capture higher-derivative terms in the effective action for the probe in the presence of shockwaves in flat space. In AdS, these are dressed by the AdS curvature corrections which are not fixed by the flat-space calculation. 

The dictionary that relates the flat-space result \eqref{eq:ds} to the EEC is as follows:
\be
\label{eq:subflatads}
(\alpha' k_1 \cdot k_2)^s \to {(-1)^s \Gamma(s+3)^2 \over 4} {(\vec n_1 \cdot \vec n_2)^s \over \lambda^{s/2}} \,.
\ee
To obtain this formula, we promote $k_i$ to $i \nabla_i$ and act with it on the shockwave profile in AdS---see formula (4.6) and the discussion around it in \cite{Hofman:2008ar}.

Uplifting the flat-space computation to AdS introduces ambiguities due to the finite AdS curvature effects. The basic observation is that these effects introduce lower powers of $(\vec n_1 \cdot \vec n_2)$ compared to the highest-spin flat-space answer. In this way there is no unique way to uplift the flat-space corrections to AdS\@. However, the highest spin contribution is uniquely fixed by the flat-space limit. We thus substitute $(\vec n_1 \cdot \vec n_2)^s \to {2^s \Gamma(s+1)^2 \over \Gamma(2s+1)} P_s(\cos \theta) + \dots $ up to subleading in spin corrections that we do not control.

We therefore obtain the following prediction for the EEC at strong coupling, derived from the flat-space limit of string scattering in AdS:
\be\label{eq:StrongExp}
\text{EEC}(\theta) = 1+\sum_{s=2}^\infty{d_s \over \lambda^{s/2}}\frac{2^{s-2}\Gamma (s+1)^2 \Gamma (s+3)^2}{\Gamma (2 s+1)} P_{s}(\cos\theta) + \dots \,,
\ee
where the flat-space limit only captures the leading in $\lambda$ behavior of every multipole. In particular, \eqref{eq:StrongExp} is in perfect agreement with \eqref{eq:c3c4}, which was obtained independently from the detector-source OPE\@.  Let us point out that these arguments readily generalize to higher-point energy correlators by considering $\langle 0 | \prod_{i=1}^n  e^{i \vec{k}_i \cdot \vec X(\sigma_i, 0)} | 0 \rangle$ in \eqref{eq:correlatorwsstring}. 

\subsection{Pad\'e model for finite coupling }\label{sec:Pade}
In this section, we introduce Pad\'{e} approximations for the EEC and for the EEC multipoles, based on the perturbative and strong coupling expansions\footnote{In this section we go back to using $g=\sqrt{\lambda}/(4\pi)$ for the coupling.}, as models for their finite-coupling behavior that will be tested against the numerical bootstrap results.  

We consider two Pad\'e models: one constructed in terms of $\gamma_{2,-1}^{(+)}(g)$, the anomalous dimension of the minimal-twist light-ray operator that controls the small-angle behavior of the EEC, and another built in terms of the cusp anomalous dimension $\Gamma_{\text{cusp}}(g)$. Both quantities are known from integrability at any coupling and provide natural variables for our problem. A key difference between the two models is that the $\Gamma_{\text{cusp}}(g)$-based one contains only integer powers $1/g^n$ in the strong-coupling expansion, while the $\gamma_{2,-1}^{(+)}(g)$-based model also includes half-integer powers $1/g^{n/2}$.

Let us consider a Pad\'e model in terms of $\gamma_{2,-1}^{(+)}(g)$, which is the anomalous dimension of the minimal-twist spin-three light-ray operator.  As an illustration, for the EEC multipole $c_2$, we make the following $(3, 3)$ Pad\'e ansatz
\be\label{eq:c2PadeAns}
c^{(\text{Pad\'e})}_{2}(g) = \frac{1}{(2+\gamma_{2,-1}^{(+)})^4}\frac{n_0+n_1\gamma_{2,-1}^{(+)}+n_2(\gamma_{2,-1}^{(+)})^2+n_3(\gamma_{2,-1}^{(+)})^3}{1+d_1\gamma_{2,-1}^{(+)}+d_2(\gamma_{2,-1}^{(+)})^2+d_3(\gamma_{2,-1}^{(+)})^3} \,,
\ee 
where the prefactor ensures that $c_2\sim 1/g^2$ at strong coupling. The weak coupling expansion of $\gamma_{2,-1}^{(+)}$ is given in \eqref{eq:LROPWeakdata} and the strong coupling expansion is \cite{Kologlu:2019mfz}
\be\label{eq:gammaExp}
\gamma_{2,-1}^{(+)} &\underset{g\to\infty}{=}\sqrt{8\pi g}-3+\OO(g^{-1/2}) \,.
\ee 
We then fix the undetermined coefficients in \eqref{eq:c2PadeAns} by imposing 
 \be
  \begin{aligned}
c^{(\text{Pad\'{e}})}_{2}(g)-\sum_{k=0}^3g^{2k}c^{(k)}_2 &\sim\OO(g^8)\,,\qquad g\to0 \,,\cr
c^{(\text{Pad\'{e}})}_{2}(g)-\frac{1}{4g^2}-\frac{\pi^2-15\zeta_3}{8\pi^3g^3} &\sim \OO(1/g^{7/2})\,,\qquad g\to\infty \,.
 \end{aligned}
 \ee
The strong coupling result is expected to hold in the planar limit, while the weak coupling expansion to three loops is also valid in the non-planar regime. The resulting Pad\'{e} model consistently lies within the numerical bounds presented in \secref{sec:bootstrap} and shown in \figref{fig:c2Intro}. Setting $n_3$ and $d_3$ to zero in \eqref{eq:c2PadeAns} and imposing only the leading $\OO(1/g^2)$ behavior at strong coupling modifies the result by less than $1\%$  at intermediate coupling.

For the higher EEC multipoles  $c_s$, we use the $(2, 2)$ Pad\'e ansatz
\be\label{eq:csPadeAns}
c^{(\text{Pad\'e})}_{s}(g) = \frac{1}{(2+\gamma_{2,-1}^{(+)})^{2s}}\frac{n_0+n_1\gamma_{2,-1}^{(+)}+n_2(\gamma_{2,-1}^{(+)})^2}{1+d_1\gamma_{2,-1}^{(+)}+d_2(\gamma_{2,-1}^{(+)})^2}\,,
\ee 
where the coefficients are fixed by matching the strong-coupling asymptotic $\OO(1/g^{s})$, and the weak-coupling expansion up to three loops. For $c_2$, we showed above that including subleading corrections does not significantly affect the Pad\'e approximation. For $c_{s\geq 3}$, such higher-order data are not currently available. In \figref{fig:planar_legendre}, we compare these Pad\'e approximations with the numerical bootstrap bounds in the planar theory. \footnote{By introducing further undetermined coefficients in the Pad\'e ansatz and comparing to the numerical bounds at strong coupling we estimate that $c_3|_{1/g^4}\approx -0.1$. It would be interesting to reproduce this estimate analytically.}

For the Pad\'e approximation of the full $\text{EEC}(z)$, we adopt the following ansatz 
\be\label{eq:eecpade}
\text{EEC}_{\text{Pad\'e}}(z) = 1+\frac{-1+n_1(z)\Gamma_{\text{cusp}}(g)}{1+d_1(z)\Gamma_{\text{cusp}}(g)+d_2(z)\Gamma_{\text{cusp}}(g)^2+d_3(z)\Gamma_{\text{cusp}}(g)^3}\,,
\ee 
where $\Gamma_{\text{cusp}}(g)$ is the cusp anomalous dimension, whose weak-coupling expansion is given in \eqref{eq:GammaCWeak}, and whose strong-coupling behavior is $\Gamma_{\text{cusp}}(g) = 2g+\cdots$ as $g\to\infty$. The parameters of the ansatz are fixed by imposing the following matching conditions:
\be
&\text{EEC}_{\text{Pad\'e}}(z)-\sum_{k=0}^3g^{2k}\text{EEC}^{(k)}(z)\sim\OO(g^8),\qquad\hspace{16mm} g\to0 \,,\cr
&\text{EEC}_{\text{Pad\'e}}(z)-(1+\frac{1}{4g^2}(1-6z+6z^2)))\sim \OO(g^{-3}),\qquad g\to\infty \,,
\ee 
which we only trust away from the kinematical endpoints $z\to0,1$ where large logarithms invalidate the perturbative expansion. The resulting expression for $\text{EEC}_{\text{Pad\'e}}(z)$ develops a spurious pole with a small residue.\footnote{Such poles with small residues due to a zero nearby are known as the Froissart doublets and are well-known in the theory of Pad\'e approximants, see e.g. \cite{Baker_Graves-Morris_1996,Costin:2020hwg,Costin:2020pcj,stahl1997convergence,buslaev2013convergence}.} After subtracting such poles, we observe a good agreement between the Pad\'e model and the bootstrap bounds. In \figref{fig:eecg04Intro}, we show the resulting Pad\'e approximation at the intermediate value of the coupling $g=0.4$ and we find good agreement with the numerical bootstrap bounds. Further comparisons between the Pad\'e approximations and the numerical bounds can be found in \figref{fig:planar_eec} for various values of the coupling.   

In conclusion, the Pad\'e approximations introduced in this section perform remarkably well when compared with the numerical bootstrap bounds presented in \secref{sec:bootstrap}, providing accurate analytic descriptions of the EEC across the entire range of the coupling.

\section{Inversion formula and large spin}\label{sec:inversion}

Even though the EEC is defined on the interval $z \in [0, 1]$, we can contemplate analytically continuing it to complex values of $z$.  Such an analytic continuation, if possible, would allow us to potentially develop dispersion relations for the EEC\@.   With this motivation in mind, there are two natural questions that we can ask about the EEC at finite coupling:\footnote{We thank Cyuan-Han Chang for discussions on these points.}
\begin{itemize}
    \item What is the analytic structure of the EEC at finite coupling in the complex $z$-plane?
    \item What is the behavior of the EEC in the limit $|z| \to \infty$?
\end{itemize}
Regarding the first question, both weak/strong and OPE results are consistent with the assumption that the EEC is analytic everywhere in the complex $z$-plane away from two branch points, one at $z=0$ and one at $z=1$. The answer to the second question is less clear. We show in \appref{app:ComplexAngle} that the large-$|z|$ limit of the EEC is related to the Regge limit of the integrated Mellin amplitude. However, we do not have first-principle arguments to constrain the behavior in this limit. 
We nevertheless conjecture that at finite coupling we have\footnote{In \secref{sec:inversionTest}, we will present a numerical test of this conjecture at finite coupling in the planar limit using the bootstrap results.}
\be
\label{eq:EECregge}
\lim_{|z| \to \infty} {\text{EEC}(z) \over |z|^{2}} = 0 \,.
\ee
Because of this assumption, we can write a dispersive representation of the EEC multipoles $c_{s \geq 2}$.\footnote{More generally, $\lim_{|z| \to \infty} {\text{EEC}(z) \over |z|^{s_0}} = 0$ implies that $c_{s \geq s_0}$ are dispersive.}

In this Section, we use this dispersive representation, together with the known asymptotics  of the EEC in the small-angle and the back-to-back limits, to determine the large-spin behavior of $c_s$. We test the dispersive representation at weak coupling, where the assumptions used in the derivation are manifestly satisfied. As we will see in \secref{sec:nonplanarbootstrap} and \secref{sec:planarbootstrap}, the large-$s$ asymptotics of the EEC multipoles is important when studying the convergence of the numerical bootstrap methods. 

\subsection{Branch points from large-spin multipoles}

As mentioned above, the EEC naturally has two branch points: in the forward limit, at $z = 0$, and in the back-to-back limit, at $z = 1$.  Contributions from finite spin $s$ are manifestly analytic in $z$, since they are polynomials. Therefore, the non-analytic behavior close to the limiting points $z=0,1$ is encoded in the large-spin behavior of $c_s$. 

As a warm-up, let us see how a nontrivial power-like behavior close to $z=0,1$, of the form $1/z^{\#}$ and $1/(1-z)^{\#}$, emerges from the sum over large-spin EEC multipoles. The relevant limit to consider is $s\to\infty$ while keeping $\rho^2=zs^2$ fixed. In this limit, the Legendre polynomials reduce to
\be
P_{s} \Big(\pm\left(1-2\frac{\rho^2}{s^2}\right) \Big)  \approx (\pm 1)^s J_0(2\rho) \,, \qquad \text{as $s \to \infty$} \,,
\ee 
where $J_0(x)$ is a Bessel function.  Now assume that for large $s$, the EEC multipoles behave as 
\be
c_s\approx s^p \Big( c_+ {1 + (-1)^s \over 2}+c_- {1 - (-1)^s \over 2} \Big) \,,
\ee
for some power $p$. For even $s$, only $c_+$ contributes, while for odd $s$, only $c_-$ does. We can then approximate the sum in the EEC multipole decomposition \eqref{eq:leg} by an integral to obtain
\be 
\label{eq:largespineectoy}
\text{EEC}(z) &\approx {1 \over 2}\left( \frac{c_++c_-}{z^{\frac{1}{2}(1+p)}}+\frac{c_+-c_-}{(1-z)^{\frac{1}{2}(1+p)}} \right) \int_0^\infty d\rho \rho^p J_0(2\rho) \cr 
&= {1 \over 2}\left( \frac{c_++c_-}{z^{\frac{1}{2}(1+p)}}+\frac{c_+-c_-}{(1-z)^{\frac{1}{2}(1+p)}} \right) \frac{\Gamma \left(\frac{p+1}{2}\right)}{2 \Gamma \left(\frac{1}{2}-\frac{p}{2}\right)} \,.
\ee 
What we learn from this expression is that the singular behavior in the small-angle region is controlled by the {\em average} over even and odd spin partial waves, whereas the singular behavior in the back-to-back region is controlled by the {\em difference} between even and odd spin partial waves. 

We consider the contribution from a twist-2 trajectory with $\tau=2+\gamma$ to the light-ray OPE, then the small angle behavior is a power law $z^{-1+\frac{\gamma}{2}}$. Comparing this power law to \eqref{eq:largespineectoy} we find that $p=1-\gamma$. In particular, when $\gamma=0$ the EEC multipoles are linear in $s$ at large $s$, consistent with the free theory result $c_{2k}=4k+1$. Note also that the EEC multipoles $c_s$ decay faster with $s$ as $\gamma$ increases, consistent with the free-theory case $\gamma=0$ being the extremal solution that maximizes $c_s$---see \appref{app:MultipoleBounds}.

Before proceeding, let us also illustrate how this mechanism works in terms of the superconformal blocks in the free theory. Taking the large-$J$ limit of the twist-two contribution, we find 
\be
(\lambda_{\tau,J}^{(0)})^2 &=\sqrt{\pi }  \frac{2^{-J-1} \sqrt{J}}{(N_c^2-1)}+\cdots \,, \cr
\alpha(\tau,J)\sin\left(\frac{\pi \tau}{2}\right)^2 \Big|_{\tau=2} &= \frac{2^{3 J+6}(N_c^2-1)}{\pi}+\cdots \,, \cr
f_{J+2,s}(\tau=2) &=\sqrt{\pi } 2^{-2 (J+2)} \sqrt{J}+\cdots \,,
\ee  
where $f_{J+2,s}(\tau)$ is defined in \eqref{eq:superblock}, and identifying $s=J+2$ at zero coupling leads to 
\be
c_{s}|_{\tau=2}=\left(1+(-1)^s \right) s+\cdots \,,
\ee 
to leading order in $s\gg 1$. In terms of the formulas above, this corresponds to $c_+ = 2$, $c_- = 0$ and $p=1$. 

\subsection{Inversion formula for EEC multipoles}

In \appref{app:FGeecmultipoles}, we derive the Froissart-Gribov inversion formula for the EEC multipoles, following the standard steps \cite{Gribov:2003nw}. The analytic continuation is done separately for even and odd  EEC multipoles, which we denote by $c_s^+$ and $c_s^-$, respectively.  We obtain (for details, see \appref{app:FGeecmultipoles}):
\be
\label{eq:inversionmultipole}
c_s^{\pm} =2 {2s+1 \over \pi} \int_{-\infty}^{0} d\tilde z\,  Q_s(1-2 \tilde z) \Big( {\rm Disc}_{z=0} \text{EEC}(\tilde z) \pm {\rm Disc}_{z=1} \text{EEC}(1- \tilde z) \Big) \,, 
\ee
where the Legendre-$Q$ functions are defined in \eqref{eq:legendreQ}. The discontinuities are defined as 
\be
{\rm Disc}_{z=0} \text{EEC}(z) &= {\text{EEC}(z-i 0) - \text{EEC}(z+i0) \over 2 i} \,, \cr
{\rm Disc}_{z=1} \text{EEC}(1-z) &= {\text{EEC}(1-z+i 0) - \text{EEC}(1-z-i0) \over 2 i} \,. 
\ee
The representation \eqref{eq:inversionmultipole} is analytic in the complex spin variable $s$. In deriving this formula, we  dropped the arc at infinity using \eqref{eq:EECregge}, and consequently the result \eqref{eq:inversionmultipole} is valid only for ${\rm Re}[s]>1$. 

Let us point out that the discontinuity of the EEC in $z$ is related to the quadruple discontinuity of the correlator discussed in \cite{Caron-Huot:2017vep}. Indeed, the EEC itself is given by the source-detector channel double discontinuity of the correlator, and its further discontinuity in $z$ vanishes for any finite sum of the source-detector channel conformal blocks, which is also a defining property of the quadruple discontinuity. It is therefore very interesting that the EEC satisfies dispersion relations, reminiscent of the Mandelstam representation for scattering amplitudes. 

\paragraph{Large-spin multipole contribution from the small angles}\mbox{}\\
Next, we consider the contribution of a simple power-like behaviour $\text{EEC}(z) \sim z^{\frac{\gamma}{2}-1}$ to the inversion integral above. In this case, the discontinuity is equal to ${\rm Disc}_{z=0}  \text{EEC}(z) \sim (\sin \frac{\pi \gamma}{2}) (-z)^{\frac{\gamma}{2}-1}$. For this case, the integral \eqref{eq:inversionmultipole} can be evaluated exactly (see Eq.~(6.22) in \cite{Correia:2020xtr}), leading to
\be\label{eq:CsExactLROPE}
 \frac{c_s^+ + c_s^-}{2} =\frac{(2 s+1) \Gamma (\frac{\gamma}{2}) \Gamma (s-\frac{\gamma}{2} +1)}{\Gamma (1-\frac{\gamma}{2} ) \Gamma (s+\frac{\gamma}{2}+1)} \,.
\ee
Setting $\gamma = 0$ reproduces the free-theory result $(c_s^+ + c_s^-)/2 = 2s + 1$. On the other hand, in the limit $s \to \infty$  we obtain
\be
\label{eq:naiveasymcs}
\frac{c_s^+ + c_s^-}{2} \approx \frac{2 \Gamma (\frac{\gamma}{2}) s^{1-\gamma }}{\Gamma (1-\frac{\gamma}{2})} \,.
\ee
However, Eq.\ \eqref{eq:naiveasymcs} cannot be valid for arbitrary $\gamma$, since the EEC multipoles $c_s$ would become negative for $\gamma>2$, in contradiction with unitarity. 
In order to see that the EEC multipoles $c_s$ are indeed positive, recall that the light-ray OPE formula in \eqref{eq:leadtraj} \cite{Kologlu:2019bco} includes a nontrivial pre-factor that we did not take into account. Restoring the pre-factor and reinstating a general three-point function, we obtain
\be
\label{eq:asymptoticcs}
\frac{c_s^+ + c_s^-}{2} \approx a_{2,-1}^{(+)}\frac{128\pi^4 \Gamma (\gamma +2) s^{1- \gamma }}{\gamma ^2 (\frac{\gamma}{2} +1)} \sin ^2 \left(\frac{\pi  \gamma}{2} \right ) \,,
\ee
and the $c_s$ are positive. We expect this formula to be reliable when $s$ is the largest parameter in the problem. In particular, in the strong-coupling limit with $s$ fixed and $g \to \infty$, Eq. \eqref{eq:asymptoticcs} no longer holds.

\paragraph{Large-spin multipole contribution from the back-to-back region}\mbox{}\\
The discontinuity in the back-to-back region is responsible for the asymmetry between odd- and even-spin contributions in \eqref{eq:inversionmultipole}. In the context of the inversion formula, we are interested in the part of the EEC that is non-analytic in $z$ near $z=1$.

Inserting the leading back-to-back-region asymptotic formula \eqref{eq:BB} into the inversion integral \eqref{eq:inversionmultipole},\footnote{We have found it easier to perform the spin-$s$ projection directly and then compared it to the evaluation of the inversion integral.} we find that the contribution of the back-to-back cut takes the form
\be \label{eq:asympback}
(c_s^+ - c_s^-)/2 =  H(g) (2s+1) e^{-2 \Gamma_{\text{cusp}}(g) \log^2(2s/b_0)} \left( {2s \over b_0} \right)^{-2 \Gamma_{\text{coll}}(g)} \,. 
\ee

\paragraph{Summary}\mbox{}\\

Adding together \eqref{eq:asymptoticcs} and \eqref{eq:asympback}, we obtain the following result for large-spin multipole coefficients of the EEC at finite coupling
\be\label{eq:LargeSFinal}
c_s^{\pm} = {(2s+1) \over 2} \Big(a_{2,-1}^{(+)}\frac{128\pi^2\Gamma (\gamma +2) s^{-\gamma }}{\gamma ^2 (\frac{\gamma}{2} +1)}  \sin ^2(\frac{\pi  \gamma}{2}) \pm H(g) e^{-2 \Gamma_{\text{cusp}}(g) \log^2(2s/b_0)} \left( {2s \over b_0} \right)^{-2 \Gamma_{\text{coll}}(g)} \Big) \,.
\ee
In the planar limit, $\Gamma_{\text{cusp}}(g)$, $\Gamma_{\text{coll}}(g)$ and $\gamma=\gamma_{2,-1}^{(+)}$ are known for finite-coupling from integrability \cite{Beisert:2006ez,Freyhult:2007pz,Freyhult:2009my,Gromov:2014bva} while $a_{2,-1}^{(+)}$ and $H(a)$ are only known to the first few orders at weak coupling. We will test this formula at finite coupling in \secref{sec:InvertedBoots} in the planar limit.

\subsection{Weak-coupling test}\label{subsec:InvWeakCoupling}
We now test Eq. \eqref{eq:LargeSFinal} explicitly in perturbation theory. The relevant data at weak coupling is given by \cite{Kotikov:2004er,Beisert:2006ez,Freyhult:2007pz,Freyhult:2009my,Fioravanti:2009xt,Eden:2012rr,Alday:2013cwa,Korchemsky:2021okt}
\be\label{eq:GammaCWeak}
H(g) &=1-\frac{2 \pi ^2}{3} g^2+\frac{8 \pi ^4}{9} g^4-64 \left(\frac{17 \zeta _3^2}{12}+\frac{197 \pi ^6}{10080}\right) g^6+\cdots \,, \cr
\Gamma_{\text{cusp}}(g)&=4 g^2-\frac{4 \pi ^2}{3}g^4+\frac{44 \pi ^4}{45} g^6-256 \left(\frac{\zeta _3^2}{8}+\frac{73 \pi ^6}{20160}\right) g^8+\cdots \,, \cr
\Gamma_{\text{coll}}(g)&=-24 \zeta _3 g^4+64 \left(\frac{\pi ^2 \zeta _3}{12}+\frac{5 \zeta _5}{2}\right)
   g^6-256 \left(\frac{7 \pi ^4 \zeta _3}{480}+\frac{5 \pi ^2 \zeta _5}{48}+\frac{175 \zeta _7}{32}\right) g^8 \,, \cr
   &{}+\cdots \,,,
\ee 
and
\be\label{eq:LROPWeakdata}
\gamma_{2,-1}^{(+)} &= 8g^2+16(-\zeta_3+\frac{\pi^2}{3}-4)g^4+64(-3\zeta_3+3\zeta_5-\frac{4\pi^2}{3}+\frac{\pi^4}{120}+16)g^6+\cdots \,,\cr
\frac{a^{(+)}_{2,-1}}{a^{(+,0)}_{2,-1}} &=1-8g^2+16 \left(\frac{\zeta _3}{2}+\frac{11 \pi ^4}{360}-\frac{2 \pi ^2}{3}+12\right) g^4\cr
&{}+64\left(\frac{3 \zeta _3^2}{4}+6 \zeta _3+\pi ^2 \left(6-\frac{7 \zeta _3}{6}\right)+12 \zeta _5-\frac{109 \pi ^6}{7560}-\frac{\pi ^4}{24}-80\right) g^6+\cdots \,,
\ee 
where $a_{2,-1}^{(+,0)}=\frac{1}{32\pi^4}$.

\begin{figure}[]
    \centering
    \begin{subfigure}{.45\linewidth}
        \centering
        \includegraphics[width=\linewidth]{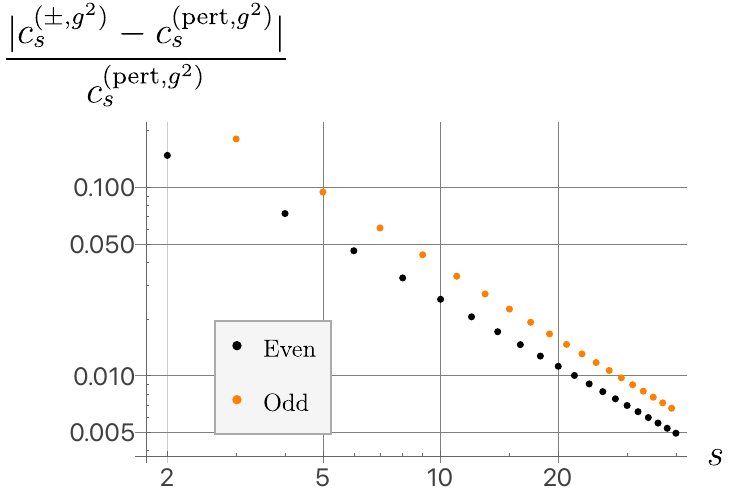}
        \caption{}
    \end{subfigure}
    \hfill
    \begin{subfigure}{.45\linewidth}
        \centering
        \includegraphics[width=\linewidth]{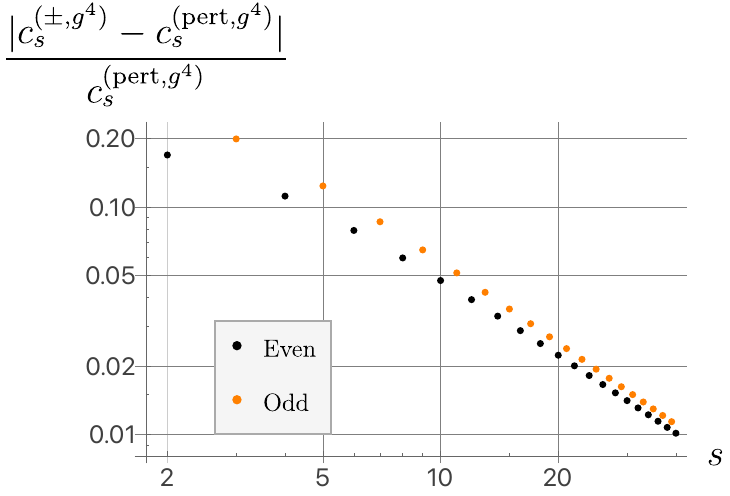}
        \caption{}
    \end{subfigure}
    \caption{Comparing the perturbative result $c_s^{(\text{pert})}$ obtained from decomposing the known results for the perturbative expansion of $\text{EEC}(z)$, including the distributional terms, with $c_s^{\pm}$ from \eqref{eq:LargeSFinal} at $\OO(g^2)$ in (a) and $\OO(g^4)$ in (b). Note that \eqref{eq:LargeSFinal} assumes large $s\gg1$ and consistent with the figures, on the other hand \eqref{eq:inversionmultipole} is valid at finite $s$. } 
    \label{fig:cs1}
\end{figure}

At zero coupling, we find that $c_s^+ = 2s+1$ and $c_s^-=0$ as expected. At small $g$, we compare in \figref{fig:cs1} the formula above against explicit results obtained from the known perturbative EEC reviewed in \secref{sec:Review}. We find good agreement up to relatively low spin.

\FloatBarrier

\section{Numerical Bootstrap}\label{sec:bootstrap}

In this section, we use the numerical conformal bootstrap to put constraints on the EEC and its multipoles in $\SU(N_c)$ $\mathcal{N} = 4$ SYM theory. In \secref{sec:bootstrap_setup}, we discuss the bootstrap inputs. 
Then, in \secref{sec:nonplanarbootstrap}, we present numerical results for $N_c=2,3,\ldots,10$, and in \secref{sec:planarbootstrap} we present numerical results in the planar limit.  The results presented in each case are as functions of the coupling $g \equiv g_\text{YM}\sqrt{N_c} / (4\pi)$. For finite $N_c$, our results are invariant (by construction) under the S-duality transformation that sends $g_\text{YM} \mapsto 4\pi/g_\text{YM}$ (or equivalently $g \mapsto N_c/(4\pi g)$).\footnote{It is also possible to obtain bounds at finite $N_c$ as a function of the complexified coupling ${\theta \over 2 \pi} + {4 \pi i \over g_{\text{YM}}^2}$, and then similarly these bounds are invariant under the full $\text{SL}(2,\mathbb{Z})$ S-duality group. We have observed that our bounds have only very small dependence on the $\theta$-angle in the $\text{SL}(2,\mathbb{Z})$ fundamental domain, so we set the $\theta$-angle to zero and study only the dependence on the Yang-Mills coupling.}

\subsection{$\cN = 4$ superconformal bootstrap}
\label{sec:bootstrap_setup}

The $\mathcal{N} = 4$ superconformal bootstrap was introduced in \cite{Beem:2013qxa,Beem:2016wfs}, where  crossing symmetry and unitarity were used to place bounds on the conformal data of $\mathcal{N} = 4$ superconformal field theories with arbitrary gauge group $G$\@.  The information about $G$ was inputted through the conformal anomaly coefficient $c = (\dim G)/4$.  The bounds of \cite{Beem:2013qxa,Beem:2016wfs} thus applied to all theories with the same value of $c$, and in particular they were not sensitive to the (complexified) Yang-Mills coupling, which is an exactly marginal parameter for any $G$\@.  These studies were later augmented with integral constraints  from supersymmetric localization \cite{Binder:2019jwn,Chester:2020dja}, which allow for more refined bootstrap bounds that depend on the Yang-Mills coupling \cite{Chester:2021aun,Chester:2023ehi}. 

In \cite{Caron-Huot:2022sdy}, it was first shown how to apply the superconformal bootstrap directly in the $\SU(N_c)$ ${\cal N} = 4$ SYM theory in the planar limit ($N_c\to\infty$), where the bootstrap constraints can also be augmented with spectral input from integrability; this work also introduced the use of dispersive sum rules in the $\mathcal{N} = 4$ bootstrap. In \cite{Caron-Huot:2024tzr}, the setup for the planar theory was extended to use integral constraints as well.

In this work, we bootstrap the EEC and its multipoles both at finite $N_c$ and in the planar limit, using the setups described in \cite{Chester:2023ehi} and \cite{Caron-Huot:2024tzr}, respectively. At finite $N_c$, we include one additional type of constraint, introduced in this work, that follows from the ANEC\@. In the planar limit, this constraint still holds but we find that including it in the numerical problem does not improve our bounds.

In the remainder of this subsection, we give brief reviews of the four types of constraints that appear in the bootstrap in this paper: crossing symmetry, the new ANEC constraint, dispersive sum rules, and integral constraints. We also review the spectral input from integrability used in the planar case. We then give the primal and dual formulations of the bootstrap problems that we solve in the following subsections.

\subsubsection{Crossing symmetry with derivative functionals}\label{sec:crossing_nonplanar}

The crossing symmetry of the four-point function $\langle \cO(x_1,y_1)\cdots \cO(x_4,y_4)\rangle$ was used in \cite{Beem:2013qxa,Beem:2016wfs} to obtain the original construction of the $\mathcal{N} = 4$ superconformal bootstrap. By combining the crossing relation \eqref{eq:h_crossing} and the OPE decomposition \eqref{eq:h_expansion}, we find the crossing equation
\begin{equation}\label{eq:nonplanar_crossing} 
    F^\text{protected}(N_c; u,v) + \sum_{\tau,J} \lambda^2_{\tau,J} F_{\tau,J}\left(u,v\right) = 0\,,
\end{equation}
where $\tau$ is the twist of the exchanged operators and the sum is over the superconformal primaries of long multiplets. The contribution of the protected operators $ F^\text{protected}(N_c;u,v)$ and the superconformal blocks $F_{\tau,J}(u,v)$ take the form 
\begin{equation}
\begin{split}
    F^\text{protected}(N_c;u,v) &\equiv u^4 v^4 \left(\cH^\text{protected}(u,v) - \cH^\text{protected}(v,u)\right) \,, \\
        F_{\tau,J}(u,v) &\equiv v^4 G_{\tau+J+4, J}(u,v) - u^4 G_{\tau+J+4, J}(v,u)\,.
\end{split}
\end{equation}
For any given spectrum of the exchanged superconformal primaries, the equation \eqref{eq:nonplanar_crossing} gives an infinite number of linear constraints, one for each $(u,v)$, on the squared OPE coefficients $\lambda^2_{\tau,J}$.

To use these constraints in practice, we must truncate the infinite set to a finite subset. We follow the standard procedure of taking derivatives with respect to $\bm{z}$ and $\bm{\bar{z}}$, defined by $u = \bm{z}\bm{\bar{z}}$ and $v=(1-\bm{z})(1-\bm{\bar{z}})$, around the crossing-symmetric point $\bm{z} = \bm{\bar{z}} = \frac{1}{2}$. We denote these derivatives by
\begin{equation}
F^{\text{protected}}_{m,n} \equiv \left. \partial_{\z}^m \partial_{\zb}^n F^\text{protected}(\z,\zb)\right|_{\z=\zb = \frac{1}{2}}\,, \qquad   F_{\tau,J;m,n} \equiv
\left.\partial_{\z}^m \partial_{\zb}^n F_{\tau,J}(\z,\zb)\right|_{\z=\zb = \frac12} \,.
\end{equation}
It is straightforward to check that $F_{\tau,J;m,n}$ is only nonzero when $m+n$ is odd, and that $F_{\tau,J;m,n} = F_{\tau,J;n,m}$. We can therefore restrict to $(m+n)$ odd and $m>n$. We then truncate the system of equations by keeping only the constraints
\begin{equation}
\label{eq:crossder}
    F^{\text{protected}}_{m,n}(N_c) + \sum_{\tau,J} \lambda^2_{\tau,J} F_{\tau,J;m,n} = 0 \,,
\end{equation}
for which $m+n \le \Lambda$, where $\Lambda$ is our truncation parameter. 

Note that in this paper we will work with $\Lambda$ up to 107, meaning we include up to 1,485 constraints of the form \eqref{eq:crossder}. This value of $\Lambda$ is larger than in many other conformal bootstrap studies; we find that this high truncation parameter is necessary to achieve convergence of the bootstrap bounds in this paper. Even despite this high order, some of the finite-$N_c$ bounds in \secref{sec:nonplanarbootstrap} are not fully converged.

\subsubsection{Constraints from ANEC}\label{sec:anec_constraint}
The ANEC is the statement that for any state $| \psi \rangle$, $\langle \psi | {\cal E}(\vec n) | \psi \rangle \geq 0$. It has been argued in \cite{Faulkner:2016mzt} and \cite{Hartman:2016lgu} that this is a property of any unitary QFT (or CFT). In \cite{Kologlu:2019bco} it was moreover argued that $[{\cal E}(\vec n_1), {\cal E}(\vec n_2)]=0$. It then follows that\footnote{Here we ignore subtleties related to the fact that ${\cal E}(\vec n)$ acts in an infinite-dimensional Hilbert space and assume that \eqref{eq:aneccondition} holds based on physical grounds.}
\be
\label{eq:aneccondition}
\langle \psi | {\cal E}(\vec n_1) {\cal E}(\vec n_2) | \psi \rangle \geq 0  \,.
\ee
For the simplest case of $| \psi \rangle$ given by the local operator acting on the vacuum, \eqref{eq:aneccondition} implies an infinite set of positivity constraints on the four-point function. These constraints \emph{do not} follow from the standard unitarity bounds on the scaling dimensions used in the conformal bootstrap. Therefore, they can be added to the standard conformal bootstrap analysis and have a potential of improving the bootstrap bounds. We will find situations in which it is indeed the case in our analysis. 

In \eqref{eq:leg} and \eqref{eq:csIntro}, we gave the expression of the EEC in terms of the OPE coefficients. The averaged null energy condition \eqref{eq:aneccondition} implies that $\text{EEC}(\theta) \ge 0$, which gives a linear inequality for the OPE coefficients at any value of $\theta$.

Instead of imposing these constraints directly, we can rearrange them to give upper bounds on each $c_s$. Indeed, we show in \appref{app:MultipoleBounds} that given the representation \eqref{eq:leg}, along with the constraint $\text{EEC}(\theta) \ge 0$, each $c_s$ has some maximum possible value, which we will call $c_{s,\text{max}}$. For example, when $s$ is even, $c_{s,\text{max}} = 2s+1$.

By combining this inequality with the OPE decomposition of the EEC \eqref{eq:csIntro}, we find
\begin{equation}\label{eq:nonplanar_anec}
    \sum_{\tau,J} \lambda^2_{\tau,J} \sin^2\frac{\pi \tau}{2} \alpha(\tau,J) f_{J+2,s}(\tau) - c_{s,\text{max}} \le 0\,.
\end{equation}
It is also possible to derive upper bounds on linear combinations of the EEC multipoles, which can written in terms of OPE coefficients in a very similar way.

When $N_c$ is finite, we are only able to obtain lower bounds on $c_s$ using crossing symmetry and integral constraints, and so the ANEC constraint \eqref{eq:nonplanar_anec} provides new information. We find that these constraints are \emph{necessary} for us to obtain bounds on a smeared EEC observable, as we will explain below in \secref{sec:nonplanarbootstrap}. By contrast, in the $N_c\to\infty$ limit, the additional inputs we include (dispersive sum rules and spectral input from integrability, described below) allow us to obtain upper bounds on $c_s$ that are already stronger than \eqref{eq:nonplanar_anec}, and so adding \eqref{eq:nonplanar_anec} to our numerical problem does not improve any bounds.

\subsubsection{Dispersive sum rules}\label{sec:dispersive}

Here, we review various families of functionals derived from CFT dispersion relations \cite{Caron-Huot:2020adz, Penedones:2019tng}. Let us note that in terms of rigor, these functionals are on less firm footing relative to the derivative functionals \eqref{eq:crossder}. The reason is that the CFT dispersion relations (to the best of our knowledge) have not been proven starting from the CFT axioms \cite{Kravchuk:2021kwe}. We will simply assume that the dispersive sum rules are swappable in the sense of \cite{Qiao:2017lkv} (see also \cite{Caron-Huot:2020adz}), in light of evidence that they lead to correct physical results in known theories, such as the 3d Ising model and ${\cal N}=4$ SYM \cite{Caron-Huot:2022sdy,Ghosh:2023onl,Caron-Huot:2024tzr}. Let us also emphasize that while we will use dispersive sum rules only in the planar limit, they are expected to hold in the finite-$N_c$ theory as well.

The dispersive functionals are the essential tools that make the conformal bootstrap in the planar limit feasible \cite{Caron-Huot:2022sdy,Caron-Huot:2024tzr}: they decouple the even-twist double-trace operators and recast crossing symmetry in a form that involves only single-trace data. 

\paragraph{Dispersive crossing}\mbox{}

We first discuss the sum rules that implement crossing symmetry in the planar limit using single-trace operators only. To this end, it is convenient to introduce the Polyakov-Regge expansion of the correlator. This representation is obtained by substituting the usual OPE decomposition \eqref{eq:h_expansion} into the CFT dispersion relations of \cite{Caron-Huot:2020adz}, which yields
\begin{equation}\label{eq:transformed_h_expansion}
    \cH(u,v)
    = \cH^\text{strong}(u,v)
    + \sum_{\tau,J} \tilde{\lambda}^2_{\tau,J}\,
      \cP^{\cN=4}_{\tau,J}(u,v)\,,
\end{equation}
where $\cP^{\cN=4}_{\tau,J}(u,v)$ denotes the Polyakov--Regge block associated with a single conformal block. It is defined by inserting that conformal block into the CFT dispersion relations, see e.g.\ (2.17) in \cite{Caron-Huot:2022sdy}. Equivalently, $\cP^{\cN=4}_{\tau,J}(u,v)$ can be computed in Mellin space using the representation given in \eqref{eq:p_integral}. The function $\cH^\text{strong}(u,v)$ is the $g\to\infty$ limit of the reduced correlator, given by $\cH^\text{strong} = -\bar{D}_{2,4,2,2}$; it is likewise obtained by plugging $\cH^\text{protected}(u,v)$ inside the dispersive integral. It can be argued, see \cite{Caron-Huot:2022sdy}, that the expansion \eqref{eq:transformed_h_expansion} is convergent for $(u,v)$ in the Euclidean region
\begin{equation}
\label{uv-euclidean}
    u,v>0 \text{ real}\quad \text{and}\quad 4uv\geq(1-u-v)^2 \,.
\end{equation}
This is the region within which $\zb=\z^*$. 

We can use the crossing relation $\cH(u,v) = u^{-4} \cH(\tfrac{1}{u},\tfrac{v}{u})$ to write the dispersive crossing sum rules.\footnote{We use this crossing because the Polyakov-Regge blocks are individually symmetric under $u\leftrightarrow v$.} Importantly, under the crossing transformation $(u,v) \to (\tfrac{1}{u},\tfrac{v}{u})$ the Euclidean region \eqref{uv-euclidean} is mapped into itself. Therefore we can apply the Polyakov-Regge expansion on both sides of the crossing relation. In this way, we obtain the dispersive crossing sum rules. 

In this paper, we will be only using them in the planar limit. In this case, we get
\begin{equation}\label{eq:constraint_X}
    \sum_{\st\tau,J} \tilde{\lambda}^2_{\tau,J} X_{\tau,J}(u,v) = 0 \,,\qquad\text{with}\qquad X_{\tau,J}(u,v) \equiv \cP^{\cN=4}_{\tau,J}(u,v) - \frac{1}{u^4} \cP^{\cN=4}_{\tau,J}(\tfrac{1}{u},\tfrac{v}{u}) \,.
\end{equation}
The essential improvement of \eqref{eq:constraint_X} over \eqref{eq:nonplanar_crossing}
is that the sum only involves the single-trace operators. The restriction to single trace operators is the consequence of the fact that the Polyakov-Regge blocks vanish for $\tau \in 2\mathbb{Z}$. In \eqref{eq:p_integral}, we give the explicit Polyakov-Regge block useful for numerical evaluation as a Mellin integral.

\paragraph{Mellin sum rules}\mbox{}

In addition to the dispersive crossing sum rules, we can use the enhanced Regge-boundedness of the reduced correlator to write CFT dispersive sum rules based on anti-subtracted dispersion relations. This  is most easily shown in Mellin space. The Mellin amplitude $M(s,t)$ is defined by
\be
\label{eq:mellinH}
\mathcal{H}(u,v) = 
\iint \frac{ds\, dt}{(4\pi i)^2}\;
u^{\frac{s}{2}-4}\,
v^{\frac{t}{2}-4}\,
\Gamma\left(4 - \frac{s}{2}\right)^{2}
\Gamma\left(4 - \frac{t}{2}\right)^{2}
\Gamma\left(4 - \frac{u}{2}\right)^{2}
\, M(s,t)\,,
\ee
where $s+t+u=16$ and the contour is taken such that the OPE is correctly reproduced \cite{Penedones:2019tng}.
The Mellin amplitude is fully crossing-symmetric:
\be
\label{eq:crossingM}
M(s,t) = M(t,s) = M(s,u) \,.
\ee
As a reference point, at leading order at strong coupling we have the supergravity contribution
\be
\label{eq:mellinstrong}
M_{\text{strong}}(s,t) = {1 \over ({s \over 2}-3)({t \over 2}-3)({u \over 2}-3)} \,. 
\ee
At finite $g$, the Mellin amplitude in the fixed-$t$ Regge limit satisfies
\begin{equation}
\label{eq:regge}
    \lim_{|s|\to\infty} M(s,t) \sim |s|^{J_* - 4} \,, \qquad\text{with}\qquad J_* < 2\,,
\end{equation}
when ${\rm Re}[t]<6$.\footnote{In the strong-coupling limit $g \to \infty$, $J_*=2$.} The crossing symmetry \eqref{eq:crossingM} implies that the same behavior applies in the $u$-channel Regge limit (where $s \to \infty$ while keeping $u$  fixed).  

Let us recall that the standard causality constraint \cite{Camanho:2014apa}, or the `bound on chaos' \cite{Maldacena:2015waa}, implies that in the planar limit $M(s,t) \lesssim s^2$. Instead, in \eqref{eq:regge} we get enhanced Regge-boundedness (an extra $-4$ in the exponent). This originates from the phenomenon of superconvergence \cite{Kologlu:2019bco}, and the fact that the four-point function of interest can be related by supersymmetry to the four-point function of stress-energy tensors that satisfies the standard Regge boundedness.

The Regge bound \eqref{eq:regge} allows one to write various sum rules; here we follow \cite{Caron-Huot:2022sdy}. For example, we can consider the following identity
\be
\label{eq:sumrule}
\oint_{\infty} {ds' \over 2 \pi i} (s'-t) M(s', s+t-s') = 0 \,,
\ee
where the contour is a large circle and we used the fixed-$u$ Regge limit. By deforming the contour away from infinity we pick up the residues from the singularities of the Mellin amplitude,
\begin{equation}\label{eq:mellin_poles}
M(s,t) \sim {\tilde \lambda^2_{\tau,J} {\cal Q}_{\tau+4,J}^m(t) \over s-\tau-4-2m}\,,
\end{equation}
which are fully fixed by the OPE data and the kinematical Mack polynomials $\mathcal{Q}^m_{\tau+4,J}(t)$. In this way we obtain the family of sum rules
\begin{equation}\label{eq:bhatt_sum}
    \hat{B}^\text{protected}_t + \sum_{\st \tau,J} \tilde{\lambda}^2_{\tau,J} \hat{B}_{\tau,J}(t) = 0\,,
\end{equation}
where, following \cite{Caron-Huot:2022sdy}, we have set $s=6$ in \eqref{eq:sumrule}, and we multiplied \eqref{eq:sumrule} by ${1 \over t-6}$. Here, $\hat{B}^\text{protected}_t = \frac{8}{(t-4)(t-6)}$ and $\hat{B}_{\tau,J}(t)$ is given explicitly in \eqref{eq:bhat_b_defs}. The protected part $\hat{B}^\text{protected}_t$ originates from the $s'=6$ pole of the strong-coupling Mellin amplitude \eqref{eq:mellinstrong}, and the sum in \eqref{eq:bhatt_sum} is over the long single-trace superconformal multiplets. We can also perform the inverse Mellin transform and write a position-space version of this constraint,
\begin{equation}\label{eq:bv_sum}
    0 = B_v^\text{protected} + \sum_{\st \tau,J} \tilde{\lambda}^2_{\tau,J} B_{\tau,J}(v)\,;
\end{equation}
here $B_v^\text{protected} = \frac{v^2 - 1 - 2v\log v}{v(1-v)^3}$, and $B_{\tau,J}(v)$ can also be found in \eqref{eq:bhat_b_defs}.

In addition, we include functionals denoted $\Psi_\ell$ and $\Phi_{\ell,\ell+2}$ that analytically bootstrap the leading-order perturbative results for the scaling dimensions and OPE coefficients of twist-two operators. These functionals are obtained by integrating $\hat{B}_t$ over $t$ against specific kernels constructed to have chosen properties;
\begin{equation}
\begin{split}
\Psi_{\ell;\tau,J} &= \delta_{\ell,J}\left( 1+ (\tau-2)\beta_\ell \right)+O((\tau-2)^2)\,,\\
 \Phi_{\ell,\ell+2;\tau,J} &= 0 + (\tau-2) \left( \delta_{\ell,J} - \frac{\Phi^\infty_\ell}{\Phi^\infty_{\ell+2}}\delta_{\ell+2,J}\right) + O((\tau-2)^2)\,.
\label{Phi-features}
\end{split}
\end{equation}
$\Phi^\infty_\ell$ and $\beta_{\ell}$ are respectively given in \eqref{eq:phiellinft} and \eqref{betal}. The constraints derived in this way take the form
\begin{equation}
\begin{split}
    \Psi^\text{protected}_\ell + \sum_{\st \tau,J} \tilde{\lambda}^2_{\tau,J} \Psi_{\ell;\tau,J} &= 0\,,\\
    \sum_{\st \tau,J} \tilde{\lambda}^2_{\tau,J}\Phi_{\ell,\ell+2;\tau,J} &= 0\,,
\end{split}
\end{equation}
where $\Psi^\text{protected}_\ell = -\frac{2\Gamma(\ell+3)^2}{\Gamma(2\ell+5)}$ and $\Psi_{\ell;\tau,J}$ and $\Phi_{\ell,\ell+2;\tau,J}$ can be found in \eqref{eq:psi_phi_defs}. In this paper we use the functionals $X_{u,v}$, $B_v$, $\hat{B}_t$, $\Psi_\ell$, and $\Phi_{\ell,\ell+2}$ to derive bounds in the planar limit. The specific list of functionals used is given in \tabref{tab:planar_fnals} of \appref{app:bootstrap}. 

The dispersive functionals we have discussed here can in principle also be used to bootstrap $\cN = 4$ SYM at finite $N_c$. We leave this direction to future work \cite{wip}.

\subsubsection{Integral constraints}\label{sec:integral_constraints}
To study the theory at a fixed coupling $g$, we need to impose additional constraints that are sensitive to it. This approach was initiated in \cite{Chester:2021aun} through the use of integral constraints on the scalar four-point function derived using supersymmetric localization \cite{Binder:2019jwn,Chester:2020dja}. These constraints provide the exact value (for any $N_c$ and $g$) of two integrals of $\cH(u,v)$ with particular supersymmetry-preserving measures. When applied to the OPE expansion \eqref{eq:h_expansion}, they take the form
\begin{equation}\label{eq:nonplanar_integral}
     I^\text{protected}_{k}(N_c) - \mathcal{F}_k(N_c, g) + \sum_{\tau,J} \lambda^2_{\tau,J} I_{k,\tau,J} = 0 \,,
\end{equation}
with $k=2,4$ labeling the number of mass derivatives of the sphere free energy needed to derive the constraint. The functions $\mathcal{F}_k(N_c, g)$ are known exactly from supersymmetric localization. They are expressed in terms of $(N_c - 1)$-dimensional integrals. An efficient method for calculating them is described in \cite{Chester:2023ehi}.

To write down the integral constraints explicitly, it is convenient to introduce the following two integral operators $I_2$ and $I_4$ \cite{Binder:2019jwn,Chester:2020dja}:
\be
        I_2[f] &\equiv -\frac{2}{\pi}\int dr\,d\theta \left.\frac{r^3 \sin^2\theta}{u^2}f(u,v)\right|_{\substack{u=1+r^2-2r\cos\theta\\v=r^2}}\,, \\
        I_4[f] &\equiv -\frac{32}{\pi}\int dr\,d\theta \left. r^3 \sin^2\theta\left(u^{-1} + u^{-2}(1+v)\right)\bar{D}_{1,1,1,1}(u,v) f(u,v)\right|_{\substack{u=1+r^2-2r\cos\theta\\v=r^2}}\,, \nonumber
\ee
where the $\bar{D}_{1,1,1,1}$ function is given by
\begin{equation}
    \bar{D}_{1,1,1,1}(u,v) = \frac{1}{\bm{z}-\bm{\bar{z}}}\left(\log(\bm{z} \bm{\bar{z}}) \log \frac{1-\bm{z}}{1-\bm{\bar{z}}} + 2 \text{Li}_2(\bm{z}) - 2\text{Li}_2(\bm{\bar{z}})\right) \,.
\end{equation}
In terms of these integrals, the quantities $I_{k,\tau,J}$ and $I_{k,N_c}^\text{protected}$ that enter \eqref{eq:nonplanar_integral} are given by
\begin{equation}\label{eq:block_integrals}
    \begin{split}
        I_{k,\tau,J} &= I_k\left\lbrack \frac{G_{\tau+J+4,J}(u,v)}{u^4}\right\rbrack \,, \\
        I_{k}^\text{protected}(N_c) &= I_k\left\lbrack u^2\cH^\text{protected}(u,v)\right\rbrack \,.
    \end{split}
\end{equation}

To use these integral constraints in the $N_c\to\infty$ limit, it is useful to express them in terms of the Polyakov-Regge expansion (defined in \eqref{eq:transformed_h_expansion}) of the reduced correlator. The advantage of such a representation is that the double-trace operators do not contribute.  Rewritten in this way, the constraints \eqref{eq:nonplanar_integral} in the $N_c \to \infty$ limit become
\begin{equation}
\label{eq:planarsusyloc}
    \sum_{\st \tau,J} \tilde{\lambda}^2_{\tau,J} I^\text{PR}_{k,\tau,J} + \left(\tilde{I}_{k}^{\text{protected}} - \tilde{\mathcal{F}}_k(g)\right) = 0\,,
\end{equation}
where the key difference is that the sum over the superconformal primaries now only involves \emph{single-trace} operators. In \eqref{eq:planarsusyloc}, we introduced the rescaled three-point functions $\tilde{\lambda}^2_{\tau,J} = \lim_{N_c \to \infty} c \lambda^2_{\tau,J} $. Similarly, we defined $\tilde{I}_k^\text{protected} = \lim_{N_c\to\infty} c I_{k,N_c}^\text{protected}$ and $\tilde{\mathcal{F}}_k(g) = \lim_{N_c\to\infty} c \mathcal{F}_k(N_c,g)$. Note that $\tilde{\mathcal{F}}_k(g)$ was first calculated in Appendix~A of \cite{Chester:2020dja}, and $\tilde{I}_k^\text{protected}$, which is the integral of the supergravity part, is given in \cite{Caron-Huot:2024tzr}. Explicitly, we have 
\begin{equation}
    \begin{split}
        \tilde{I}_{2}^\text{protected} &= \frac{1}{4}\,, \qquad \tilde{I}_{4}^\text{protected} = 48\zeta_3-24\,,\\
        \tilde{\mathcal{F}}_2(g) &= \int_0^\infty \frac{te^{-t}\,dt}{(1-e^{-t})^2}\left( J_1(2gt)^2-J_2(2gt)^2\right)\,, \\
        \tilde{\mathcal{F}}_4(g) &= 48\zeta(3)-\frac{8}{g^2} \int_0^\infty \frac{te^{-t}\,dt}{(1-e^{-t})^2}J_1(2gt)^2 \\
&\phantom{=}-\frac{192}{g}\int_0^\infty  \frac{t e^{-t}J_1(2gt)\,dt}{(1-e^{-t})^2} \int_0^\infty\frac{t'e^{-t'}J_1(2gt')\, dt' }{(1-e^{-t'})^2} 
\left( \frac{t J_0(2gt)J_1(2gt')-(t\leftrightarrow t')}{t'^2-t^2}\right)\,.
    \end{split}
\end{equation}
The relevant integrated Polyakov-Regge blocks $I^\text{PR}_{k,\tau,J}$ are given explicitly in \eqref{eq:I_int}.

\subsubsection{Spectral input in the planar limit}
\label{sec:spectral_input}

In the $N_c\to\infty$ planar limit, $\cN = 4$ SYM is integrable \cite{Minahan:2002ve} (for a review, see \cite{Beisert:2010jr}), and this has been leveraged to construct the quantum spectral curve (QSC) framework \cite{Gromov:2023hzc, Hegedus:2016eop} for calculating the scaling dimensions of single-trace operators as a function of the 't Hooft coupling. 

\begin{figure}
    \centering
    \includegraphics[width=0.9\linewidth]{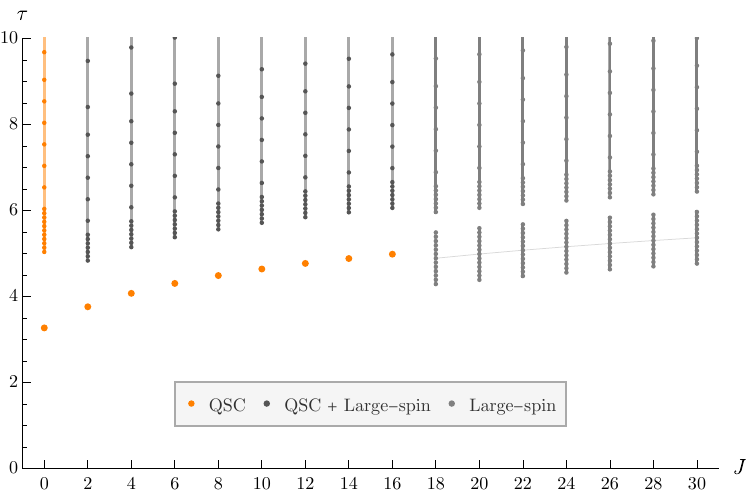}
    \caption{The discretization of the allowed twists $\tau(g,J)$ that we use in the bootstrap of the planar theory for $g<1$, here shown explicitly for $g=0.4$. For $J = 0$, the leading-twist $\tau_0(g,J)$ and the subleading twist $\tau_1(g,J)$ both come from the quantum spectral curve (QSC) \cite{Gromov:2023hzc,Hegedus:2016eop}. For $J=2,\ldots,16$, $\tau_0(g,J)$ comes from the QSC and we set $\tau_1$ using the gap $\tau_\text{gap}(g,J)$ in the $J\to\infty$ limit. For $J>16$, we estimate $\tau_0(g,J)$ from large-spin asymptotics (shown in the light gray line), and add a window around it to account for deviations from this estimate at finite spin; we again use $\tau_\text{gap}(g,J\to\infty)$ to estimate $\tau_1$. }
    \label{fig:spectrum_example}
\end{figure}

In practice, we cannot obtain the data of all single-trace operators. In bootstrap studies, we typically deal with an unknown operator spectrum by imposing positivity of a functional for all twists allowed by unitarity; this is the approach we take at finite $N_c$, as we describe in \secref{sec:bootstrap_assembly}. In the planar limit, we will use QSC results combined with large-spin asymptotic formulas to construct a smaller set $\tau(g,J)$ of possible twists of single-trace operators at spin $J$ for the theory at coupling $g$. Concretely, this allowed set takes the form
\begin{equation}
    \tau(g,J) = \{\tau_0(g,J)\} \cup \{\tau \,\mid\, \tau \ge \tau_1(g,J)\}\,. 
\end{equation}
Here $\tau_0(g,J)$ and $\tau_1(g,J)$ are the leading and subleading twist respectively. The actual set of single-trace spin-$J$ operators at coupling $g$ all have twists $\tau \in \tau(g,J)$.

The way in which we calculate or estimate $\tau_0(g,J)$ and $\tau_1(g,J)$ depends on the spin and the coupling. For $J = 0$, both $\tau_0$ and $\tau_1$ are known at any coupling from the quantum spectral curve \cite{Hegedus:2016eop,Gromov:2023hzc, Caron-Huot:2024tzr}. For $0.1\le g\le 1$, we also use $\tau_0$ calculated from the quantum spectral curve for $J = 2,4,\ldots,16$.\footnote{We thank Julius Julius and Nika Sokolova for providing this specific data using the solver introduced in \cite{Gromov:2023hzc}.} We complement this with large-spin asymptotic formulas \cite{Basso:2010in, Basso:2013aha} for $\tau_0(g,J)$ and $\tau_1(g,J)$, which can be written in terms of $\Gamma_\text{coll}(g)$ and $\Gamma_\text{cusp}(g)$ \cite{Beisert:2006ez}; see for instance (2.38)--(2.40) of \cite{Caron-Huot:2022sdy}.

For the leading trajectory, we use the large-spin asymptotic formula for  $J>16$, and we add a small window around the estimated value of $\tau_0(g,J)$ to account for any deviations between this formula and the true value.\footnote{For $J>140$ we do not add such a window, as we expect the large-spin formula to be very accurate. We have checked that fine details---such as the size of the window, or the precise value of the spin cutoff above which we do not include it---do not significantly affect any results} We use the large spin formula to estimate\footnote{
From integrability, we know the value of $\tau_\text{gap}(g,J\to\infty)$ is less than the value of $\tau_\text{gap}(g,J = 0)$ for any coupling. If we assume that $\tau_\text{gap}(g,J)$ decreases monotonically with $J$, then we can consider $\tau_\text{gap}(g,J\to\infty)$ to be a conservative estimate of the gap at finite spin.} the gap $\tau_\text{gap}(g,J) \equiv \tau_1(g,J) - \tau_0(g,J)$ using its value in the $J\to\infty$ limit for all $J>0$.
In \figref{fig:spectrum_example}, we give an example of the spectral assumptions we input at $g = 0.4$, and indicate which of them come from the quantum spectral curve and which come from large-spin asymptotic formulas. 

For $g>1$, the QSC data for $\tau_0$ that has been computed using the tools of \cite{Gromov:2023hzc} is limited\footnote{The QSC solver becomes prohibitively slow at large spin and coupling. It would be interesting to extend the QSC results further, but as we describe below, suitable approximations to requisite data are available and so this is not a limiting factor in our analysis.} to spins 0, 2, 4, and 6. In \cite{Caron-Huot:2024tzr}, it was explained how to interpolate between the low-spin QSC results and asymptotic formulas using an expression of the form 
\begin{equation}\label{eq:gkplike}
    \Delta_{\text{GKP-like}}(J)=\Delta^{g\to \frac{1}{2}\Gamma_\text{cusp}(g)}_{\text{GKP}}(J)+C(g) \,,
\end{equation}
where $\Delta_{\text{GKP}}(J)$ is defined parametrically by\footnote{These integrals can be evaluated in terms of elliptic integrals of the second kind:
\begin{equation}
\begin{split}
    \Delta_\text{GKP}(\omega) &= 8g\frac{i}{\omega^2 - 1}\left(E\left(i\rho_0 \mid 1-\omega^2\right) -\omega^2 F\left(i\rho_0 \mid 1-\omega^2\right)\right)\,,\\
    J(\omega) &= 8g\frac{i\omega}{\omega^2 - 1}\left(E\left(i\rho_0 \mid 1-\omega^2\right) - F\left(i\rho_0 \mid 1-\omega^2\right)\right)\,.
\end{split}
\end{equation}
}
\begin{equation}
    \begin{split} \label{eq:GKP}
       \Delta_{\text{GKP}}(\omega) &= 8g\int^{\rho_0}_0 d\rho\,\frac{\cosh^2\rho}{\sqrt{\cosh^2\rho-\omega^2\sinh^2\rho}}\,,\\
        J(\omega) &= 8g\int^{\rho_0}_0 d\rho\,\frac{\omega\sinh^2\rho}{\sqrt{\cosh^2\rho-\omega^2\sinh^2\rho}}\,.
    \end{split}
\end{equation}
The formulas \eqref{eq:GKP}, derived by Gubser, Klebanov, and Polyakov (GKP) in \cite{Gubser:2002tv}, give the scaling dimension and spin of operators dual to classical strings in $\text{AdS}_5\times S^5$ that lie at the equator of $S^3$ and have azimuthal angular velocity $\omega$; $\rho$ is the radial coordinate of global AdS$_5$, and its upper limit is given by $\rho_0 = \coth^{-1}\omega$. This formula is derived in the $g\to\infty$ limit.  The formula \eqref{eq:gkplike} is a modification of \eqref{eq:GKP} in which one replaces $g\to \frac{1}{2}\Gamma_\text{cusp}(g)$ (recall that $\Gamma_\text{cusp}(g) \approx 2g$ at large $g$) and one tunes the additive constant $C(g)$ so that \eqref{eq:gkplike} matches with the large-spin asymptotic formula \eqref{eq:twist2dataLargeJ}. This fixes $C(g)$ to 
\be 
C(g) = 2\Gamma_{\text{cusp}}(g)(1+\log\left(e^{\gamma_E}\Gamma_{\text{cusp}}(g)/4\right))+2+\Gamma_{\text{coll}}(g) \,.
\ee 
That the model \eqref{eq:gkplike} works well at finite $g$ can be seen in Figure~2 of \cite{Caron-Huot:2024tzr}.  
In the following section, we explain how we impose positivity constraints on a bootstrap functional for a discretization of the twist spectrum $\tau(g,J)$ discussed in this section, to obtain bounds on the energy-energy correlator and its multipoles in the planar limit.

\subsubsection{Primal and dual bootstrap problems}\label{sec:bootstrap_assembly}

Finally, we will show how we assemble these different constraints into bootstrap problems, both for finite $N_c$ and in the planar limit. We defer more details to \appref{app:bootstrap}.

At finite $N_c$, we have the crossing constraints in \secref{sec:crossing_nonplanar}, the integral constraints in \secref{sec:integral_constraints}, and the ANEC constraints in \secref{sec:anec_constraint}. The primal bootstrap problem is given by\footnote{Note that we can also include ANEC constraints on linear combinations of EEC multipoles; these are completely analogous to the ones written here, but for the sake of simpler notation we have not written them explicitly.}
\begin{equation}\label{eq:primal_lp_nonplanar}
    \begin{array}{ll@{}ll}
    \text{min/max} \quad & \displaystyle\sum &\beta_{\tau,J} \lambda^2_{\tau,J} &\\
    \text{subject to}& \displaystyle\sum   &\lambda^2_{\tau,J}F_{\tau,J;m,n}  + F^{\text{protected}}_{m,n}(N_c) = 0 \,,\quad &\quad m+n\text{ odd}, m<n\\
    & \displaystyle\sum   &\lambda^2_{\tau,J}I_{k,\tau,J} + (I_{k,N_c}^\text{protected} - \mathcal{F}_k(N_c,g_\text{YM})) = 0 \,,  &\quad k=2,4\\
    &\displaystyle\sum &\lambda^2_{\tau,J} \sin^2\frac{\pi \tau}{2} \alpha(\tau,J) f_{J+2,s}(\tau) - \max c_s \le 0\,, &\quad s=2,3,\ldots \\
                     &                                                
     &\lambda^2_{\tau,J} \ge 0 &\quad  \forall J=0,2,\ldots,\,\tau \ge J+2
    \end{array}
\end{equation}
where $\beta_{\tau,J}$ are objective coefficients. Using \eqref{eq:eecrep} and \eqref{eq:cs1}, we see that we can choose these coefficients as follows to bound the EEC or its multipoles, respectively:
\begin{equation}\label{eq:objective_coefficients}
\begin{split}
    \text{EEC}(z):\qquad \beta_{\tau,J} &= \sin^2\left(\frac{\pi \tau}{2}\right) \alpha(\tau,J) \cP_{J+2}(\tau,1-2z)\,,\\
    c_s:\qquad \beta_{\tau,J} &= \sin^2\left(\frac{\pi \tau}{2}\right) \alpha(\tau,J) f_{J+2,s}(\tau)\,.
\end{split}
\end{equation}

For the planar theory, the primal bootstrap problem takes a similar form, with the dispersive crossing constraints, other dispersive constraints, and integral constraints.\footnote{Note that these constraints are not all linearly independent. However, we find it useful to include instances of all the types of constraints we have listed, since their different behaviors in different limits allows them to complement one another and improve overall convergence.} We have\footnote{Note that $\beta_{\tau,J}$ contains a factor of $c$, which can be moved onto $\lambda^2_{\tau,J}$ to write the objective function in terms of $\tilde{\lambda}^2_{\tau,J} = \lim_{N_c\to\infty} c \lambda^2_{\tau,J}$.}
\begin{equation}\label{eq:primal_lp_planar}
    \begin{array}{ll@{}ll}
    \text{min/max} \quad & \displaystyle\sum &\beta_{\tau,J} \lambda^2_{\tau,J} & \\
    \text{subject to}& \displaystyle\sum   &\tilde{\lambda}^2_{\tau,J}X_{\tau,J}(u,v) = 0 \,,\quad &\quad (u,v)\text{ Euclidean \eqref{uv-euclidean}}\,,\\
    &\displaystyle\sum   &\tilde{\lambda}^2_{\tau,J}B_{\tau,J}(v) + B_v^\text{protected} = 0 \,,\quad &\quad v>0\text{ real}\,,\\
    &\displaystyle\sum   &\tilde{\lambda}^2_{\tau,J}\hat{B}_{\tau,J}(t) + \hat{B}_t^\text{protected} = 0\,,\quad &\quad 4<\Re t< 6\,,\\
    &\displaystyle\sum   &\tilde{\lambda}^2_{\tau,J}\Psi_{\ell;\tau,J} + \Psi_\ell^\text{protected} = 0\,,\quad &\quad \ell=0,2,\ldots\,,\\
    &\displaystyle\sum   &\tilde{\lambda}^2_{\tau,J}\Phi_{\ell,\ell+2;\tau,J} = 0\,,\quad &\quad \ell=0,2,\ldots\,,\\
    & \displaystyle\sum   &\tilde{\lambda}^2_{\tau,J}I_{k,\tau,J} + (I_{k}^\text{strong} - I_k(g)) = 0\,,  &\quad k=2,4\,,\\
    
                     &                                                
     &\tilde{\lambda}^2_{\tau,J} \ge 0 &\quad  \forall J=0,2,\ldots,\,\tau \in \tau(g,J)\,.
    \end{array}
\end{equation}

To solve the problems \eqref{eq:primal_lp_nonplanar} and \eqref{eq:primal_lp_planar} numerically, we need to truncate the infinite set of primal constraints, and also truncate the infinite set of primal variables $\lambda^2_{\tau,J}$ to a finite set. At finite $N_c$, the only truncation is the one we described in \secref{sec:crossing_nonplanar}: we keep all derivatives with $m+n \le \Lambda$, and so our bootstrap analysis becomes more powerful as we increase $\Lambda$. In the planar limit, we have to choose some finite subset of each of the various types of constraints; this is detailed in \tabref{tab:planar_fnals} of \appref{app:bootstrap}.

In both cases, to truncate the infinite set of primal variables, we make three truncations: a cutoff on the maximum spin we include, a cutoff on the maximum twist we include,\footnote{In the planar theory, we add additional constraints to impose asymptotic positivity at large twists. In the non-planar theory, the large twist regime is somewhat subtle, because of the behavior of the integrated blocks. Both of these points are discusssed in \appref{sec:large_twist}.} and a discretization of the space of possible twists. These truncations are chosen so that further refinement does not change our bounds by any significant amount; detailed examples of this are shown in \tabref{tab:nonplanar_convergence_check} of \appref{app:bootstrap} at finite $N_c$, and in \tabref{tab:planar_convergence_check} of \appref{app:bootstrap} in the planar limit.

We also pass from the primal problem to the dual problem \cite{dantzig}, as is standard in the numerical bootstrap (see \cite{Guerrieri:2021tak} for a related application of duality). The primal constraints become dual variables, denoted $\alpha_c$, where $c$ indexes the primal constraints in the truncated problem of interest. The primal variables $\lambda^2_{\tau,J}$ become the dual constraints that impose positivity at each $(\tau,J)$. The less common aspect of our primal problem is the inequality constraints from the averaged null-energy condition; the fact that these are inequalities means that the coefficients of their corresponding functionals must be non-negative. Concretely, we can take the dual of our primal problems using the following recipe:
\begin{equation}\label{eq:primal_to_dual}
    \begin{tikzpicture}[baseline=-3cm]
        \node at (0,0) (top) {
        $\begin{array}{ll@{}ll}
    \text{max} \quad & \displaystyle\sum_{\tau,J} &\beta_{\tau,J} \lambda^2_{\tau,J} &\\
    \text{subject to}\quad & \displaystyle\sum_{\tau,J}   &\lambda^2_{\tau,J}A_{c;\tau,J} + B_c = 0\,,\quad & \quad c=1,2,\ldots \\
    & \displaystyle\sum_{\tau,J}   &\lambda^2_{\tau,J}a_{c';\tau,J} + b_{c'} \le 0\,,\quad & \quad c'=1,2,\ldots\\           
     & &\lambda^2_{\tau,J} \ge 0 &\quad  \forall\, \tau, J
    \end{array}$};
    \node at (0,-5) (bottom) {
        $\begin{array}{ll@{}ll}
    \text{min} \quad & \alpha_0 -{} &1 \\
    \text{subject to}\quad & \alpha_0 +{} &\displaystyle\sum_c B_c \alpha_c + \displaystyle\sum_{c'} b_{c'} \alpha_{c'} = 1\\
    & &\displaystyle\sum_{c}   \alpha_c A_{c;\tau,J} + \displaystyle\sum_{c'} \alpha_{c'} a_{c',\tau,J} \ge \beta_{\tau,J}\,,\quad & \quad \forall\, \tau, J \\
    & & \alpha_{c'} \ge 0 & \quad \forall c'
    \end{array}$};
    \draw[ultra thick,->] (top) -- (bottom);
    \end{tikzpicture}
\end{equation}
To solve the dual program in this form, we use the SDPB solver \cite{Simmons-Duffin:2015qma}; details of the numerical setup can be found in \appref{app:bootstrap}.\footnote{Although SDPB is a semidefinite program solver, and we only need to solve linear programs (a special case of semidefinite programs), it is still convenient to use SDPB because it supports arbitrary numerical precision. Since our problems are by nature ill-conditioned (with dual constraints at nearby twists being nearly degenerate), it is important that we use extended-precision arithmetic.}

\subsection{Results at finite $N_c$}
\label{sec:nonplanarbootstrap}

\begin{figure}
    \centering
    \includegraphics[width=\linewidth]{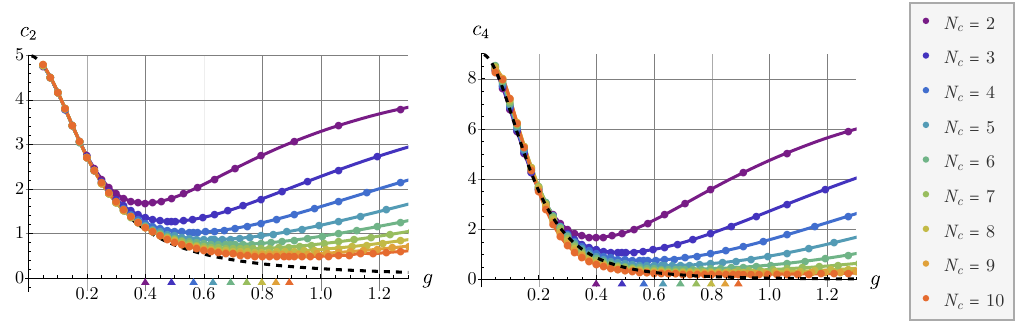}
    \caption{We plot lower bounds on the EEC multipoles $c_2$ and $c_4$ for $N_c=2,\ldots,10$ as a function of coupling. For $c_2$ we extrapolate to $\Lambda\to\infty$ from $\Lambda=35,43,\ldots,75$, and for $c_4$ we extrapolate using $\Lambda=67,75,\ldots,107$ (see Appendix~\ref{app:bootstrap_extrapolation} for extrapolation details). In both cases, the extrapolated bounds appear to follow the planar Pad\'e approximant (black dashed line) for a range of $g$ that grows with $N_c$. The values of $g$ corresponding to the self-dual coupling for each value of $N_c$ are indicated with colored triangles at the bottom of each plot.}
    \label{fig:c24_nonplanar}
\end{figure}

For the theory at finite $N_c$, our best-converged bootstrap results are lower bounds for the even EEC multipoles; we presented our bounds on $c_2$ in \figref{fig:c2Intro_nonplanar}. In \figref{fig:c24_nonplanar}, we show these bounds again along with lower bounds on $c_4$. 

To obtain these results, we derived bounds at many values of the truncation parameter $\Lambda$: for $c_2$, $\Lambda = 35,43,\ldots,75$, and for $c_4$, $\Lambda = 67,75,\ldots,107$. Even at these large truncation parameters, the bounds are not fully converged (especially for $c_4$ at larger values of $N_c$), and so we extrapolate to $\Lambda\to\infty$. Details of the extrapolation procedure can be found in \appref{app:bootstrap_extrapolation}.

\begin{figure}
    \centering
    \includegraphics[width=0.6\linewidth]{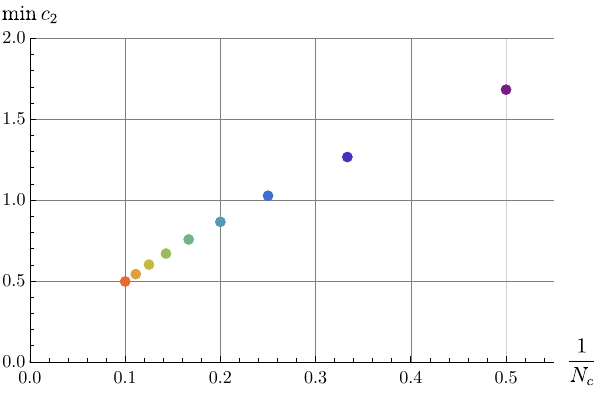}
    \caption{We plot the minimum lower bound on $c_2$, obtained at the self-dual coupling values in \figref{fig:c24_nonplanar}, as a function of $\frac{1}{N_c}$.}
    \label{fig:c2_sd}
\end{figure}

The finite-$N_c$ regime is distinguished from the large-$N_c$ regime by S-duality, which requires observables to be invariant under $g \to N_c/(4\pi g)$. This means in particular that there is a self-dual coupling $g = \sqrt{\frac{N_c}{4\pi}}$. In \figref{fig:c24_nonplanar}, we see that our lower bounds on $c_2$ and $c_4$ take their minimal value at the self-dual coupling. In \figref{fig:c2_sd}, we plot our lower bound on $c_2$ at the self-dual coupling as a function of $N_c$, showing how this bound decreases towards zero as we increase $N_c$.\footnote{It would be interesting to explore in more detail the regime where $N_c \to \infty$ at fixed Yang-Mills coupling.  This regime was referred to as the ``very strong coupling limit'' in \cite{Binder:2019jwn, Chester:2019jas, Chester:2020vyz}.}

For $c_3$, we can also obtain lower bounds as a function of coupling at finite $N_c$, but they are qualitatively different from the bounds on $c_2$ or $c_4$. They do not appear to converge towards the Pad\'e approximant of $c_3$ in the planar limit as we increase $N_c$. Thus, it is not clear whether our bounds on $c_3$ are close to being saturated for any values of $N_c$ or $g$. Additionally, they converge very slowly as a function of $\Lambda$. We leave the improvement of bounds on odd EEC multipoles at finite $N_c$ to future work.

\begin{figure}
    \centering
    \includegraphics[width=0.8\linewidth]{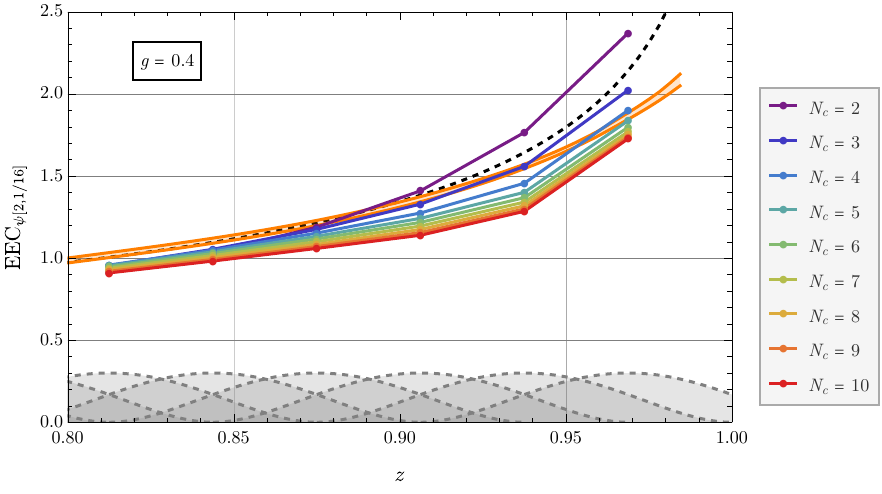}
    \caption{We plot lower bounds on $\text{EEC}_{\psi[2,1/16]}(z)$ in the back-to-back region for $N_c = 2,\ldots,10$ at $g = 0.4$ using a truncation parameter $\Lambda = 107$. The black dashed line is the Pad\'e approximant in the planar limit, smeared as in \eqref{eq:smearing}, and the orange region is an interpolation of the bootstrap bounds in the planar limit, also smeared accordingly. The smearing kernel $\psi_{2,\frac{1}{16}}$, defined in \eqref{eq:smearing_kernel}, is depicted in the gray dashed curves.}
    \label{fig:smeared_eec_nonplanar}
\end{figure}

At finite $N_c$, we are not able to obtain any bounds on $\text{EEC}(z)$ at a particular value of $z$. However, if we integrate $\text{EEC}(z)$ against a smearing kernel, we can do somewhat better. We define
\begin{equation}\label{eq:smearing}
    \text{EEC}_{\psi[n,\delta]}(z) \equiv \frac{\int_{0}^{1}dz'\, \psi_{n,\delta}(z-z') \text{EEC}(z') }{\int_{0}^{1}dz'\, \psi_{n,\delta}(z-z')} \,,
\end{equation}
where
\begin{equation}\label{eq:smearing_kernel}
    \psi_{n,\delta}(z) = \Theta(|z| \le \delta) (z-\delta)^n (z+\delta)^n\,.
\end{equation}
The results are not especially sensitive to the details of the kernel, but for concreteness we will take $n = 2$ and $\delta = \frac{1}{16}$. We can then obtain lower bounds\footnote{Note that for these bounds, it is important that we use the ANEC constraint described in \secref{sec:anec_constraint}.} of $\text{EEC}_{\psi[2,1/16]}(z)$ for $z\gtrsim 0.8$. In \figref{fig:smeared_eec_nonplanar}, we show the lower bounds we obtain on the smeared EEC in this region at $g = 0.4$. The lower bounds do appear to converge towards a large-$N_c$ curve, but this curve is somewhat below the Pad\'e approximant to the EEC in the planar theory at this coupling. This may be related to the behavior of our bounds on $c_3$ and qualitatively similar behavior for other odd EEC multipoles; these odd multipoles contribute to the smeared EEC, and so again our lower bounds may not be saturated.

\subsection{Results in the planar limit}
\label{sec:planarbootstrap}

By solving \eqref{eq:primal_lp_planar} with the objective coefficients given in \eqref{eq:objective_coefficients}, we can obtain upper and lower bounds on the EEC multipoles $c_s$ as well as the EEC. 

\begin{figure}
    \centering
    \begin{subfigure}{\linewidth}
        \centering
        \includegraphics[width=\linewidth]{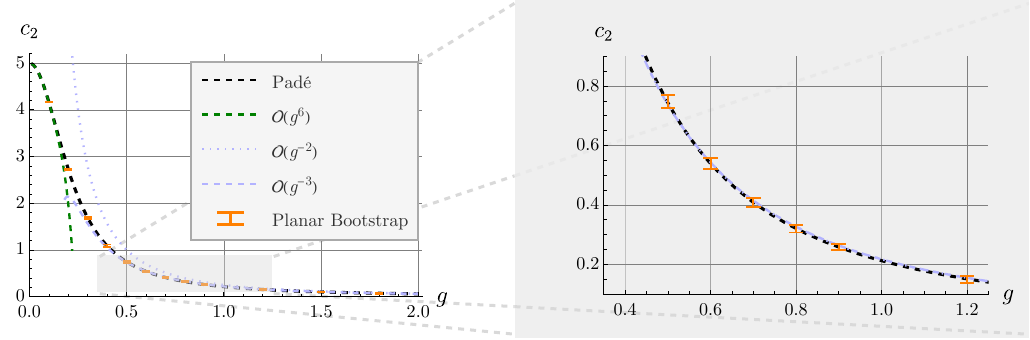}
        \caption{}
        \label{fig:planar_legendre:a}
    \end{subfigure}

    \vspace{1em}

    \begin{subfigure}{.45\linewidth}
        \centering
        \includegraphics[width=\linewidth]{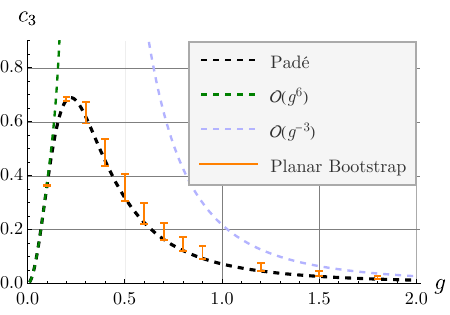}
        \caption{}
        \label{fig:planar_legendre:c}
    \end{subfigure}
    \hfill
    \begin{subfigure}{.45\linewidth}
        \centering
        \includegraphics[width=\linewidth]{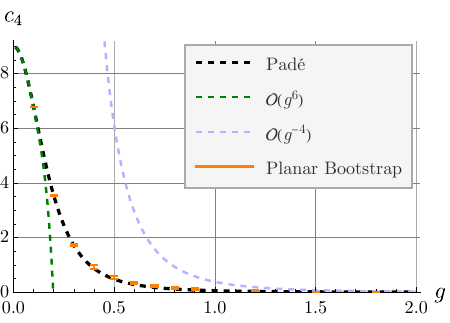}
        \caption{}
        \label{fig:planar_legendre:d}
    \end{subfigure}

    \caption{Bootstrap bounds on the EEC multipoles $c_s$ as a function of $g$ in the planar limit. The bounds are obtained using 40 dispersive functionals listed in \tabref{tab:planar_fnals} of \appref{app:bootstrap}. The bounds follow Pad\'e and match with weak and strong coupling expansions in the relevant regime,}
    \label{fig:planar_legendre}
\end{figure}

In \figref{fig:planar_legendre}, we plot bounds on $c_2$, $c_3$, and $c_4$; the bounds for $c_2$ appeared also in \figref{fig:c2Intro}. In each case, we compare with weak- and strong-coupling expansions given in \secref{sec:Review} as well as Pad\'e approximants described in \secref{sec:Pade}. The Pad\'e approximants are seen to lie within our bounds in each case. The details of the grid (twist and spin cutoff) are discussed in \appref{app:bootstrap}. The bounds in this section are derived with a fixed set of functionals detailed in table \tabref{tab:planar_fnals}. We check that the bounds remain stable against further refinement in grid (see \tabref{tab:planar_convergence_check}) and varying the number of functionals (see \figref{fig:fnal-convg-planar} for the example of $c_2$ at $g=0.4$). 

\begin{figure}
    \centering
    \begin{subfigure}{.45\linewidth}
        \centering
        \includegraphics[width=\linewidth]{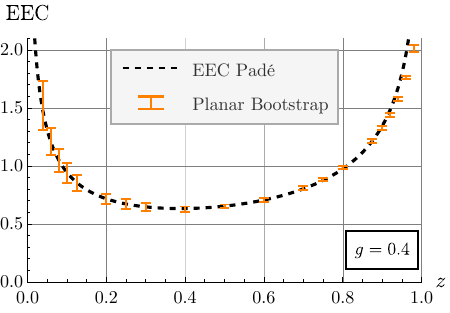}
        \caption{}
        \label{fig:planar_eec:a}
    \end{subfigure}
    \hfill
    \begin{subfigure}{.45\linewidth}
        \centering
        \includegraphics[width=\linewidth]{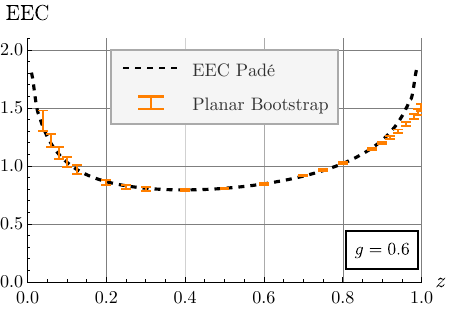}
        \caption{}
        \label{fig:planar_eec:b}
    \end{subfigure}

    \vspace{1em}

    \begin{subfigure}{.45\linewidth}
        \centering
        \includegraphics[width=\linewidth]{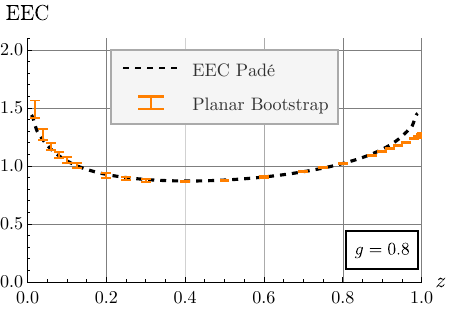}
        \caption{}
        \label{fig:planar_eec:c}
    \end{subfigure}
    \hfill
    \begin{subfigure}{.45\linewidth}
        \centering
        \includegraphics[width=\linewidth]{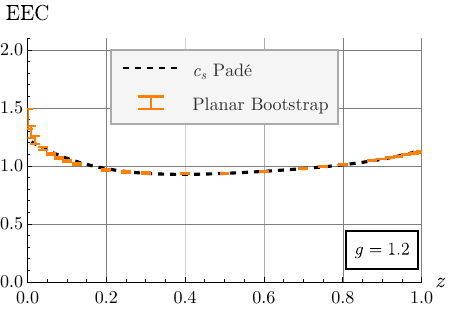}
        \caption{}
        \label{fig:planar_eec:d}
    \end{subfigure}

    \vspace{1em}

    \begin{subfigure}{.45\linewidth}
        \centering
        \includegraphics[width=\linewidth]{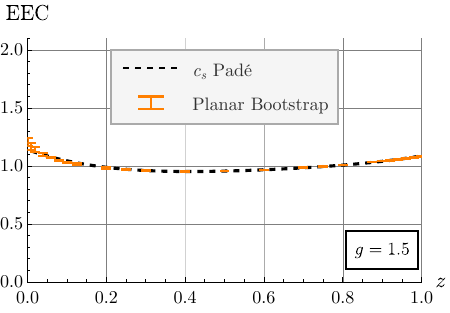}
        \caption{}
        \label{fig:planar_eec:e}
    \end{subfigure}
    \hfill
    \begin{subfigure}{.45\linewidth}
        \centering
        \includegraphics[width=\linewidth]{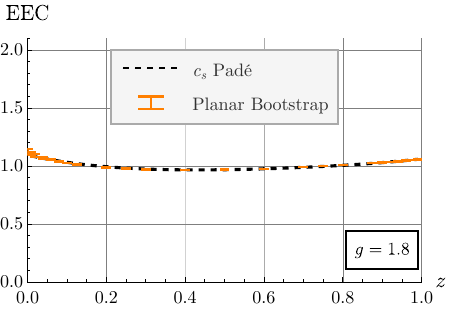}
        \caption{}
        \label{fig:planar_eec:f}
    \end{subfigure}

    \caption{Bootstrap bounds on $\text{EEC}(z)$ in the planar limit for several different values of $g$. As a consistency check for $g=1.2,1.5,1.8$, when the EEC is regular as $z\to0$, we can estimate $c_0$ and $c_1$ from interpolations of our bounds and find that they differ from 1 and 0 respectively by less than 0.01.}
    \label{fig:planar_eec}
\end{figure}

We plot bounds on the EEC for multiple couplings $g$ from $0.4$ to $1.8$ in \figref{fig:planar_eec}. For $g < 1$, we compare directly with the EEC Pad\'e derived in \secref{sec:Pade}; for $g>1$, we compare with an approximation $1+\sum_{s=2}^4 c_s(g) P_s(1-2z)$, where $c_s(g)$ comes from the Pad\'e approximants derived in \secref{sec:Pade}. In each case, we observe good agreement; our bounds become tighter and the agreement becomes more precise as we increase $g$. We also see that the EEC becomes more flat, consistent with the schematic expectation in \figref{fig:review}. It is worth noting that the current numerical setup in the planar limit (detailed in \appref{app:bootstrap}), does not yield tight bounds for EEC at $g\leq 0.3$. Improving our numerical techniques required to tighten the bounds in this weak coupling regime is an interesting direction that we leave to future work.

\section{Back-to-back limit, light-ray OPE, and the inversion formula at finite coupling}\label{sec:InvertedBoots}
In this section, we complete our analysis by combining the bootstrap data with analytic methods in the planar limit. Our goal is to extract (i) $H(g)$ and $H_0(g)$ in the back-to-back limit, (ii) the coefficient $a^{+}_{2,-1}(g)$ in the collinear limit, and (iii) the EEC itself. Lastly, we will use our bounds on low-spin EEC multipoles together with the inversion formula from \secref{sec:inversion} to reconstruct the high-spin contribution to the EEC. In particular, we bridge the gap described in \secref{sec:bootstrap} between the perturbative regime $g \lesssim 0.2$ and the regime $g \gtrsim 0.4$. This intermediate region is particularly challenging to constrain numerically due to the enhanced large–spin contributions to the EEC, and it lies beyond the reach of standard perturbation theory.

\subsection{Back-to-back limit and numerical bounds in the planar limit}\label{sec:bbFits}
\begin{figure}[!htbp]
    \centering
    \begin{subfigure}{.45\linewidth}
        \centering
        \includegraphics[width=\linewidth]{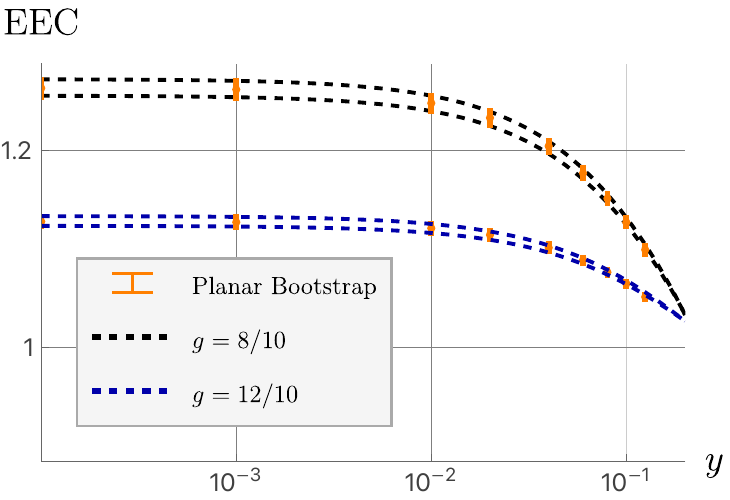}
        \caption{}
        \label{fig:BBFits1}
    \end{subfigure}
    \hfill
    \begin{subfigure}{.45\linewidth}
        \centering
        \includegraphics[width=\linewidth]{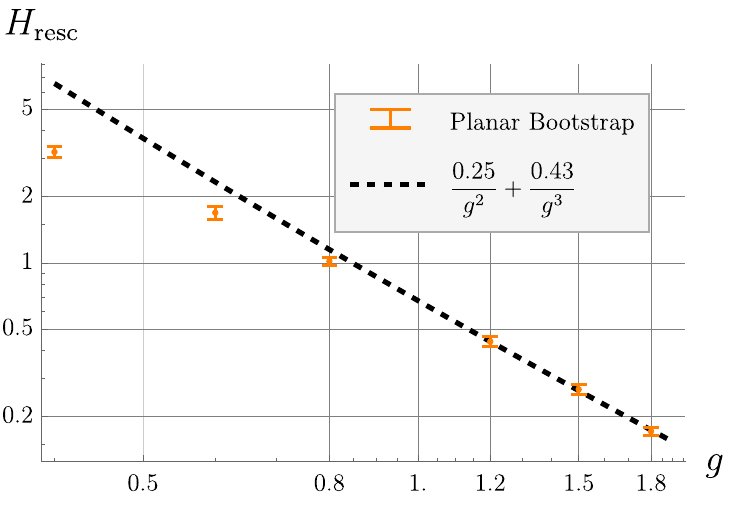}
        \caption{}
        \label{fig:BBFits2}
    \end{subfigure}

     \vspace{1em}

    \begin{subfigure}{.45\linewidth}
        \centering
        \includegraphics[width=\linewidth]{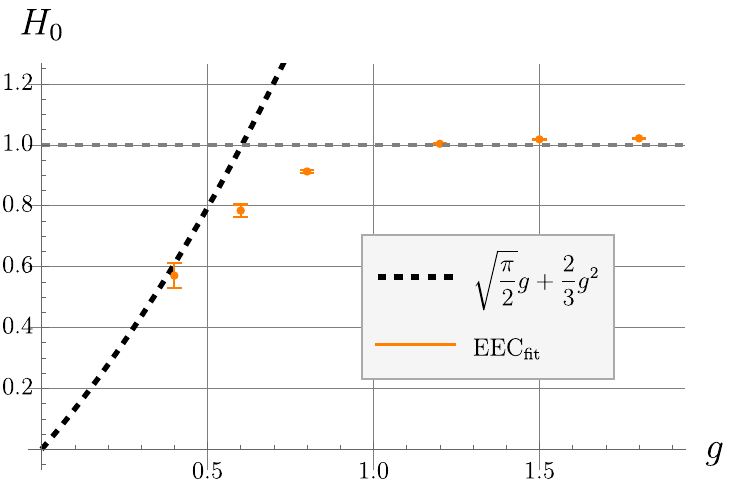}
        \caption{}
        \label{fig:H0Fit}
    \end{subfigure}
    \hfill    
    \caption{In (a), we compare bootstrap bounds on $\text{EEC}(y\equiv 1-z)$ in the back-to-back limit with a two-parameter fit to the model \eqref{eq:BBtoFit}. In (b) and (c), we plot the best-fit parameters $H(g)$ and $H_0(g)$ as functions of $g$ (with $H(g)$ rescaled to $H_\text{resc}(g)$ as described in \eqref{eq:hresc}). In (b) we see that $H_\text{resc}(g)$ decays as a power of $1/g$ at strong coupling, which means $H(g)$ decays exponentially in $g$ at strong coupling. In (c) we see that $H_0(g)$ approaches the perturbative expansion \eqref{eq:btbhighertwists} at weak coupling and becomes close to 1 at strong coupling, as we expect.  
    }
    \label{fig:BBFits}
\end{figure}

To describe the back-to-back limit we perform two-parameter fits using the twist-2 contribution \eqref{eq:BB} and an undetermined constant to model the twist-$4$ contribution:
\be\label{eq:BBtoFit}
\text{EEC}_{\text{fit}}(y \equiv 1- z)&= {H(g) \over 4y} \int_0^\infty db\,  b J_0(b)\cr
&\times\exp \Big[-{1 \over 2} \Gamma_{\text{cusp}}(g) \log^2(b^2/(y b_0^2))-\Gamma_{\text{coll}}(g) \log(b^2/(y b_0^2)) \Big]\cr
&+H_0(g) \,.
\ee
Here, $H(g)$ and $H_0(g)$ are the two parameters that we will be fitting, while $(\Gamma_{\text{cusp}}(g),\Gamma_{\text{coll}}(g)))$ are known in the planar limit from integrability. In \figref{fig:BBFits1}, we show examples of the resulting fits for the EEC in the back-to-back limit, i.e.\ a fit of \eqref{eq:BBtoFit} against the numerical bounds shown in \figref{fig:planar_eec}, at $g=0.8$ and $g=1.2$. These fits follow the numerical bounds well. We perform similar fits for all values of $g\in\{0.4,0.6,0.8,1.2,1.5,1.8\}$. In \figref{fig:BBFits2}, we plot our estimates of $H(g)$, rescaled as
\begin{equation}\label{eq:hresc}
\begin{split}
    H_\text{resc}(g) &\equiv H(g) e^{\frac{(-1+\Gamma_{\text{coll}}(g))^2}{2\Gamma_{\text{cusp}}(g)}} \\
    &\approx H(g) e^{4g(\log(g/2)+\gamma_E+1)+\cdots}\,, \qquad\text{for $g \gg 1$}.
\end{split}
\end{equation}
Since $H_\text{resc}(g)$ decays as a power law in $1/g$ at strong coupling (as shown in \figref{fig:BBFits2}), we conclude that $H(g)$ does in fact decay exponentially in $g$, as suggested by \eqref{eq:HStrong}. We also plot our estimates of $H_0(g)$ in \figref{fig:H0Fit}, and find good agreement with our expectations at both weak and strong coupling.

\subsection{Inversion at finite coupling}\label{sec:inversionTest}
In \secref{sec:inversion} we derived an inversion formula for the EEC multipoles in terms of the discontinuities around $z=0$ and $z=1$. These discontinuities can be approximated by the light-ray OPE \eqref{eq:leadtraj} and the back-to-back formula \eqref{eq:BB} respectively, which we tested in perturbation theory in \secref{subsec:InvWeakCoupling}. Here, we test the inversion formula \eqref{eq:inversionmultipole} at finite $g$ using the bootstrap results. For concreteness, we consider $g=0.4$. 

To apply the inversion formula, we need to know the discontinuity of the EEC around $z=0$ and $z=1$. These are not available at finite $g$. To proceed, we will approximate the discontinuity around $z=0$ by a single celestial block \eqref{eq:leadtraj}, and the discontinuity around $z=1$ using \eqref{eq:BBtoFit}. Even with this approximation we still have two unknown parameters: $a_{2,-1}^{(+)} \Big|_{g=0.4}$ and $H(g=0.4)$. The latter parameter was obtained in the previous subsection, and it can be read off from \figref{fig:BBFits} to be $H(g) \Big|_{g=0.4} \approx 0.466$.  To estimate the contribution from the light-ray OPE, we fit the EEC data \figref{fig:planar_eec:a} with a celestial block. This fit gives $a_{2,-1}^{(+)} \Big|_{g=0.4} \approx 3 \times 10^{-4}$.

\begin{figure}
    \centering
        \includegraphics[width=0.7\linewidth]{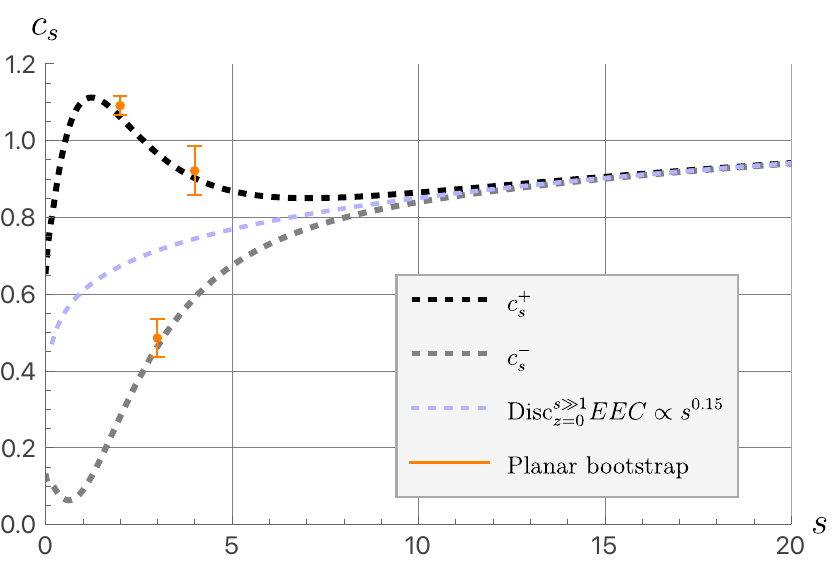}
    \caption{Here we show the result at $g=0.4$ for $c_s^{\pm}$ using the inversion formula \eqref{eq:inversionmultipole} in black(gray). It agrees well the bootstrap bounds in orange. In blue we have also plotted the large-spin expansion of the light-ray OPE whose power is determined by $s^{1-\gamma_{2,-1}^{(+)}(g=0.4)}$.}
    \label{fig:cs04Inversion}
\end{figure}

We then plug the discontinuities into the inversion formula \eqref{eq:inversionmultipole} and calculate $c_s$. We present our results for $c_s^{\pm}$, $s\in\mathbb{R}$ in \figref{fig:cs04Inversion}. For integer $s$, we compare these results with the bootstrap bounds (shown in orange in \figref{fig:cs04Inversion}) and find that they are consistent. This suggests that i) the conjecture \eqref{eq:EECregge} on the Regge behaviour of the EEC holds at this value of the coupling and ii) that the leading contribution to the discontinuity of the EEC is well-approximated by the leading contribution from the light-ray OPE and the back-to-back formula \eqref{eq:BB}. As the coupling increases, this approximation will get worse, since  subleading corrections to the EEC discontinuities become more important; in principle, it is also possible for the assumption \eqref{eq:EECregge} to break down. Indeed, we do find that when repeating this exercise at $g = 0.6$, the estimates of the low-spin EEC multipoles are slightly outside the bootstrap bounds. We expect this deviation is due to larger subleading corrections rather than to a failure of \eqref{eq:EECregge}. In the next section, we will, however, consider smaller values of the coupling where we expect the discontinuities of the EEC to be even better approximated by their leading $z\to 0$ and $z\to 1$ behavior.

\subsection{Bridging the gap: EEC at small but finite $g$}
In \secref{sec:planarbootstrap}, we obtained numerical bootstrap bounds for $\text{EEC}(z)$ at couplings $g\geq 0.4$. For smaller values of the coupling, large-spin contributions become more important (see e.g.\ \eqref{eq:LargeSFinal} and \figref{fig:lrOPEPred}). However, sharp bounds on low-lying EEC multipoles were obtained in this region. In this section, we will use these bounds together with the inversion formula to obtain predictions for the light-ray OPE coefficient $a_{2,-1}^{(+)}$ and the hard function $H(g)$. With these in hand, we obtain the $z\to 0,1$ behaviors of $\text{EEC}(z)$.  Furthermore, we use the inversion formula \eqref{eq:inversionmultipole} to obtain a prediction for $c_s^{\pm}$ and reconstruct the EEC for all angles. 

\begin{figure}
    \centering
    \begin{subfigure}{.5\linewidth}
        \centering
        \includegraphics[width=\linewidth]{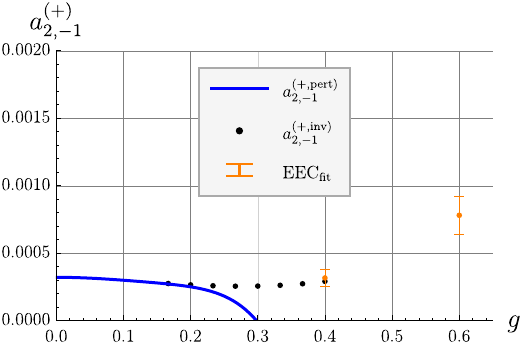}
        \caption{}
        \label{fig:LROPEDisp}
    \end{subfigure}
    \hfill
    \begin{subfigure}{.48\linewidth}
        \centering
        \includegraphics[width=\linewidth]{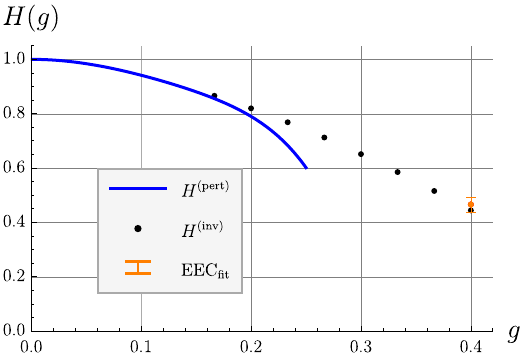}
        \caption{}
        \label{fig:HDisp}
    \end{subfigure}
    \caption{In black points, we plot our estimates for $a_{2,-1}^{(+)}$ (in (a)) and $H(g)$ (in (b)) obtained by using the inversion formula \eqref{eq:inversionmultipole} and imposing \eqref{eq:fixaandH}. In each case we see that the results approach the perturbative expansions at weak coupling, plotted in blue. In orange, we further show estimates extracted from the bootstrap bounds in \figref{fig:planar_eec}. In particular, at $g=0.4$ we see that $a_{2,-1}^{(+)}$ and $H$ are consistent with the value extracted from the numerical bootstrap bounds in \figref{fig:planar_eec:a} and \figref{fig:BBFits}. Given this data, we can reconstruct $c_s^{\pm}$.}
    \label{fig:discRecon}
\end{figure}

In order to reconstruct the EEC, we first obtain new predictions for $a_{2,-1}^{(+)}$ and $H(g)$. We do this by requiring that 
\be\label{eq:fixaandH}
c_{3}^{-}(g) &= c_3^{(\text{Pad\'e})}(g) \,, \cr
c_{4}^{+}(g) &= c_4^{(\text{Pad\'e})}(g)\,,\qquad 0.1\leq g\leq 0.4 \,,
\ee 
where we use \eqref{eq:inversionmultipole} for $c_s^\pm$, and we again approximate the discontinuities by the leading behaviour due to the light-ray OPE \eqref{eq:leadtraj} and the back-to-back formula \eqref{eq:BB} as described in \secref{sec:inversionTest}, and then solve for $(a_{2,-1}^{(+)},H(g))$. 
The result of this computation is shown in \figref{fig:discRecon}, where we further show the perturbative predictions that our results tend to at weak coupling. We also find good agreement with the value of $H(g=0.4)$ that we obtained independently in \secref{sec:bbFits} by fitting against the EEC, and so this serves as a consistency check of both methods. 

Given the values of $(a_{2,-1}^{+},H(g))$, we can reconstruct $c_{s}^{\pm}$ and from there the full EEC, in particular at intermediate values of the coupling. However, at large spin the EEC multipoles behave as $s^{1-\gamma_{2,-1}^{(+)}}$, and so unless we are at a large enough coupling for which this power is sufficiently negative, we cannot accurately approximate the EEC by summing Legendre polynomials up to some spin cutoff $s_{\text{max}}$.  Instead, in practice we will use the following approximation: 
\be\label{eq:EECInv}
\text{EEC}^{(\text{inv})}(z) &\approx \frac{8\pi^4 a_{2,-1}^{(+)} \Gamma\Big(3+\gamma_{2,-1}^{(+)}\Big)z^{-1+\gamma_{2,-1}^{(+)}/2}}{\Gamma\left(2+\frac{\gamma_{2,-1}^{(+)}}{2}\right)^3\Gamma\Big(-1-\frac{\gamma_{2,-1}^{(+)}}{2}\Big)}\\
&+\sum_{s=0}^{s_{0}}\left\lbrack c_{s}^{+}\frac{1+(-1)^s}{2}+c_{s}^{-}\frac{1-(-1)^s}{2}-c_s^{(z^{-1+\gamma/2})}\right\rbrack P_s(1-2z) \,,\nonumber\\
c_s^{(z^{-1+\gamma/2})} &= \frac{64 \pi ^2 (2 s+1) a^{(+)}_{2,-1} \sin ^2\Big(\frac{\pi  \gamma^{(+)}_{2,-1}}{2}\Big) \Gamma\Big(\gamma^{(+)}_{2,-1}+2\Big) \Gamma \Big(s-\frac{\gamma
^{(+)}_{2,-1}}{2}+1\Big)}{\Big(\frac{\gamma^{(+)}_{2,-1}}{2}+1\Big) \Big(\gamma^{(+)}_{2,-1}\Big)^2 \Gamma\Big(s+\frac{\gamma^{(+)}_{2,-1}}{2}+1\Big)} \,,
\ee 
with $c_s^{\pm}$ given by \eqref{eq:inversionmultipole} for $s\geq 2$. Here we further use $c_0^{+}=1$ and $c_1^{-}=0$ to exactly impose the Ward identities. The subtraction of $c_s^{(z^{-1+\gamma/2})}$, calculated in \eqref{eq:CsExactLROPE}, ensures that for $s\leq s_0$ the EEC multipoles are given by $c_{s}^{\pm}$ for $s$ even(odd). For $s>s_{0}$, we approximate it by the contribution from the light-ray OPE and $s_{\text{max}}$ is chosen such that for $s>s_{0}$ the contribution from the back-to-back cut becomes negligible compared to the contribution from the light-ray OPE. The large-$s$ behaviour of summand in \eqref{eq:EECInv} is significantly improved compared to a sum over coefficients $c_s^{\pm}$ (as can be seen in \figref{fig:cs04Inversion}).

\begin{figure}
        \centering
        \includegraphics[width=0.7\linewidth]{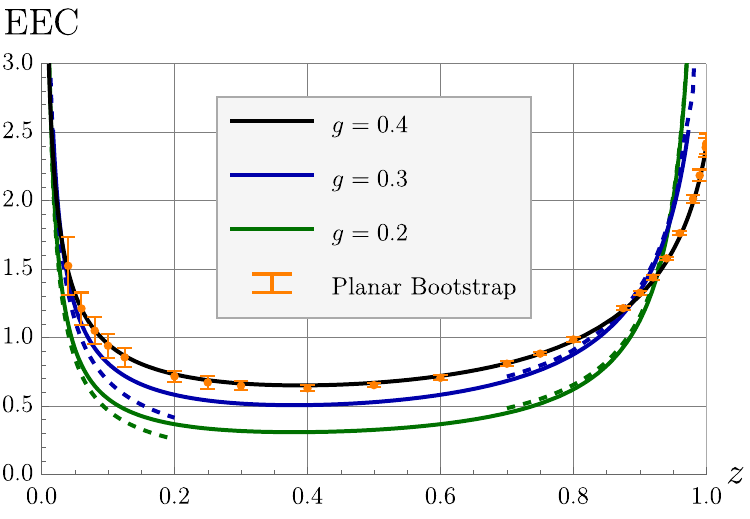}
    \caption{Here we have reconstructed the full EEC at $g = 0.2$, 0.3, and 0.4 using \eqref{eq:EECInv}, with $c_s^\pm$ estimated using the inversion formula \eqref{eq:inversionmultipole} as shown in \figref{fig:cs04Inversion}. For $g = 0.4$, the reconstruction is consistent with the planar bootstrap bounds from \figref{fig:planar_eec:a}. For lower values of $g$, we do not have bootstrap bounds, but we can check that the reconstruction agrees with the light-ray OPE formula \eqref{eq:smallgsmallz} at small angles, with the Pad\'e approximant from \secref{sec:Pade} at intermediate angles, and with the back-to-back formula \eqref{eq:BBtoFit} at large angles. In particular, by equating $\text{EEC}_{g=0.4}^{(\text{inv})}(z=1)$ with the value at $z=1$ predicted by \eqref{eq:BBtoFit}, with $H(0.4)$ given by \figref{fig:BBFits}, we read off $H_0(0.4)\approx0.55$, consistent with the bounds obtained in \figref{fig:H0Fit}.}
    \label{fig:ReconstructEEC}
\end{figure}

In \figref{fig:ReconstructEEC}, we show the results of this reconstruction for $g = 0.2$, 0.3, and 0.4. For $g = 0.4$, we obtain good agreement with the bootstrap bounds from \figref{fig:planar_eec:a}. For $g = 0.2$ and 0.3, we are not able to obtain two-sided bounds on the EEC using the numerical bootstrap, and so this method is our only way of calculating the EEC that is reliable at all angles. We can check that the result is consistent with the light-ray OPE formula \eqref{eq:smallgsmallz} at small angles, with the Pad\'e approximant from \secref{sec:Pade} at intermediate angles, and with the back-to-back limit \eqref{eq:BBtoFit} at large angles. It would be interesting to attempt the same exercise as done in this section at finite $N_c$.

\section{Discussion}\label{sec:discussion}

Conformal field theories offer an arena to explore relativistic Lorentzian dynamics of quantum fields, where the strongly coupled regime can be treated in a controlled way. This exploration is made possible thanks to powerful nonperturbative conformal bootstrap methods, which tightly constrain the dynamics. To this day, work in this area has mostly concentrated on Euclidean observables, such as the scaling dimensions and three-point functions of local operators. On the other hand, the same methods are directly applicable to Lorentzian collider observables, such as the charge- and energy- correlators, obtained from integrated Lorentzian insertions of the conserved currents. Apart from their intrinsic theoretical interest, these observables received a lot of attention recently in the phenomenological and experimental communities (see \cite{Moult:2025nhu} for a review), which further motivated our study.  

We considered a state obtained by acting on the vacuum with a scalar operator carrying definite momentum $p^\mu = (p^0, \vec 0)$. The energy–energy correlator in this state is expressed as an integral transform of the four-point function $\langle {\cal O}^\dagger T_{\mu \nu} T_{\rho \sigma} {\cal O} \rangle$. It can be decomposed in multipoles as
\be
\langle {\mathcal E}(\vec{n}_1){\mathcal E}(\vec{n}_2) \rangle = \left( \frac{p^0}{4\pi} \right)^2 \Big( 1 + \sum_{s=2}^\infty c_s P_s(\cos \theta)  \Big) \,,
\ee
where we denoted $\vec n_1\cdot \vec n_2 = \cos \theta$. As explained in \cite{Kologlu:2019bco}, we can use the source-detector OPE channel ${\cal O} \times T_{\rho \sigma}$ to express $c_s$ in terms of the OPE data, in such a way that the contribution of each primary operator to $c_s$ is manifestly nonnegative \cite{Mecaj2025}. We can then use conformal bootstrap techniques to constrain the multipoles $c_s$, or the energy–energy correlator itself. 

The discussion above applies to an arbitrary CFT\@. In this work, we have applied analytical methods and numerical conformal bootstrap techniques to study the energy–energy correlator in arguably the simplest four-dimensional gauge theory, in regimes that were previously inaccessible to perturbation theory or the strong-coupling expansion. The main simplifications in studying ${\mathcal N}=4$ SYM theory are twofold. First, supersymmetry relates the energy–energy detector correlator to a technically simpler scalar–scalar detector correlator with simpler kinematics. In particular, as a kinematical input to the bootstrap analysis, we used the source-detector OPE, generalized from \cite{Kologlu:2019bco} to the case of superconformal blocks. Second, this theory and its underlying four-point correlator are constrained by a wealth of available data, including results from integrability and supersymmetric localization.

In the planar limit, integrability, the dispersive sum rules, and supersymmetric localization provide the necessary dynamical input for the numerical conformal bootstrap. With these inputs, it was possible to obtain sharp upper and lower bounds on the EEC multipoles and the EEC itself---see \figref{fig:planar_legendre} and \figref{fig:planar_eec}, correspondingly. The basic physical picture that we observed as we transitioned from weak to strong coupling was a smooth interpolation between highly-peaked jet-like distribution at weak coupling (with single-trace operators dominating the light-ray OPE) to the homogeneous distribution at strong coupling  (with double-trace operators dominating the light-ray OPE). Using Pad\'{e} approximations based on the analytical strong-coupling predictions obtained in this work, we were able to also capture this transition analytically.  We found that a Pad\'{e} model led to very accurate results at intermediate couplings (see \eqref{eq:csPadeAns} and \eqref{eq:eecpade}).  

Supersymmetric localization and the work in \cite{Chester:2021aun,Chester:2023ehi} analogously allowed us to obtain lower bounds on various observables for gauge groups with finite rank (see \figref{fig:c24_nonplanar}). We presented some evidence that the bounds on even-spin EEC multipoles are close to being saturated and thus capture the physics of the theory. We found that at small $g_{\text{YM}}$, the small $N_c$ results follow well the weakly-coupled planar results, but they start departing significantly from them as we approach the self-dual point $g_{\text{YM}}^2 = 4 \pi$. Assuming that our lower bounds are close to saturation, we found that observables at the self-dual point for small $N_c$ look relatively weakly coupled if we compare them to the ones in the planar theory. We therefore expect that a collider experiment in $\SU(2)$ ${\cal N}=4$ SYM at the self-dual point will exhibit jets. Based on our results, see for example \figref{fig:c2_sd}, we expect that for $N_c \gtrsim 10$ a collider experiment at the self-dual point in the laboratory will exhibit strongly coupled features characteristic of an emergent gravitational dual. We have further found it easier to obtain bounds on the even-spin multipoles, which are the ones that at weak coupling receive large contributions from twist-two operators, in contrast with the odd-spin multipoles. It would be interesting to further understand why this is the case.

We further discussed properties of the EEC at complex $z$. Assuming maximal analyticity and polynomial boundedness in the complex-$z$ plane, in \secref{sec:inversion} we derived a Froissart-Gribov inversion formula for the EEC multipoles $c_s$ in terms of the discontinuities around $z=0$ and $z=1$. We have explored this representation both analytically at weak coupling as well as numerically at finite values of the coupling, and found good agreement. Moreover, at finite but small coupling, with sharp jets and large-spin contributions dominating, in \secref{sec:InvertedBoots} we reconstructed the EEC for all angles in a region of coupling space that is challenging numerically due to the truncation in spin. 

Let us list a few immediate future directions:
\begin{itemize}
\item It would be very interesting to use bootstrap techniques to study collider observables in other CFTs, such as, for example, the 3d Ising and $O(N)$ models, or the  Aharony-Bergman-Jafferis-Maldacena (ABJM) theory \cite{Aharony:2008ug}.
\item In the planar limit, we obtained two-sided bounds on both the EEC and the EEC multipoles, in contrast to the finite-$N_c$ regime, where we obtained only lower bounds. The extra input in the planar limit came from spectral results from integrability and the use of dispersive sum rules. It would therefore be very interesting to explore the use of dispersive functionals at finite $N_c$, and explore how they would affect the bounds, both for the energy-energy correlator considered in this work and more general observables \cite{wip}. 
\item It would be very interesting to rigorously establish the analytic properties of the EEC and its limiting behavior in the complex-$z$ plane. 
\item In our quest for understanding finite coupling signatures in CFT collider observables, it would be very interesting to go beyond two detectors. Considering three detectors, for instance, is technically challenging because it requires a detailed understanding of the CFT five-point function. Here we would like to point out the simple fact that higher-point energy correlators can still be computed using the OPE in the comb source-detector-...-detector-source channel (see, for example, \cite{Rosenhaus:2018zqn,Harris:2024nmr,Poland:2023fivept,Poland:2024fivept,Antunes:2022higherpt,Antunes:2025Polyakov}).
\item It would also be interesting to explore energy correlators in more complicated states.  For instance, a particularly interesting setup is to study collider observables in heavy states that are created by source operators with scaling dimension $\Delta \gg 1$. Such sources are expected to universally produce a homogeneous distribution of energy to leading order in $1/\Delta$ \cite{Chicherin:2023gxt,Firat:2023lbp}.  However, the subleading corrections are non-universal and exhibit interesting emergent phenomena, such as the sound jets observed in \cite{Cuomo:2025pjp} when studying energy correlators in the state created by a large charge operator in $O(2)$ Wilson-Fisher CFT\@. In holographic CFTs, it would further be interesting to understand the leading correction in heavy states dual to black holes/thermal states.
\end{itemize}

\section*{Acknowledgements}
We thank Simon Caron-Huot for collaboration at early stages of this project, and for useful comments and insights. We also thank Cyuan-Han Chang, Hao Chen, Frank Coronado, Johan Henriksson,  Nima Lashkari, Ian Moult, and Hua Xing Zhu for useful discussions. The authors thank the Yukawa Institute for Theoretical Physics at Kyoto University for its hospitality during the workshop “Progress of Theoretical Bootstrap''. This project has received funding from the European Research Council (ERC) under the European Union’s Horizon 2020 research and innovation program (grant agreement number 949077). RK is supported by the Titchmarsh Research Fellowship at the Mathematical Institute and by the Walker Early Career Fellowship at Balliol College. RD is supported by a Pappalardo Fellowship in Physics at MIT\@. SSP is supported in part by the U.S.~Department of Energy under Award No.~DE-
SC0007968. ZZ is supported by a Hawking Fellowship at Perimeter Institute. For the purpose of open access, the authors have applied a CC BY public copyright licence to any Author Accepted Manuscript (AAM) version arising from this submission.

\appendix

\section{Further details on the weak coupling expansion}\label{app:contacterms}

Here we give some further details on the weak-coupling expansion.  As explained in the main text, the terms in the small coupling expansion of the EEC in \eqref{eq:EECExpansion} contain distributional pieces localized at $z=0$ and $z=1$.  The complete expressions, including these distributional pieces, are necessary for obtaining the EEC multipoles. 

The complete one-loop weak-coupling result including distributional terms is given by 
\be\label{eq:1loop}
    \text{EEC}^{(1)}(z)=& -\frac{2(z+(1+z)\ln(1-z))}{z^2}+2\Big(-\delta(z)-\zeta_2\delta(1-z)+\Big[\frac{1}{z}\Big]_0\cr
    &-\Big[\frac{\ln(1-z)}{1-z}\Big]_1\Big) \,,
\ee 
where $[\cdot]_{z_*}$ denotes the plus distribution defined by $\int_0^1 dz f(z)[X(z)]_{z_*} = \int_0^1 dz (f(z)-f(z_*))X(z)$ for any test function $f(z)$.  The complete expression at two loops is given by   \cite{Korchemsky:2019nzm,Kologlu:2019mfz}
\be \label{eq:twoloop}
  \text{EEC}^{(2)}(z) &= \text{EEC}^{(2)}_{\text{reg}} (z)+8\Big(\Big[\frac{1}{z}\Big]_0(\frac{\pi^2}{6}-\frac{1}{2}\zeta_3-3)+\Big[\frac{\ln z}{z}\Big]_0+\frac{\zeta_3}{2}\Big[\frac{1}{1-z}\Big]_1\cr
    {}&+\frac{\pi^2}{4}\Big[\frac{\ln (1-z)}{1-z}\Big]_1+\frac{1}{2}\Big[\frac{\ln^3(1-z)}{1-z}\Big]_1+c^{(2)}_0\delta(z)+c^{(2)}_1\delta(1-z)\Big)\,,\cr
 c^{(2)}_0 &=\frac{11\pi^4}{45}-4\pi^2+56 \,, \cr
 c^{(2)}_1 &=\frac{4\pi^4}{9} \,,
\ee 
where the regular part of the EEC is defined by 
\be 
\text{EEC}^{(2)}_{\text{reg}}(z)= \text{EEC}^{(2)}(z)-\frac{1}{z}\text{Res}_{z=0}\text{EEC}^{(2)}(z)-\frac{1}{z-1}\text{Res}_{z=1}\text{EEC}^{(2)}(z) \,.
\ee 

\section{Details on superconformal detector blocks}\label{app:superconformal}
In this appendix we include some further details related to the superconformal block decomposition of the EEC in the source-detector channel.

\subsection{From superconformal primaries to conformal primaries}
In the main text, we gave the decomposition \eqref{eq:normBlocks} (see also \cite{Dolan:2001tt}), which we repeat here: 
\be\label{eq:normBlocks2}
{v^2 \over u^2} G_{\tau+4, J}(z, \bar z) = \sum_{k=0}^8\sum_{p=-4}^{4} d_{k,p}(\tau,J)G_{\tau+k-p, J+p}(z, \bar z) \,.
\ee 
In this expression, the non-zero coefficients $d_{k,p}(\tau,J)$ are (all the others are zero):
\begingroup
\allowdisplaybreaks
\makeatletter\def\f@size{9}\check@mathfonts
\def\maketag@@@#1{\hbox{\m@th\large\normalfont#1}}
\be
&d_{0,0}=1\,,\qquad d_{1,-1} = -\frac{1}{2}\,,\qquad d_{1,1}=-2\,,\qquad d_{2,0}=1\,,\cr
&d_{2,-2} = \frac{3 J^2-6 J (J+\tau +1)+3 (J+\tau )^2+6 (J+\tau )-8}{32 (\tau +3) (\tau-1)}\,,\cr
&d_{2,2}= \frac{3 J^2+6 J (J+\tau +3)+3 (J+\tau )^2+18 (J+\tau )+16}{2 (2 J+\tau +1) (2 J+\tau +5)}\,,\cr
&d_{3,-3}=-\frac{(\tau +2)^2}{128 (\tau+3) (\tau-1)}\,,\qquad d_{3,3}=-\frac{(2 J+\tau +4)^2}{2 (2 J+\tau +3) (2 J+\tau +5)}\,,\cr
&d_{3,-1}=\frac{-3 \tau ^2-6 \tau +8}{16 (\tau -1) (\tau +3)}\,,\cr
&d_{3,1}=\frac{-12 J^2-12 J \tau -36 J-3 \tau ^2-18 \tau -16}{4 (2 J+\tau +1) (2 J+\tau +5)},\qquad d_{4,2}=\frac{(2 J+\tau +4)^2}{4 (2 J+\tau +3) (2 J+\tau +5)}\,,\cr
&d_{4,-4}=\frac{(\tau +2)^2 (\tau +4)^2}{4096 (\tau+5) (\tau+1) (\tau +3)^2}\,,\qquad 
d_{4,-2}=\frac{(\tau +2)^2}{64 (\tau +1) (\tau +3)}\,,\cr
&d_{4,0}=\frac{(3 \tau  (\tau +2)-8) \left(12 J^2+12 J (\tau +3)+3 \tau  (\tau +6)+16\right)}{64 (\tau -1) (\tau +3) (2 J+\tau +1) (2 J+\tau +5)}\nonumber\,,\cr
&d_{4,4}= \frac{(2 J+\tau +4)^2 (2 J+\tau +6)^2}{16 (2 J+\tau +3) (2 J+\tau +5)^2 (2 J+\tau +7)}\,,\cr
&d_{5,-3}=-\frac{\left(\tau ^2+6 \tau +8\right)^2}{2048 (\tau +3)^2 \left(\tau ^2+6 \tau +5\right)}\,,\cr
&d_{5,-1}=-\frac{(\tau +2)^2 \left(12 J^2+12 J (\tau +3)+3 \tau  (\tau +6)+16\right)}{256 (\tau +1) (\tau +3) (2 J+\tau +1) (2 J+\tau +5)}\,,\cr 
&d_{5,1}=-\frac{(3 \tau  (\tau +2)-8) (2 J+\tau +4)^2}{64 (\tau -1) (\tau +3) (2 J+\tau +3) (2 J+\tau +5)}\,,\cr 
&d_{5,3} =-\frac{(2 J+\tau +4)^2 (2 J+\tau +6)^2}{32 (2 J+\tau +3) (2 J+\tau +5)^2 (2 J+\tau +7)}\,,\cr 
&d_{6,-2}=\frac{(\tau +2)^2 (\tau +4)^2 \left(12 J^2+12 J (\tau +3)+3 \tau  (\tau +6)+16\right)}{8192 (\tau +1) (\tau +3)^2 (\tau +5) (2 J+\tau +1) (2 J+\tau +5)}\,,\cr
&d_{6,0}=\frac{(\tau +2)^2 (2 J+\tau +4)^2}{256 (\tau +1) (\tau +3) (2 J+\tau +3) (2 J+\tau +5)}\,,\cr
&d_{6,2}=\frac{(3 \tau  (\tau +2)-8) (2 J+\tau +4)^2 (2 J+\tau +6)^2}{512 (\tau -1) (\tau +3) (2 J+\tau +3) (2 J+\tau +5)^2 (2 J+\tau +7)}\,,\cr 
&d_{7,-1}=-\frac{(\tau +2)^2 (\tau +4)^2 (2 J+\tau +4)^2}{8192 (\tau +1) (\tau +3)^2 (\tau +5) (2 J+\tau +3) (2 J+\tau +5)}\,,\cr
&d_{7,1}=-\frac{(\tau +2)^2 (2 J+\tau +4)^2 (2 J+\tau +6)^2}{2048 (\tau +1) (\tau +3) (2 J+\tau +3) (2 J+\tau +5)^2 (2 J+\tau +7)}\,,\cr
&d_{8,0}=\frac{(\tau +2)^2 (\tau +4)^2 (2 J+\tau +4)^2 (2 J+\tau +6)^2}{65536 (\tau +1) (\tau +3)^2 (\tau +5) (2 J+\tau +3) (2 J+\tau +5)^2 (2 J+\tau +7)}\,.
\ee
\endgroup

\subsection{OPE for EEC multipoles}\label{app:OPEMultipoles}
  
Here, we provide more details for the decomposition of
\be\label{eq:blockdecomp1} 
{\cal P}_{J+2}(\tau, 1-2z) = \frac{1}{z^2}\sum_{r=0}^{J+4} a_{r,J}(\tau)P_r(1-2z)
\ee 
into a sum of Legendre polynomials that was used in the main text around \eqref{eq:KDef}. Let us rewrite this by defining $\tilde{P}_r(1-2z)=P_r(1-2z)-(1-r(r+1)z)$ (such that $\tilde{P}_r\sim z^2$ as $z\to 0$). Equation \eqref{eq:blockdecomp1} becomes 
\be\label{eq:cc}
\frac{1}{z^2}\sum_{r=0}^{J+4} a_{r,J}(\tau)P_r(1-2z) = \frac{1}{z^2}\sum_{r=2}^{J+4} a_{r,J}(\tau)\tilde{P}_r(1-2z) \,,
\ee
where we used that the blocks ${\cal P}_{J+2}(\tau, 1-2z)$ are regular as $z\to0$.\footnote{The statement that there are no singularities in \eqref{eq:cc} as $z\to 0$ is enforced by the sum rules $\sum_r a_r(\tau,J)=\sum_r r(r+1)a_r(\tau,J)=0$.} 

We now consider the decomposition of \eqref{eq:cc} in terms of Legendre polynomials.  We decompose each term as
 \be
   \frac{\tilde{P}_r (1 - 2z)}{z^2} = \sum_s K_{r, s} P_s(1-2z) \,,
 \ee 
for some coefficients $K_{r, s}$.  Using the orthogonality of the Legendre polynomials, we can find the coefficients $K_{r, s}$ using
\be
K_{r,s}\equiv (2s+1)\int_0^1 dz\, P_s(1-2z)\frac{\tilde{P}_r(1-2z)}{z^2} \,.
\ee 
We find that $K_{r,s}$ are given by 
\begin{equation}
    K_{r,s} = \begin{cases} 2(2s+1)\left\lbrack s(s+1) - r(r+1) + (r(r+1) + s(s+1))(H_r - H_s)\right\rbrack & r \ge s + 2 \\
    0 & \text{otherwise} \end{cases}\,.
\end{equation}

\section{Details on the stringy correction}\label{app:stringy}

In this appendix, we present some of the calculations necessary to obtain the stringy corrections to the EEC multipoles discussed in \secref{sec:stringysubsection}.

In particular, we need to compute sums of the form 
\be\label{eq:genCs} 
c_s = \sum_{\delta=1}^\infty \sum_{J=0,2,\ldots}^{2(\delta-1)}\frac{f(\delta,J;\lambda)}{(\tau(\delta,J))^{2s}}\Big[p_0(J)+\frac{p_1(J)}{\tau(\delta,J)}+\frac{p_2(J)}{(\tau(\delta,J))^2}+\cdots\Big] \,,
\ee 
where $p_i(J)$ are polynomials in $J$, and where we have expanded the OPE decomposition of $c_s$ at large $\tau$. Further expanding at large $\lambda$, and recalling that $\tau(\delta,J)=2\sqrt{\delta}\lambda^{1/4}+\cdots$, it is clear that we need to do perform sums of the form $\sum_{\delta=1}^\infty \sum_{J=0,2,\ldots}^{2(\delta-1)} \delta^{-s-k}p(J)X(\delta,J) $, where $p(J)$ is a polynomial in $J$ and $X(\delta,J)$ is a product of various $f_i(\delta,J)$ and $\tau_i(\delta,J)$ factors. The key point is that sums of this form were beautifully understood in \cite{Alday:2022uxp,Alday:2022xwz,Alday:2023mvu}, whose results and notation we use. For further details we refer the reader to these papers. In particular, let us first consider the sum over spin and make the following definitions
\be\label{eq:someDefs}
F_{m}^{(0)}(\delta)&=\frac{4^m}{\Gamma(2m+2)}\sum_{J=0,2,\ldots}^{2(\delta-1)}(J-m+1)_m(J+2)_m f_0(\delta,J) \,,\nonumber\\
T_0^{(2)}(\delta) &= \sum_{J=0,2,\ldots}^{2(\delta-1)}\sqrt{\delta}f_0(\delta,J)\tau_2(\delta,J)\,,\\
F_0^{(2)}(\delta) &= \sum_{J=0,2,\ldots}^{2(\delta-1)}\Big[ \delta f_2(\delta,J)-\frac{39J}{4}f_0(\delta,J)\Big]\,.\nonumber
\ee 
We will express the subleading correction to $c_2$ and the leading-order result for any $c_{s}$ in terms of the quantities in \eqref{eq:someDefs}, summed over $\delta$. 
The sums over $\delta$ of the functions in \eqref{eq:someDefs} were performed in \cite{Alday:2022uxp,Alday:2022xwz,Alday:2023mvu}.

The $\OO(\lambda^{-3/2})$ contribution to $c_2$ is found using \eqref{eq:stringyOps} and \eqref{eq:StringysumRulec2}, and it is given by 
\be\label{eq:c2Exp}
c_2|_{\lambda^{-3/2}}= \sum_{\delta=1}^\infty\sum_{J=0}^{2(\delta-1)}\Big[-\frac{48f_0(\delta,J)\tau_2(\delta,J)}{\delta^{5/2}}+\frac{24f_2(\delta,J)}{\delta^2}-f_0(\delta,J)\frac{9(52 J (J+4)+103)}{4 \delta ^3}\Big]\,.
\ee  
To derive this relation, we used $\tau_1(\delta,J) = -J-2$ and $f_1(\delta,J)=f_0(\delta,J)\frac{3J+\frac{23}{4}}{\sqrt{\delta}}$.  The expression \eqref{eq:c2Exp} can now be written in terms of the quantities in \eqref{eq:someDefs} as 
\be\label{eqc2Eq1}
c_2|_{\lambda^{-3/2}}=\sum_{\delta=1}^\infty \delta^{-3}\Big[24F_0^{(2)}(\delta)-48T^{(2)}_0(\delta)-\frac{351}{2}F_{1}^{(0)}(\delta)-\frac{927}{4}F_0^{(0)}(\delta)\Big] \,.
\ee 
We can simplify this expression as follows.  It was shown in \cite{Alday:2022xwz} that $T_0^{(2)}(\delta)$ can also be expressed as
\be
T_0^{(2)}(\delta) &=  2+\frac{1}{4}\delta Z_1(\delta-1)+\delta^2Z_2(\delta-1) \,,
\ee 
where $Z_i$ are Euler-Zagier sums.  When summed over $\delta$, the Euler-Zagier sums lead to multiple zeta values: 
\be
\zeta(s,s_1,s_2,\ldots) =\sum_{\delta=1}^\infty \frac{Z_{s_1,s_2,\ldots}(\delta-1)}{\delta^s} \,.
\ee 
Next, $F_0^{(2)}(\delta)$ can be expressed as 
\be 
F_0^{(2)}(\delta) &= -2\delta^3 Z_3+2\delta^2Z_2+\frac{89}{8}\delta Z_1+\frac{405}{32}+2\delta^3\zeta(3) \,,
\ee
where $Z_{i}=Z_i(\delta-1)$.  Lastly, 
 \be
    F_0^{(0)}(\delta)=\sum_{J=0,2,\ldots}^{2(\delta-1)} f_0(\delta,J)=1 \,.
 \ee
With these identities, we obtain
\be \label{eq:c2lam32}
c_2|_{\lambda^{-3/2}}=-120\zeta(3)+48\sum_{\delta=1}^\infty(\zeta(3)-Z_3(\delta-1)) \,,
\ee
where we use $\sum_{\delta=1}^\infty F^{(0)}_1(\delta)\delta^{-6-2s}= 2\zeta(2s+5,1)$ at $s=-\frac{3}{2}$, as well as $\zeta(2,1)=\zeta(3)$. Using the large $\delta$ asymptotics of $Z_{s>1}$, namely
\be 
Z_s = \zeta(s)-\frac{1}{(s-1)\delta^{s-1}}+\cdots \,,
\ee 
we see that the remaining sum in \eqref{eq:c2lam32} converges. The final result is 
\be\label{eq:c2Strong}
c_2= \frac{4\pi^2}{\lambda}+\frac{8\pi^2-120\zeta(3)}{\lambda^{3/2}}+\cdots \,,
\ee 
which includes the subleading correction to the leading result obtained in \cite{Hofman:2008ar}.

Let us do the same exercise for the leading correction to $c_3$ which will be of $\OO(\lambda^{-3/2})$. An analogous calculation to the one for $c_2$ gives 
\be\label{eq:strongc3}
c_3 &= (N_c^2-1)\sum_{\tau,J}\lambda_{\tau,J}^2\frac{15 J (J+1) (J+2) 2^{J+2 \tau +19} \sin ^2\left(\frac{\pi  \tau }{2}\right)}{\pi ^3\tau ^{12}}+\cdots,\cr 
&= \frac{120}{\lambda^{3/2}}\sum_{\delta=1}^\infty\sum_{J=0}^{2(\delta-1)}\frac{J(J+2)}{\delta^3} f_0(\delta,J)+\cdots \,,
\ee 
where in the second line we have inserted the OPE data \eqref{eq:stringyOps}. The sum over $J$ can again be done by noting that it reduces to the sum over $F_1^{(0)}(\delta)$ from which we arrive at
\be\label{eq:strongc3Final} 
c_3 =\frac{360\zeta(3)}{\lambda^{3/2}}+\cdots\,.
\ee 
Similarly, we find for $c_4$
\be
c_4 &= (N_c^2-1)\sum_{\tau,J}\lambda_{\tau,J}^2\frac{27 (J-1) J (J+1) (J+2) (J+3) 2^{J+2 \tau +24} \sin ^2\left(\frac{\pi  \tau }{2}\right)}{7 \pi ^3 \tau ^{14}}+\cdots\,,
\ee 
and analogously to the computation above, by inserting the data \eqref{eq:stringyOps} and using the tools of \cite{Alday:2022xwz} we find the following leading-order contribution to $c_4$
\be\label{eq:strongc4}
c_4 =\frac{684\pi^4}{7\lambda^2}+\cdots\,.
\ee 

\section{EEC in Mellin space}\label{app:ComplexAngle}

In this appendix, we use the Mellin representation of the EEC to study its behavior at large $|z|$ and to provide an alternative derivation of the source–detector OPE expansion.

\subsection{Large-$z$ limit}

We now discuss the analytic continuation of the EEC in the limit $|z|\to\infty$, since this limit is required to justify dropping the arc contribution in the derivation of the inversion formula for the Legendre coefficients $c_s$ in \secref{sec:inversion}.

We begin with the Mellin representation of the EEC \cite{Belitsky:2013xxa}, 
\begin{align} \label{EEC_Mellin}
\mbox{EEC}(z) &=- \frac{1}{4 z^3} \int_{- i \infty}^{+ i \infty} \frac{du}{2 \pi i}
{\pi \over 2 \sin {\pi u \over 2}} \left(\frac{z}{1-z}\right)^{\frac{u}{2}-1}
\widetilde M(u)\,,
\\
\widetilde M(u)&\equiv \left({u \over 2} -2 \right)^2 \left({u \over 2} -3 \right)^2 \int_{- i \infty}^{i \infty} \frac{ds}{2\pi i} M(s,16-s-u)\,,
\label{Mu_Mellin}
\end{align}
where the Mellin amplitude $M(s,t)$ was introduced in \eqref{eq:mellinH}, and $\widetilde M(u)$ is the integrated Mellin amplitude with respect to only one of the arguments.  In \eqref{EEC_Mellin}, the integration contour is chosen such that $6>{\rm Re}(s),{\rm Re}(u),{\rm Re}(16-s-u)>4$.

In the weak- and strong-coupling limits, $\widetilde M(u)$ reduces to
\be
\widetilde M_{\text{weak}}(u) = {8 g^2 \over 2-{u \over 2} }  \,, \qquad 
\widetilde M_{\text{strong}}(u) =2 \left(2-{u \over 2}\right)\left(3-{u \over 2}\right) \,. 
\ee
These formulas originate from 
\be
M_{\text{weak}}(s,t) &= \frac{-2 g^2 }{\left(\tfrac{s}{2}-3\right)^2\left(\tfrac{t}{2}-3\right)^2\left(\tfrac{u}{2}-3\right)^2} \,, \cr
M_{\text{strong}}(s,t) &= \frac{1}{\left(\tfrac{s}{2}-3\right)\left(\tfrac{t}{2}-3\right)\left(\tfrac{u}{2}-3\right)} \,,
\ee
upon doing the integral \eqref{Mu_Mellin}. 

We now show that the large-$|z|$ behavior of the EEC is governed by the large-$u$ asymptotics of $\widetilde M(u)$. For definiteness, we continue $z$ to the upper half–plane and rewrite the EEC as
\be
\label{eq:ReggeasymptMellin}
\mbox{EEC}(z) &=- \frac{1}{4 z^3} \int_{- i \infty}^{+ i \infty} \frac{du}{2 \pi i}
{\pi e^{i \pi(u/2-1)}\over 2 \sin {\pi u \over 2}} \left(1-{1 \over z}\right)^{1-\frac{u}{2}}
\widetilde M(u) \,. 
\ee
The integrand is dominated by the region $u \to -i\infty$. For definiteness, let us assume that in this limit
\be
\label{eq:reggintegr}
\lim_{u \to -i\infty} \widetilde M(u) \sim u^{J_0} e^{J_1 u} \,.
\ee
A priori there is no reason to expect that $\tilde M(u)$ is polynomially-bounded. If $J_1 \neq 0$ and real, the integral \eqref{eq:ReggeasymptMellin} converges for $|z| \to \infty$ implying that $z^3 \mbox{EEC}(z)$ is bounded by $\cO(1)$. For $J_1 =0$, we instead find
\be
\lim_{|z| \to \infty} \mbox{EEC}(z) \sim z^{J_0 - 2} \,. 
\ee
It would be interesting to constrain the large-$u$ behavior \eqref{eq:reggintegr} directly from first principles.

\subsection{The source-detector OPE}

Using dispersion relations in Mellin space, one can derive the following representation for the source–detector OPE blocks (see formula~(7.18) in~\cite{Korchemsky:2021okt}, evaluated at $\omega_2=0$) 
\be \label{eq:alphaPMellin}
\alpha(\tau,J) {\cal P}_{J+2}(\tau,1-2z) &= -{N_c^2-1 \over 16} {1 \over z^3} \int_{- i \infty}^{i \infty} {d u \over 2 \pi i} {\pi \over 2 \sin {\pi u \over 2}} \left({u \over 2} -2 \right)^2 \left({u \over 2} -3 \right)^2 \nn \\
&\times \left({z \over 1-z} \right)^{{u \over 2}-1} \sum_{m=0}^\infty {{\cal Q}_{J,m}^{\tau+4}(u-8)  \over ({\tau \over 2}+m)^2 ({\tau \over 2}+m-1)^2} \ ,
\ee
where we set $j_{\text{there}} = 2 - \tfrac{u}{2}$, shifted $\tau \to \tau + 4$, and we introduced an overall normalization chosen to match the conventions of the present paper. The Mack polynomials $\mathcal{Q}_{J,m}^{\tau+4}$ are given in~\cite{Costa:2012cb}. They are evaluated for $d=4$ and external dimensions $\Delta_i=2$ in the expression above.

To evaluate the integral over $u$ and the sum over descendants $m$, it is convenient to factorize them using Eq.~(C.6) of~\cite{Caron-Huot:2022sdy}.\footnote{We thank Simon Caron-Huot for pointing this out to us.} The integral over $u$ can then be evaluated straightforwardly
\be
\label{eq:mellinopeint}
- {1 \over z^3} \int_{- i \infty}^{i \infty} {d u \over 2 \pi i} {\pi \over 2 \sin {\pi u \over 2}} \left({u \over 2} -2 \right)^2 \left({u \over 2} -3 \right)^2  \left({z \over 1-z} \right)^{{u \over 2}-1} \left({8 - u \over 2} \right)_k \cr
=\Gamma(k+3) \partial_z^2 \left[ z^{k+2} (1-z)^2 \right] \,. 
\ee
Likewise, the sum over descendants in \eqref{eq:alphaPMellin} can be evaluated explicitly, and it can be expressed in terms of $\Gamma$-functions. The resulting expression provides a representation of the source–detector OPE in terms of the polynomials in~\eqref{eq:mellinopeint} and the functions $\left[Q_{\tau,J}\right]_{q,k}$ defined in~\cite{Caron-Huot:2022sdy}, which can be computed recursively. Moreover, \eqref{eq:mellinopeint} makes it manifest that the blocks ${\mathcal{P}}_{J+2}(\tau,1-2z)$ are orthogonal to $P_0(1-2z)$ and $P_1(1-2z)$. This representation has proven particularly convenient for evaluating the large-$J$ blocks that enter the numerical bootstrap analysis.

\section{The Froissart-Gribov inversion formula for EEC multipoles}
\label{app:FGeecmultipoles}

In this appendix we derive the dispersive representation for the EEC multipoles $c_s$. The argument is analogous to the derivation of the Froissart-Gribov representation for the partial waves of scattering amplitudes \cite{Gribov:2003nw}, and we review it here.

It is convenient to consider
\be
f(x) \equiv \text{EEC} \Big({1-x \over 2} \Big)-1 = \sum_{s=2}^\infty c_s P_s(x) \,, \qquad x\equiv \cos \theta \,.
\ee
By performing the multipole projection, we can invert the formula above
\be \label{eq:gotcs}
c_s = {(2s+1) \over 2} \int_{-1}^{1} dx\, f(x) P_s(x) \,. 
\ee
Next, we define the $Q_s(x)$ function
\be
\label{eq:legendreQ}
Q_s(x) = \frac{\sqrt{\pi } 2^{-s-1} x^{-s-1} \Gamma (s+1) \,
	_2F_1\left(\frac{s+1}{2},\frac{s+2}{2};s+\frac{3}{2};\frac{1}{x^2}\right)}{\Gamma
	\left(s+\frac{3}{2}\right)} \,.
\ee
It has a cut for $x \in [-1,1]$ such that
\be
{Q_s(x+i 0)-Q_s(x-i 0) \over 2 i} = - {\pi \over 2} P_s(x) \,.
\ee
We can, therefore, rewrite the integral \eqref{eq:gotcs} as 
\be
c_s = (2s+1)\oint_{[-1,1]} {dx \over 2 \pi i}\,  f(x) Q_s(x) \,,
\ee
where the integral goes counterclockwise. We then deform the contour and drop the arc at infinity to get
\be
\label{eq:projectionopened}
c_s = {2s+1 \over \pi} \left( \int_{1}^{\infty} dx\,  Q_s(x) {\rm Disc}_{x=1} f(x) + \int_{-\infty}^{-1} dx\,  Q_s(x) {\rm Disc}_{x=-1}f(x) \right) \,.
\ee
We can use the identity
\be
Q_s(-z) = (-1)^{s+1} Q_s(z)
\ee
to rewrite \eqref{eq:projectionopened} as
\be
c_s^{\pm} = {2s+1 \over \pi} \int_{1}^{\infty} dx\, Q_s(x) \Big( {\rm Disc}_{x=1} f(x) \mp {\rm Disc}_{x=-1} f(-x) \Big) \ . 
\ee
Let us now carefully switch to the $z$ variable defined by $x=1-2z$, and define
\be
{\rm Disc}_{z} X(z) &= {X(z+i 0) - X(z-i0) \over 2 i} \,.
\ee
We then obtain the final formula
\be
\label{eq:inversionmultipole2}
c_s^{\pm} =2 {2s+1 \over \pi} \int_{-\infty}^{0} d\tilde z\, Q_s(1-2 \tilde z) \Big( {\rm Disc}_{z=0} \text{EEC}(\tilde z) \pm {\rm Disc}_{z=1} \text{EEC}(1- \tilde z) \Big) \,. 
\ee
As usual, this formula is amenable to analytic continuation in $s$ to complex values. In deriving this formula, we dropped the arc at infinity. At large $|z|$, we have $Q_s(1-2z) \sim {1 \over z^{s+1}}$. Therefore, the representation above is valid to $s=2$ only if $\lim_{|z| \to \infty} {\text{EEC}(z)  \over z^2} = 0$. It would be interesting to understand if this is always the case; here, we simply assume it.

At large $s$,  we can approximate $Q_s(x)$ (see, for instance, a related discussion in \cite{Correia:2020xtr}) with 
\be
Q_s(x) \simeq \frac{\sqrt{\pi } \kappa (x)^{-s}}{\sqrt{s} \sqrt{\kappa (x)^2-1}} \,, \qquad \kappa(x) = x+\sqrt{x^2-1} \,.
\ee
Therefore, the integral \eqref{eq:inversionmultipole2} is effectively localized close to the upper integration limit $\tilde z=0$, which is controlled by $z=0,1$ regions of the EEC\@. This fact makes precise the intuition that the large-$s$ limit is controlled by the $z \to 0$ and $z \to 1$ limits of the EEC\@. 

\section{Universal bounds on EEC multipoles}\label{app:MultipoleBounds}

This appendix explores universal bounds on the EEC multipoles $c_s$. As already mentioned in the introduction, these are nonnegative due to unitarity, as they can be expressed as a norm of a certain state in QFT 
\be
c_s \propto \int_{S^2} d \Omega_{\vec n_1} d \Omega_{\vec n_2} Y_{s,m}^*(\vec n_1) \langle {\cal E}(\vec n_1) {\cal E}(\vec n_2) \rangle Y_{s,m}(\vec n_2) \geq 0 \,,
\ee
where $Y_{s,m}^*(\vec n)$ are the usual spherical harmonics. Energy conservation implies that $c_0=1$ and in gapless theories, such as CFTs, we also have $c_1=0$.\footnote{We could have $c_1 \geq 0$ for gapped theories, such as QCD\@.} In addition to this property, we should also require that EEC itself to be nonnegative for every angle 
\begin{equation}\label{eq:multipole_primal}
{\rm EEC}(z) = 1+\sum_{s=2}^\infty c_s P_s(1-2z)\geq 0 \,. 
\end{equation}
In this section, we derive universal bounds on $c_s$ that follow from this condition.  

\subsection{Upper bound on $c_s$}

\begin{figure}[!t]
    \centering
    \includegraphics[width=0.5\linewidth]{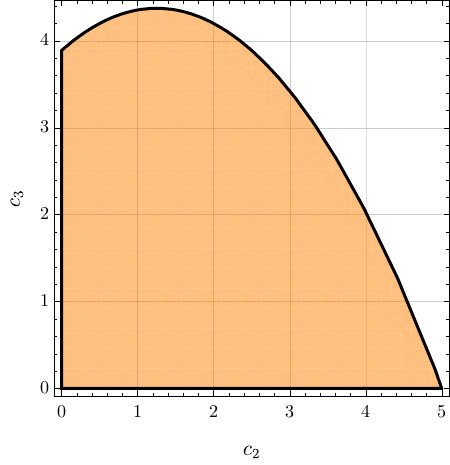}
    \caption{The allowed region in the multipole coefficients $c_2$ and $c_3$, given analytically in \eqref{eq:c2c3}.}
    \label{fig:c2c3}
\end{figure}

\begin{figure}[!t]
    \centering
    \includegraphics[width=0.7\linewidth]{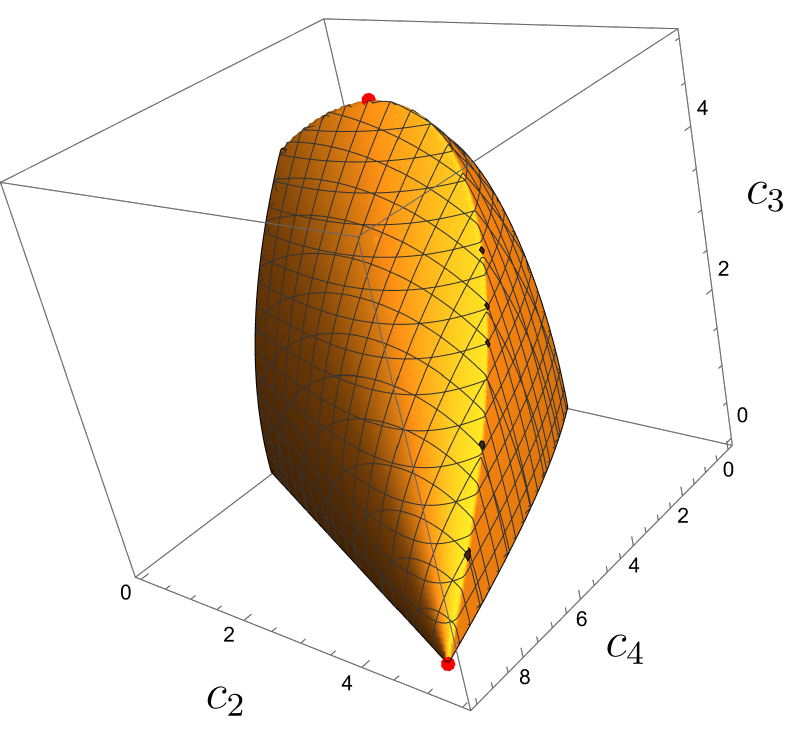}
    \caption{The allowed region in the space of multipole coefficients $(c_2,c_3,c_4)$. \figref{fig:c2c3} is a two-dimensional projection of this three-dimensional shape. Red dots correspond to maximal value solutions for $c_2$ and $c_4$ $(5,0,9)$, and the one for $c_3$ is given by $(5/4, 35/8, 81/64)$ correspondingly.}
    \label{fig:c2c3c4}
\end{figure}

To derive an upper bound on $c_s$, we can start by writing a primal linear program. The variables will be the values of $\text{EEC}(z)$ at every $z\in [0,1]$, and the objective will be $c_s = (2s+1)\int_0^1 \text{EEC}(z)P_s(1-2z)\,dz$. The constraints will be $\text{EEC}(z) \ge 0$ along with $\int_0^1 \text{EEC}(z)\,dz = 1$ and $\int_0^1 z\,\text{EEC}(z)\,dz = \frac{1}{2}$; the latter two impose $c_0 = 1$ and $c_1 = 0$. This linear program,
\begin{equation}
    \begin{array}{ll@{}ll}
    \text{max} \quad & (2s+1)\int_0^1 \text{EEC}(z)P_s(1-2z)\,dz &\\
    \text{subject to}\quad & \int_0^1 \text{EEC}(z)\,dz - 1 = 0\,,\quad & \\
     & \int_0^1 \text{EEC}(z) z\,dz - \frac{1}{2} = 0\,,\quad & \\          
     & \text{EEC}(z) \ge 0 &\quad  \forall\, z\in [0,1]\,,
    \end{array}
\end{equation}
is of the form we considered in \eqref{eq:primal_to_dual}. We can use the recipe given there to form a dual linear program:
\begin{equation}\label{eq:max_cs_dual}
    \begin{array}{ll@{}ll}
    \text{min} \quad & \alpha_0 + \frac{1}{2}\alpha_1 &\\
    \text{subject to}\quad & \alpha_0 + \alpha_1 z \ge (2s+1)P_s(1-2z)\,, & \quad \forall z\in [0,1]\,.
    \end{array}
\end{equation}

The dual program is easier to solve. For even values of $s$, we have $P_s(1) = P_s(-1) = 1$, and so the optimal solution has $\alpha_0 = 2s+1$ and $\alpha_1 = 0$, with objective value $2s+1$. The fact that the dual constraints are saturated only at $z = 0$ and $z = 1$ means that the primal optimizer has support only at these values; that is, the bounds on even $c_s$ are all saturated by
\begin{equation}
    \text{EEC}(z) = \frac{1}{2}\Big(\delta(z) + \delta(1-z)\Big)\,.
\end{equation}
Note that this EEC has odd multipoles all equal to 0, so it satisfies $c_s \ge 0$ for all $s$; thus, we would not obtain better bounds by imposing $c_s \ge 0$ as an additional constraint in our primal problem.

For odd values of $s$, the optimal values of $\alpha_0$ and $\alpha_1$ depend on $s$. We can interpret \eqref{eq:max_cs_dual} graphically: we need to find a linear function that upper bounds a Legendre polynomial everywhere, and then minimize its value at $z = \frac{1}{2}$. For example, the optimal linear functions for bounding $c_3$ and $c_5$ are shown below:
\begin{center}
    \scalebox{.75}{\begin{tikzpicture}
    \begin{axis}[
        axis lines = middle,
        xlabel = $z$,
        ylabel = {$7P_3(1-2z)$},
        xmin = -.1, xmax = 1.1,
        ymin = -8.4, ymax = 8.4,
        xtick = {0.5,1}, ytick = {-7,4.375,7},
        yticklabel style={/pgf/number format/fixed,
                  /pgf/number format/precision=3},
        samples = 200,
        domain = 0:1
    ]
        \addplot[thick, black] {7/2 * (5 * (1-2*x)^3 - 3 * (1-2*x))};
        \addplot[thick, blue, dashed] {7*(1 - (3/4)*x)};
        \addplot[mark=none, gray, dashed] coordinates {(0,4.375) (.5,4.375) (.5,0)};
    \end{axis}
    \end{tikzpicture}}
    \hspace{1in}
    \scalebox{.75}{\begin{tikzpicture}
    \begin{axis}[
        axis lines = middle,
        xlabel = $z$,
        ylabel = {$11P_5(1-2z)$},
        xmin = -.1, xmax = 1.1,
        ymin = -13.2, ymax = 13.2,
        xtick = {.5,1}, ytick = {-11,7.39424,11},
        yticklabel style={/pgf/number format/fixed,
                  /pgf/number format/precision=5},
        samples = 200,
        domain = 0:1
    ]
        \addplot[thick, black] {11/8 * (63 * (1-2*x)^5 - 70 * (1-2*x)^3 + 15 * (1-2*x))};
        \addplot[thick, blue, dashed] {11*(0.671227 + 0.328773*(1-2*x))};
        \addplot[mark=none, gray, dashed] coordinates {(0,7.39424) (.5,7.39424) (.5,0)};
    \end{axis}
    \end{tikzpicture}}
\end{center}
For $c_3$, the upper bound is exactly $c_3 \le \frac{35}{8}$; for odd multipoles with $s>3$, we can solve the problem numerically and find, for instance,
\begin{equation}
    c_5 \le 7.39424\,, \qquad c_7 \le 10.2895\,, \qquad c_9 \le 13.1451\,.
\end{equation}
Like in the even multipole case, we see that the dual constraints are saturated at only two values of $z$, $z = 0$ and another $s$-dependent value which we can call $z_s$. In terms of this, the primal optimal solution is given by
\be
{\rm EEC}(z) = \left(1-{1 \over 2 z_s}\right)\delta(z) + {1 \over 2 z_s} \delta(z-z_s) \,.
\ee
It is straightforward to check that this solution has $c_s \ge 0$ for all $s$, so again we would not improve by imposing this as an additional constraint from the outset.

At large odd values of $s$, the upper bound on $c_s$ grows like\footnote{This can be derived using the asymptotic formula $z_s = 1 - \frac{j_{1,1}}{4(s+1/2)^2} + \cdots$ for the position of the rightmost maximum of $P_s(1-2z)$, which can be found in Section 18.16 of the DLMF \cite{NIST:DLMF}, along with a formula due to Rayleigh \cite{10.1112/plms/s1-9.1.61}, $\lim_{s\to\infty} P_s\left(\cos \frac{\rho}{s}\right) = J_0(\rho)$.}
\begin{equation}
    \max c_s = \frac{1-J_0(j_{1,1})}{2}(2s+1) + O\left(s^{-1}\right) \approx 0.701(2s+1) + O\left(s^{-1}\right) \,,
\end{equation}
where $J_0$ is a Bessel function and $j_{1,1}$ is the first zero of the Bessel function $J_1$.

\subsection{Joint bounds}

We can also solve similar programs to derive joint bounds on different multipole coefficients. In \figref{fig:c2c3}, we plot the allowed region of the coefficients $c_2$ and $c_3$. In this case, the constraints can be determined analytically:
\begin{equation}\label{eq:c2c3}
    c_2\cos\chi + c_3\sin\chi \le \begin{cases} 5\cos\chi & \cot\chi \ge 7/3 \,,\\
    \frac{5}{56}\sin\chi\left(40+14\cot\chi+9\csc^2\chi\right) & \cot\chi < 7/3 \,.
    \end{cases}
\end{equation}
It is also not hard to derive the shape for the allowed values of the triple $(c_2,c_3,c_4)$ which we present in \figref{fig:c2c3c4}. 

\section{Bootstrap implementation details}\label{app:bootstrap}
In this section, we give further details on the numerical bootstrap implementation. 

\subsection{Calculation of bootstrap inputs}

For the finite-$N_c$ bootstrap, we need to calculate derivatives of conformal blocks, integrals of conformal blocks, derivatives and integrals of the protected contribution to the reduced correlator, and the functions $\mathcal{F}_k(N_c, g)$.

For the derivatives of conformal blocks, we use the \texttt{scalar\_blocks} code \cite{scalar_blocks} and work up to order 200 in the $r$-expansion. The integrals of conformal blocks are calculated numerically after restricting the integration region using crossing symmetry, as discussed in Section~3.1 of \cite{Chester:2023ehi}; see \appref{sec:large_twist} for a discussion of a resulting subtlety at large twist. The derivatives and integrals of the protected contribution can be evaluated (analytically and numerically, respectively) from the exact expression for the protected contribution given in \cite{Beem:2016wfs}. The method we use for calculating $\mathcal{F}_k(N_c, g)$ is given in \cite{Alday:2023pet}, and it is also described briefly in Section~2.1 of \cite{Chester:2023ehi}. Example values of the quantities $F_{\tau,J,m,n}$, $F^\text{protected}_{m,n}(N_c)$, $I_{k;\tau,J}$, $I_k^\text{protected}(N_c)$, and $\mathcal{F}_k(N_c,g)$ are given in \tabref{tab:nonplanar_example_values}.

\begin{table}
    \centering
    \begin{tabular}{cc|cc}
        \toprule
        $F_{5,0;4,3}$ & $0.324768$ & $\{I_{2;5,0},I_{4;5,0}\}$ & $\{-0.353254,41.9432\}$ \\
        $F_{5,10;4,3}$ & $9.37301$ & $\{I_{2;5,10},I_{4;5,10}\}$ & $\{3.07731\times 10^{-2},-3.45350\}$ \\
        $F_{105,0;4,3}$ & $5.82227\times 10^{-8}$ & $\{I_{2;105,0},I_{4;105,0}\}$ & $\{-0.876660,98.6568\}$ \\
        $F_{105,10;4,3}$ & $2.82475\times 10^{-8}$ & $\{I_{2;105,10},I_{4;105,10}\}$ & $\{1.41991,-159.782\}$ \\
        \midrule
        $F^\text{protected}_{4,3}(2)$ & $2.65386\times 10^{-2}$ & $\{I^\text{protected}_2(2),I^\text{protected}_4(2)\}$ & $\{0.565655,-73.4277\}$ \\
    & & $\{\mathcal{F}_2(2,0.4),\mathcal{F}_4(2,0.4)\}$ & $\{0.270370,-36.6921\}$ \\
    \bottomrule
    \end{tabular}
    \caption{Example values of quantities used to build the linear program \eqref{eq:primal_lp_nonplanar} for finite $N_c$.}
    \label{tab:nonplanar_example_values}
\end{table}

For the bootstrap in the planar limit, there are several more quantities we need to compute (in addition to various simpler pieces already described in the main text): $X_{\tau,J}(u,v)$, $B_{\tau,J}(v)$, $\hat{B}_{\tau,J}(t)$, $\Psi_{\ell;\tau,J}$, $\Phi_{\ell,\ell+2;\tau,J}$, and $I_{k;\tau,J}$. For convenience, we will give expressions for these quantities here. Details and derivations can be found in \cite{Caron-Huot:2022sdy,Caron-Huot:2024tzr}.

The formulas for these quantities will involve Mack polynomial coefficients $[Q_{\tau,J}]_{q,k}$, described in \appref{app:ComplexAngle}. For $B$ and $\hat{B}$ we will also need the quantity
\begin{equation}
    S^q_{\tau,J} \equiv \frac{2(-1)^q \Gamma(5+J-q) \Gamma(\tau+2J-1)\Gamma(\tau+2J)}{\Gamma\left(\frac{8-\tau-2q}{2}\right)^2\Gamma\left(\frac{4+\tau+2J}{2}\right)^2\Gamma\left(\frac{\tau+2J}{2}\right)^4}\,.
\end{equation}
Using this, we can define
\begin{equation}
    \tilde B_{k;\tau,J} \equiv \sum_{q=0}^{J+1-k} \Big\lbrack \left(2(\tau+2q-k)S^q_{\tau,J} - 4 S^{q+1}_{\tau,J}\right)[Q_{\tau+4,J}]_{q,k} -2 S^q_{\tau,J}[Q_{\tau+4,J}]_{q,k-1}\Big\rbrack\,.
\end{equation}
We then have
\begin{equation}\label{eq:bhat_b_defs}
    \hat{B}_{\tau,J}(t) = \sum_{k=0}^{J+1} \tilde B_{k;\tau,J} [\hat{B}_t]_k\,,\qquad B_{\tau,J}(v) = \sum_{k=0}^{J+1} \tilde B_{k;\tau,J}[B_v]_k\,,
\end{equation}
where
\begin{equation}
    [\hat{B}_t]_k \equiv \frac{\left(\frac{t}{2}-1\right)_k}{t-6}\,,\qquad [B_v]_k \equiv -\frac{\Gamma(k+3)}{2(k+2)_3}{_2F_1}(3,k+3;k+5;1-v)\,.
\end{equation}

The $\Psi$ and $\Phi$ functionals can be calculated using similar formulas:
\begin{equation}\label{eq:psi_phi_defs}
    \Psi_{\ell;\tau,J} = \sum_{k=0}^{J+1} \tilde B_{k;\tau,J} [\Psi_\ell]_k\,, \qquad 
    \Phi_{\ell,\ell+2;\tau,J} = \sum_{k=0}^{J+1} \tilde B_{k;\tau,J}[\Psi_{\ell,\ell+2}]_k\,.
\end{equation}
To define $[\Psi_\ell]_k$ and $[\Psi_{\ell,\ell+2}]_k$, we need some auxiliary quantities. Let
\begin{equation}
\begin{split}
    a_\ell(t) &\equiv \frac{\Gamma(2\ell+6)}{4\Gamma(\ell+3)^2} {}_3F_2\left(-\ell, \ell+5, \frac{t-2}{2}; 3,3; 1\right)\,,\\
    c_\ell &\equiv \frac{\Gamma(\ell+3)^4}{\Gamma(2\ell+6)^2} (\ell+1)_4 (2\ell+5)\,,\\
    \Phi_\ell^\infty &\equiv \frac{\Gamma(\ell+3)^2}{\Gamma(2\ell+5)} H_{\ell+2}\,,
\end{split}
\label{eq:phiellinft}
\end{equation}
where $H_n$ is a harmonic number. We then have
\begin{equation}
    [\Phi_{\ell,\ell_2}]_k \equiv [\Phi_\ell]_k - \frac{\Phi_\ell^\infty}{\Phi_{\ell+2}^\infty} [\Phi_{\ell+2}]_k\,,
\end{equation}
where
\begin{equation}
    [\Phi_\ell]_k \equiv \frac{\pi}{64i}\int_{-\infty}^\infty \frac{dy}{\cosh^2(\pi y/2)} c_\ell a_\ell(5+iy) \left(\int_i^y dy'\,(1+y'^2)\left\lbrack \left(\frac{1+i y'}{2}\right)_{k+1}-\left(\frac{1-i y'}{2}\right)_{k+1}\right\rbrack\right)\,.
\end{equation}
The inner integrand is a polynomial, and the outer integral can be evaluated using $\int_{-\infty}^\infty \frac{y^n\,dy}{\cosh^2(\pi y/2)} = \frac{4(-1)^{n/2}}{\pi}(2^n-2)B_n$, where $B_n$ are Bernoulli numbers. The expression for $[\Psi_\ell]_k$ is
\begin{equation}
\begin{split}
    [\Psi_\ell]_k &\equiv \frac{\Gamma(\ell+3)^2 \Gamma(k+2)}{4\Gamma(2\ell+5)}\left(\frac{(k-\ell)_{\ell+3}}{(k+1)_{\ell+4}}(k+7+\ell(\ell+5))-1\right)\\
    &\quad + \beta_{\ell}[\Phi_\ell]_k  - \sum_{n=0}^{\lfloor k/2\rfloor} I_{\ell,2n} [\Phi_{2n}]_k \,,
\end{split}
\end{equation}
where
\begin{equation}
    I_{\ell,J} \equiv \begin{cases}
        \frac{1}{(J-\ell)(\ell+J+5)}\frac{(\ell+2)_2}{(J+2)_2} \frac{\Gamma(2\ell+6)\Gamma(\ell+3)^2}{\Gamma(2\ell+5)\Gamma(J+3)^2} & 0 \le \ell < J\,,\\
        2H_{2\ell+4} - 3H_{\ell+2} + \frac{13+8\ell+\ell^2}{(\ell+1)(\ell+3)(2\ell+5)} & \ell = J\,,\\
        \frac{1}{(J+1)(J+4)}\frac{\Gamma(2J+6)\Gamma(\ell+3)^2}{\Gamma(2\ell+5)\Gamma(J+3)^2} & \ell > J\,,
    \end{cases}
\end{equation}
and 
\begin{equation}
\label{betal}
   \beta_{\ell}= H_{2\ell+4} - H_{\ell+2}+\frac{H_r^{(2)}}{2H_{\ell+2}} \,.
\end{equation}
For the Polyakov-Regge blocks, we need an additional quantity $T^{s,q}_{\tau,J}$ defined by the recursion
\begin{equation}
\begin{split}
    T^{s,q}_{\tau,J} \equiv{} &\left(q - 1 + \frac{\tau-s}{2}\right)T^{s,q-1}_{\tau,J} + \frac{1}{2}S^{q-1}_{\tau,J}\,,\\
    T^{s,0}_{\tau,J} \equiv{} &\frac{2\Gamma(\tau+2J)\Gamma(\tau+2J-1)}{\Gamma\left(\frac{8-\tau}{2}\right)^2\Gamma\left(\frac{\tau+2J}{2}\right)^4 \Gamma(\tau+J-1)(s-\tau)}\\
    {}&\times {_3F_2}\left(\frac{\tau-6}{2},\frac{\tau-6}{2},\frac{\tau-s}{2};\frac{\tau-s}{2}+1,\tau-J-1;1\right)\,.
\end{split}
\end{equation}
We then have\footnote{As explained in \cite{Caron-Huot:2022sdy}, this integral (and the similar one in \eqref{eq:I_int}) can be efficiently evaluated as a Riemann sum using the parametrization $s = 5+i\sinh(x)$.}
\begin{equation}\label{eq:p_integral}
    \mathcal{P}^{\mathcal{N}=4}_{\tau,J}(u,v) = \int_{5-i\infty}^{5+i\infty} \frac{ds}{4\pi i} \Gamma\left(4-\frac{s}{2}\right)^2\sum_{q,k=0}^J T^{s,q}_{\tau+4,J} \left(u^{\frac{s}{2}-4}[\hat{\mathcal{P}}_{s,v}]_k + v^{\frac{s}{2}-4}[\hat{P}_{s,u}]_k\right)[Q_{\Delta+4,j}]_{q,k}\,,
\end{equation}
where
\begin{equation}
    [\hat{\mathcal{P}}_{s,v}]_k \equiv \frac{\Gamma\left(\frac{s}{2}\right)^2\Gamma\left(k + \frac{s}{2}\right)^2}{\Gamma(k+s)}{_2F_1}\left(\frac{s}{2}, \frac{s}{2}+k,s+k,1-v\right)\,.
\end{equation}

The crossing objects $X_{\tau,J}(u,v)$ are then defined in terms of $\mathcal{P}_{\tau,J}(u,v)$ and $\hat{\mathcal{P}}_{\tau,J}(s,t)$ via
\begin{equation}
    X_{\tau,J}(u,v) \equiv \mathcal{P}^{\mathcal{N}=4}_{\tau,J}(u,v) - \frac{1}{u^4} \mathcal{P}^{\mathcal{N}=4}_{\tau,J}(1/u,v/u)\,.
\end{equation}

For the integral constraints, we define the quantities
\begin{equation}
\begin{split}
    [\tilde I_{2,s}]_k &\equiv -\frac{\left(\frac{s}{2}\right)_k}{2(k+2)(k+3)}\Gamma\left(\frac{s}{2}-2\right)\Gamma\left(\frac{s}{2}\right)\,,\\
    [\tilde I_{4,s}]_k &\equiv \frac{1}{2}\Gamma\left(\frac{s}{2}-2\right)\Gamma\left(\frac{s}{2}\right)\Bigg\lbrack -\frac{768(s-5)^2 k!}{(s-6)^2(s-4)(s-2)} + \frac{48\left(\frac{s}{2}\right)_k}{(k+1)_3(s-6)}\times\\
     &\hspace{-1cm} \Big(2\frac{(k+3)(s-5)(s-4)}{2k+s-2}+8(k+s-4)\\
     &\hspace{-.8cm}+((k+5)(s-4)-4)\left(H_{s/2+k-1}-H_{s/2-4}-2H_{k+3}\right)\Big)\Bigg\rbrack\,,
\end{split}
\end{equation}
where $H_n$ is a harmonic number. We then have
\begin{equation}\label{eq:I_int}
    I^\text{PR}_{p;\tau,J} = 2\int_{5-i\infty}^{5+i\infty} \frac{ds}{4\pi i} \Gamma\left(4-\frac{s}{2}\right)^2\sum_{q,k=0}^J T^{s,q}_{\tau+4,J} [Q_{\tau+4,J}]_{q,k}[\tilde{I}_{p,s}]_k\,.
\end{equation}

For example values of the quantities that enter into our bootstrap setup, see \tabref{tab:example_values}.

\begin{table}
    \centering
    \begin{tabular}{ccccc}
        \toprule
        Functional & $(\tau = 5,J = 0)$ & $(\tau = 5,J = 10)$ & $(\tau = 105, J = 0)$ & $(\tau = 105, J = 10)$ \\
        \midrule
        $\hat{B}_{\tau,J}(5+1/10)$ & $5.20170$ & $120496.$ & $1.35902\times 10^{51}$ & $7.53546\times 10^{57}$ \\
        $B_{\tau,J}(1+1/20)$ & $0.192734$ & $3851.66$ & $4.66769\times 10^{49}$ & $2.67514\times 10^{56}$ \\
        $\Psi_{0;\tau,J}$ & $0.275647$ & $1433.44$ & $6.99961\times 10^{49}$ & $3.92635\times 10^{56}$ \\
        $\Phi_{0,2;\tau,J}$ & $0.140702$ & $-4451.84$ & $4.07578\times 10^{49}$ & $2.15790\times 10^{56}$ \\
        $X_{\tau,J}(3/4,5/8)$ & $5.00878\times 10^{-4}$ & $-30.3537$ & $8.43107\times 10^{39}$ & $-4.67200\times 10^{48}$ \\
        $I^\text{PR}_{2;\tau,J}$ & $-5.68492\times 10^{-3}$ & $-85.4995$ & $-1.86890\times 10^{43}$ & $-7.94388\times 10^{49}$ \\
        $I^\text{PR}_{4;\tau,J}$ & $-0.737032$ & $5653.35$ & $-2.39198\times 10^{45}$ & $-1.02688\times 10^{52}$ \\        
        \bottomrule
    \end{tabular}
    \caption{Example values of quantities used to build the linear program \eqref{eq:primal_lp_planar} in the planar limit.}
    \label{tab:example_values}
\end{table}

\subsection{Finite truncations}\label{app:bootstrap_truncation}

As discussed in the main text, in order to solve our bootstrap problems we have to truncate the infinite set of primal constraints as well as the infinite set of primal variables. The set of constraints controls the power of the bootstrap; adding more constraints monotonically improves the bounds. By contrast, truncating the set of variables involves a loss of rigor, and we must be careful to check that our bounds are robust against adding additional variables.

At finite $N_c$, we include crossing derivatives with $m+n\le \Lambda$, with $\Lambda$ as large as 107 in some cases (details are given in \secref{sec:nonplanarbootstrap}). We also include both integral constraints \eqref{eq:nonplanar_integral}. When bootstrapping $\text{EEC}_{\psi}(z)$ (defined in \eqref{eq:smearing}), we include 31 constraints from the ANEC: 18 to give a linear relaxation of the joint $c_2$ and $c_3$ bounds in \figref{fig:c2c3}, and 13 more to give the upper bounds on $c_4,c_5,\ldots,c_{16}$.

When working at truncation parameter $\Lambda$, we truncate the spins to $J=\{0,2,\ldots,\Lambda+9\}$ and include the following values of twist at each spin:
\begin{equation}
    \tau \in \{2,2+1/32,\ldots,4,4+1/8,\ldots,8,8+1/2,\ldots,10,11,\ldots,110\} \,.
\end{equation}
In \tabref{tab:nonplanar_convergence_check}, we give evidence that this truncation does not significantly affect our bounds, using as an example our lower bound on $\text{EEC}_{\psi[2,\frac{1}{16}]}(13/16)$ for $N = 10$ and $g = 0.4$, computed at $\Lambda = 107$.

\begin{table}
    \centering
    \begin{tabular}{cccc}
        \toprule
         $\max J$ & $\max \tau$ & $\min(\tau\text{ spacing})$ & $\min \left(\text{EEC}_{\psi[2,\frac{1}{16}]}(13/16)\right)$ at $N_c = 10$, $g = 0.4$ \\
         \midrule
         116 & 110 & 1/32 & 0.909596 \\
         136 & ~ & ~ & 0.909594 \\
         ~ & 130 & ~ & 0.909518 \\
         ~ & ~ & 1/64 & 0.909065 \\
         \bottomrule
    \end{tabular}
    \caption{The effects of the maximum spin, maximum twist, and twist spacing cutoffs on one of our bounds (chosen as a bound which is one of the most sensitive to cutoffs). We see that extending any of these cutoffs does not substantially change the bound.}
    \label{tab:nonplanar_convergence_check}
\end{table}

In the planar limit, we include several types of dispersive functionals discussed in \secref{sec:dispersive}, as well as the integral constraints discussed in \secref{sec:integral_constraints}. \tabref{tab:planar_fnals} gives the exact set of dispersive functionals we used to bound EEC and $c_j$ multipoles. The same set of functionals were used for all the planar bounds.

\begin{table}
    \centering
    \begin{tabular}{cccccc}
        \toprule
         $\Psi_{\ell}$ & $\Phi_{\ell_1,\ell_2}$& $X(u,v)$ & $X(u,v)$ (cont.) & $\hat{B}(t)$ & $B(v)$ \\
         \midrule
         0 & (0,2) & $(1/2,25/7)$ & $(9/11,47/9)$ & 23/5& 101/100 \\
         ~ & (2,4) & $(4/13,22/9)$ & $(1/4,26/11)$ & 24/5& 21/20 \\
         ~ & (4,6) & $(5/8,23/7)$ & $(5/9,25/8)$ & 5& 101/10 \\
         ~ & (6,8) & $(16/17,62/11)$ & $(1/2,23/5)$ & 26/5& 50 \\
         ~ & ~ & $(7/8,23/6)$ & $(1/7,17/8)$ & 27/5& ~ \\
         ~ & ~ & $(17/18,47/10)$ & $(1/6,15/7)$ & 28/5& ~ \\
         ~ & ~ & $(2/7,21/5)$ & $(1/2,31/9)$ & 29/5& ~ \\
         ~ & ~ & $(9/11,67/10)$ & $(7/10,19/4)$ & 25/6 & ~\\
         ~ & ~ & $(14/15,35/9)$ & $(1/4,26/11)$ & 13/3 &~ \\
         ~ & ~ & $(5/7,69/11)$  & & 9/2 & ~\\
         \bottomrule
    \end{tabular}
    \caption{The explicit set of functionals used for the bootstrap in the $N_c\to\infty$ limit.}
    \label{tab:planar_fnals}
\end{table}

\begin{figure}
    \centering
    \includegraphics[width=0.6\linewidth]{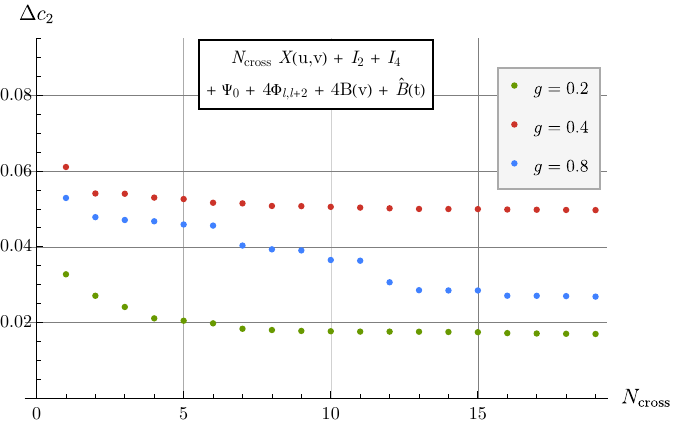}
    \caption{Here we show the convergence of the difference between the upper and lower bounds on $c_2$ as we increase the number of dispersive crossing functionals. For each of the couplings $g = 0.2$, 0.4, and 0.8 shown here, the upper and lower bounds are rather close but do not become substantially closer as we add additional crossing functionals.}
 \label{fig:fnal-convg-planar}
\end{figure}

Our bounds are not substantially improved by adding additional dispersive crossing functionals. We illustrate this in \figref{fig:fnal-convg-planar}, by showing how the upper and lower bounds on $c_2$ at couplings $g = 0.2$, 0.4, and 0.8 vary as we increase the number of $X_{u,v}$ functionals we include. We see that in this example, including $\sim 20$ crossing functionals is sufficient for the bounds to be nearly converged.

\begin{table}
    \centering
    \begin{tabular}{cccc}
        \toprule
         $\max J$ & $\max \tau$ & $\text{ grid size}$ & $\min c_2$ at $g = 0.4$ \\
         \midrule
         170 & 17 &1461& 1.06538 \\
         260 & 17 & 4000 & 1.06535 \\
         170 & 50 & 4000 & 1.06538 \\
         170 & 17 &17900  & 1.06529  \\
         \bottomrule
    \end{tabular}
    \caption{The effects of the maximum spin, maximum twist, and grid size on one of our bounds (chosen as a bound which is one of the most sensitive to cutoffs). We see that extending any of these cutoffs does not substantially change the bound. In the last row of the table, we increase the grid size by including more points in between $\tau_1(g,J)$ and $\tau_{\text{max}}$ for each $J$, thus shrinking the twist spacing considerably. Note that the finite-twist constraints up to the maximum twist values shown here are supplemented by constraints in the large-twist limit, as discussed in \appref{sec:large_twist}.}
    \label{tab:planar_convergence_check}
\end{table}

We keep spins $J\in \{0,2,\ldots,80\}\cup \{86,92,\ldots,170\}$, and we additionally use $J\in \{176,182,\ldots,260\}$ when $g>1$. For each value of $J$, we include twists
\begin{equation}
    \tau \in \{\tau_0(g,J),\tau_1(g,J),\tau_1(g,J)+1/10\ldots\}\,,
\end{equation}
where $\tau_0(g,J)$ is the twist of single-trace operators on leading trajectory with spin $J$, and $\tau_1(g,J)$ is the lower bound for single-trace operators on subleading trajectory $J \geq 2$ and the exact value of the twist for the second lightest spin-0 operator. Near the subleading trajectory we have a twist spacing of $1/10$, but we increase this spacing at larger twists, and eventually cut off the set at some maximum twist beyond which we rely on the additional constraints described in \appref{sec:large_twist}. For each value of $J$ we include, we have constraints for roughly 15 values of twist above the subleading trajectory. We can check that the truncation in spin and twist and the density of the points in our $(\tau,J)$ grid do not change our bounds significantly. We show an example of this for our lower bound on $c_2$ at $g=0.4$ in \tabref{tab:planar_convergence_check}.

\subsection{Positivity at large twist}\label{sec:large_twist}

In both the finite-$N_c$ and planar cases, we have to be careful about the positivity of bootstrap functionals at large twists. We discuss a subtlety in the finite-$N_c$ case along with additional large-twist constraints in the planar case below.

\subsubsection*{Finite-$N_c$: spin oscillations and integration regions}

To perform the integrals $I_{k,\tau,J}$ given in \eqref{eq:block_integrals}, we divide the integration region into three domains that are all exchanged by crossing symmetry, and then integrate over only one of the regions, as explained in \cite{Chester:2021aun,Chester:2023ehi}. This is necessary because the OPE decomposition of the correlator only converges in part of the integration region.

There remains a choice of how exactly to split up the integration region; crucially, the most straightforward choice of the reduced integration region will lead to sign oscillations of the form $(-1)^{J/2}$ in the integrated blocks at large twist. Since the integrated blocks grow more quickly than the block derivatives at large twist, these oscillations would force us not to include the integral constraints in a positive functional.

As explained in \cite{Chester:2021aun,Chester:2023ehi}, there are two resolutions to this problem at different levels of rigor. A more rigorous approach is to include the integral constraints twice, using two different integration regions, each adjusted in order to remove the sign oscillations at large twist. A less rigorous approach is to simply use the more straightforward integration region, and only impose positivity of the functional up to some large but finite maximum twist. The two approaches produce essentially identical bounds in practice, and this is justified by noticing that the twist cutoff can be taken larger for larger truncation parameter $\Lambda$, eventually reaching infinity in the $\Lambda\to\infty$ limit. The latter, less rigorous approach is less susceptible to numerical difficulties, and so we employ that approach in this paper.

\subsubsection*{Planar limit: Regge-limit constraints}

For the bootstrap in the planar limit, we cut off our constraints at finite $(\tau,J)$ at a relatively small maximum twist, as detailed in \tabref{tab:planar_convergence_check}. In order to enforce the dual positivity condition 
\begin{equation}
    \sum_c \alpha_c A_{c;\tau,J} + \sum_{c'} \alpha_{c'} a_{c';\tau,J} \ge \beta_{\tau,J}
\end{equation}
of \eqref{eq:primal_to_dual} for all allowed $(\tau,J)$, we will additionally impose this condition at large $\tau$. Specifically, we will take a simultaneous large-spin and large-twist limit, with the ratio parametrized by
\begin{equation}
    \frac{\tau}{J} = \sqrt{\frac{\eta + 1}{\eta - 1}}-1\,.
\end{equation}
The ratios allowed by the unitarity bound $\tau \ge 2$ correspond to $\eta \in [1,\infty)$.

Using the methods of \cite{Caron-Huot:2021enk}, it is shown in \cite{Caron-Huot:2024tzr} that in this limit the $B$, $\hat{B}$, $\Psi$, and $\Phi$ functionals of \secref{sec:dispersive} dominate. They scale as
\begin{equation}
    \lim_{\substack{J\to\infty,\tau\to\infty\\\tau/J\text{ fixed}}} \frac{\hat{B}_{\tau,J}(t)}{2\sin^2\left(\frac{\pi \tau}{2}\right)} = \frac{2^{2\tau+2J - 1/2}}{J^7} \frac{(\eta-1)^3}{\pi\sqrt{1+\eta}} \hat{B}_\infty (\eta, t)\,,
\end{equation}
with analogous formulas for $B$, $\Psi$, and $\Phi$.\footnote{ On the right hand side we have factored out some of the $\eta$-dependence so as to match the conventions of \cite{Caron-Huot:2024tzr}. Note that the precise limiting procedure described there differs from the one here, and theirs has an improved convergence rate. Here we have just given the simplest expression that reproduces the same limit.}

In \cite{Caron-Huot:2024tzr}, a closed-form formula for $\hat{B}_\infty$ is derived:
\begin{equation}
    \hat{B}_\infty(\eta, t) = \frac{2^{12}\times 3}{\pi^2 (6-t)\eta^3} \left(x\, {_2}F_1(4,\tfrac{5-t}{2};\tfrac{3}{2};x) + x^2 \,{_2}F_1(4,\tfrac{7-t}{2};\tfrac{5}{2};x)\right)_{x=1-\eta^{-2}}\,.
\end{equation}
The other limiting functions are given in \cite{Caron-Huot:2024tzr} as integral transforms of this one. For $B_\infty$, we have
\begin{equation}\label{eq:b_regge}
    B_\infty(\eta, v) = \int_{-\infty}^\infty \frac{dy}{4\pi} \underbrace{\frac{\pi^2 (1+y^2)^2}{32 \cosh^2\left(\frac{\pi y}{2}\right)} v^{\frac{iy-3}{2}}}_{[B_v]_y}\hat{B}_\infty(\eta, 5+iy)\,.
\end{equation}
For $\Phi_{\ell_1,\ell_2}$, a similar formula holds with integration kernel
\begin{equation}
    [\Phi_{\ell,\ell+2}]_y = [\Phi_{\ell}]_y - \frac{\Phi_\ell^\infty}{\Phi_{\ell+2}^\infty}[\Phi_{\ell+2}]_y\,,
\end{equation}
where
\begin{equation}
    [\Phi_\ell]_y = -\frac{i\pi^2}{16}(1+y^2)^2 \int_y^\infty \frac{dy'}{\cosh^2(\pi y'/2)}c_\ell a_\ell(5+iy')\,.
\end{equation}
Likewise, for $\Psi_\ell$, we can use \eqref{eq:b_regge} with the integration kernel replaced by
\begin{equation}
    [\Psi_\ell]_y = \frac{\pi^2 (1+y^2)^2}{32 \cosh^2\left(\frac{\pi y}{2}\right)} c_\ell a_\ell(5 + iy)  - \frac{i\pi^2 (1+y^2)^2}{16}\int_y^\infty \frac{dy'\,\tilde{\Psi}_\ell(y')}{\cosh^2\left(\frac{\pi y'}{2}\right)}\,,
\end{equation}
where
\begin{equation}
\begin{split}
    \tilde{\Psi}_\ell(y) ={}& c_\ell a_\ell(5+iy) \left(\frac{1}{2}H_{\frac{-1+iy}{2}} + \frac{1}{2}H_{\frac{-1-iy}{2}} + H_{2\ell+4}-H_{\ell+2} + \frac{\sum_{n=1}^{\ell+2}1/n^2}{2H_{\ell+2}}\right) \\
    &+ \sum_{\substack{j = 0\\j\text{ even}}}^{\ell-2} \frac{\Gamma(2j+6) \Gamma(\ell+3)^2}{\Gamma(2\ell+5)\Gamma(j+3)^2}\frac{c_j a_j(5+iy)}{(l-j)(l+j+5)}\,.
\end{split}
\end{equation}

\subsection{Numerics and extrapolations}\label{app:bootstrap_extrapolation}

We use SDPB \cite{Simmons-Duffin:2015qma} to solve the semidefinite programs appearing in our numerical boostrap analysis. The settings we use in the finite-$N_c$ case and in the $N_c\to\infty$ limit are listed in \tabref{tab:numerical_parameters}.

\begingroup
\renewcommand{\arraystretch}{1.1}
\begin{table}
	\centering
	\begin{tabular}{l|p{4cm}p{4cm}}
        \toprule
		 & Finite $N_c$ & $N_c\to\infty$ \\
		\midrule
		\texttt{precision} & 1024 & 1024 \\
		\texttt{dualityGapThreshold} & $10^{-10}$ & $10^{-30}$  \\
		\texttt{primalErrorThreshold} & $10^{-30}$ & $10^{-50}$ \\
		\texttt{dualErrorThreshold} & $10^{-30}$ & $10^{-50}$ \\
		\texttt{initialMatrixScalePrimal} & $10^{40+\Lambda}$ & $10^{50}$ \\
		\texttt{initialMatrixScaleDual} & $10^{40+\Lambda}$ & $10^{50}$ \\
		\texttt{feasibleCenteringParameter} & $0.1$ & $0.1$ \\		
		\texttt{infeasibleCenteringParameter} & $0.3$ & $0.3$ \\
		\texttt{stepLengthReduction} & $0.7$ & $0.7$ \\
		\texttt{maxComplementarity} & $10^{200+5\Lambda}$ &  $10^{100}$ \\
        \bottomrule
	\end{tabular}
	\caption{Parameters used in SDPB for the plots in this paper. Here $\Lambda$ is the truncation parameter used in the finite-$N_c$ case.}
	\label{tab:numerical_parameters}
\end{table}
\endgroup

\begin{figure}
    \centering
    \includegraphics[width=\linewidth]{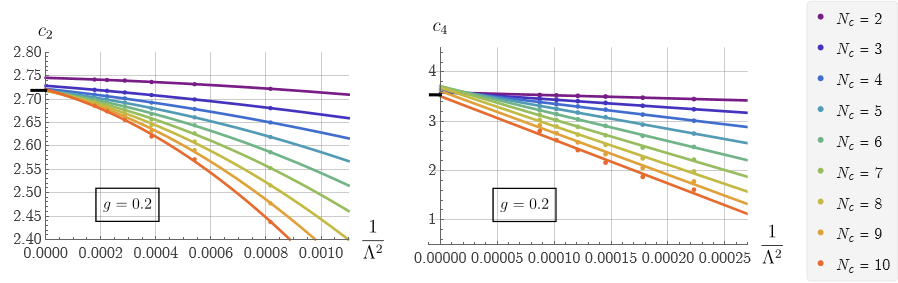}
    \caption{We show how our lower bounds on $c_2$ and $c_4$ at $g = 0.2$ for $N_c = 2,\ldots,10$ converge as we increase the truncation parameter $\Lambda$. For $c_2$, we can reliably extrapolate using a model $a_0 + a_2/\Lambda^2 + a_4/\Lambda^4$, and we see that the extrapolated bounds exhibit weak dependence on $N_c$ as they converge towards the value in the $N_c\to\infty$ limit (indicated with a black tick on the $y$-axis using the value from the Pad\'e approximant in \secref{sec:Pade}). For $c_4$, the convergence is far slower; we extrapolate using a model $a_0 + a_2/\Lambda^2$, and incur larger extrapolation errors, but still we see a qualitatively similar trend in which the bounds converge towards values that have little dependence on $N_c$ at this coupling.}
    \label{fig:extrapolation}
\end{figure}

In Figures~\ref{fig:c2Intro_nonplanar} and \ref{fig:c24_nonplanar}, we have extrapolated lower bounds on $c_2$ and $c_4$ obtained at several values of $\Lambda$ to the $\Lambda\to\infty$ limit. In \figref{fig:extrapolation}, we give examples of this extrapolation for $c_2$ and $c_4$ at $g = 0.2$, for $N_c = 2,\ldots,10$. In the case of $c_2$, we fit a model of the form $a_0 + \frac{a_2}{\Lambda^2} + \frac{a_4}{\Lambda^4}$ to bounds at $\Lambda = 35,43,\ldots,75$. This model fits our bounds very well, and we see that the extrapolated bounds converge towards the value of $c_2$ from the Pad\'e approximation in the $N_c\to\infty$ limit. In the case of $c_4$, we fit a model of the form $a_0 + \frac{a_2}{\Lambda^2}$ to bounds at $67,75,\ldots,107$; these higher values of $\Lambda$ are needed because the bounds converge much more slowly, and even at these values we are not close enough to the $\Lambda\to\infty$ scaling limit to safely introduce another term into the polynomial model. Thus, in this case, there is significantly more extrapolation error; nevertheless, we still see evidence that the bounds at this coupling are converging towards values that have relatively little dependence on $N_c$.

\newpage
\bibliographystyle{JHEP}
\bibliography{mybib}

\end{document}